%% file: dgk.tex
\def\lhls{${\cal L}_h/{\cal L}_s$}
\documentclass[runningheads,natbib]{svjour2}
\bibpunct{[}{]}{;}{n}{}{,} 
\smartqed  
\usepackage{graphicx}
\journalname{Astronomy and Astrophysics Review}
\begin{document}

\title{Modelling the behaviour of accretion flows in X-ray binaries}
\subtitle{Everything you always wanted to know about
accretion but were afraid to ask}

\titlerunning{Accretion flows in XRB}        

\author{Chris Done     \and
  Marek Gierli{\'n}ski \and 
  Aya Kubota
}

\authorrunning{Done, Gierli{\'n}ski \& Kubota} 

\institute{ C. Done and M. Gierli{\'n}ksi \at
              Department of Physics \\
	      University of Durham \\
	      South Road, Durham, DH1 3LE, UK.\\
              \email{Chris.Done@durham.ac.uk,
              Marek.Gierlinski@durham.ac.uk}           \\
           \and
           A. Kubota \at
              Institute of Physical and Chemical Research (RIKEN), \\
	      2-1 Hirosawa, Wako, Saitama 351-019, Japan\\
              \emph{Present address: Electronic Information Systems, Shibaura Institute of Technology \\
307 Fukasaku, Minuma-ku, Saitama-shi, Saitama 337-8570, Japan\\
	      \email{aya@shibaura-it.ac.jp} 
                   }
}

\date{Received: date}
\maketitle

\input{abst/abstract}

\input{intro/intro}
\input{accretion/accretion}
\input{spec/spec}
\input{ns/ns}

\input{jet/jet}
\input{timing/timing}

\input{winds/winds}

\input{se/se}

\input{conclusions/conclusions}
\input{thanks/thanks}

\input{refs/refs}
\end{document}

%% file: abst/abstract.tex
\begin{abstract}
We review how the recent increase in X-ray and radio data from black
hole and neutron star binaries can be merged together with
theoretical advances to give a coherent picture of the physics of
the accretion flow in strong gravity. Both long term X-ray light
curves, X-ray spectra, the rapid X-ray variability and the radio jet
behaviour are consistent with a model where a standard outer
accretion disc is truncated at low luminosities, being replaced by a
hot, inner flow which also acts as the launching site of the jet.
Decreasing the disc truncation radius leads to softer spectra, as
well as higher frequencies (including QPO's) in the power spectra,
and a faster jet. The collapse of the hot flow when the disc reaches
the last stable orbit triggers the dramatic decrease in radio flux,
as well as giving a qualitative (and often quantitative) explanation
for the major hard--soft spectral transition seen in black holes.
The neutron stars are also consistent with the same models, but with
an additional component due to their surface, giving implicit
evidence for the event horizon in black holes. We review claims of
observational data which conflict with this picture, but show that
these can also be consistent with the truncated disc model. We also
review suggested alternative models for the accretion flow which do
not involve a truncated disc. The most successful of these converge
on a similar geometry, where there is a transition at some radius
larger than the last stable orbit between a standard disc and an
inner, jet dominated region, with the X-ray source associated with a
mildly relativistic outflow, beamed away from the disc. However, the
observed uniformity of properties between black holes at different
inclinations suggests that even weak beaming of the X-ray emission
may be constrained by the data.

After collapse of the hot inner flow, the spectrum in black hole
systems can be dominated by the disc emission. Its behaviour is
consistent with the existence of a last stable orbit, and such data
can be used to estimate the black hole spin. By contrast, these
systems can also show very different spectra at these high
luminosities, in which the disc spectrum (and probably structure) is
strongly distorted by Comptonization. The structure of the accretion
flow becomes increasingly uncertain as the luminosity approaches
(and exceeds) the Eddington luminosity, though there is growing
evidence that winds may play an important role. We stress that these
high Eddington fraction flows are key to understanding many
disparate and currently very active fields such as ULX, Narrow Line
Seyfert 1's, and the growth of the first black holes in the Early
Universe.

\keywords{accretion, accretion discs \and black hole physics \and
  X-rays: binaries}
\end{abstract}

%% file: intro/intro.tex
\section{Introduction}
\label{sec:intro}

Understanding black hole accretion is important for its own sake.
However, it also has much wider physical and astrophysical
implications. Luminous accretion flows light up their surroundings,
regions of strongly curved spacetime. The distortions imprinted on
the observed spectrum from special and general relativity depend on
the velocity and geometry of the luminous material as well as on the
shape of the potential (Cunningham 1975; Fabian et al.\ 1989; Fabian
et al.\ 2000).  Thus observational tests of Einstein's gravity
depend on understanding these properties of the accretion flow. On
larger scales, accretion is now realized to be a key to
understanding the growth of structure in the Universe.

Most galaxies have a central black hole, the growth of which is
somehow linked to the growth of the galaxy as a whole, as the mass
of the stellar bulge on kpc scales correlates with the mass of the
black hole on subparsec scales (see e.g. Tremaine et al.\ 2002).
There is strong evidence that the mechanism for this is linked to
feedback from the accretion flow, which reheats gas in clusters of
galaxies, preventing runaway cooling and so regulating star
formation (e.g. di Matteo, Springel \& Hernquist 2005).  The coeval
growth of galaxies and their central black holes is also shown by
the similar redshift for peak activity for both quasars and star
formation at $z\sim 2$ (e.g. Boyle \& Terlevich 1998), while at much
higher redshifts the radiation from these first objects (accreting
black holes and their host galaxies) re-ionizes the Universe (Fan,
Carilli \& Keating 2006).

There are many new observational constraints on accretion flows in
strong gravity which have led to a revolution in our understanding of
these systems.  Much of this has come from the RXTE satellite, with
its huge database of X-ray observations of the accretion flow in
galactic binary systems. Concurrently there has also been a huge
increase in radio data for these objects, tying the jet emission into
the accretion flow. Last, but certainly not least, this has driven
major advances in our theoretical understanding of these
systems. There is now an emerging picture which can provide a
framework for understanding the bewildering variety of spectra and
variability (including the multiple Quasi Periodic Oscillations,
hereafter QPO's) seen from accretion flows in strong gravity.  The
focus of this review is to describe how these current models of the
changing nature and geometry of the accretion flow can give a
framework in which to broadly explain the majority of the behaviour
seen from these objects.

While this review concentrates on stellar remnant compact objects, the
physical processes should be fairly scale invariant. Understanding
accretion in strong gravity anywhere should give us pointers to
understanding it everywhere, with the accretion flows in Galactic
binaries giving a baseline model for the accretion flows onto higher
mass black holes in ultra-luminous X-ray sources, active galaxies and
quasars.

%% file: accretion/accretion.tex
\section{Accretion discs: Spectra and stability}
\label{sec:accretion}

Astrophysical black holes are very simple objects, possessing only
mass and spin.  In steady state the accretion flow should be
completely determined by these parameters, together with the mass
accretion rate.  Much of the dependence on mass can be removed by
scaling the accretion rate to the Eddington accretion rate, as sources
emitting at similar fractions of the Eddington luminosity, $L/L_{\rm
Edd}$, should have similar accretion flows. The only other parameters
which can affect the observational appearance of these systems are the
inclination angle, and any non-stationary effects from variability
behaviour. The problem is then simply to determine how the properties
of the accretion flow depend on this very restricted physical
parameter set.

Mass (and inclination) is potentially observable from the binary
orbit. Spin is more difficult to measure. Unlike mass it does not
leave a discernable mark on the external spacetime at large
distances, making a significant difference only close to the event
horizon. Mass accretion rate is also not trivial to observe.  The
luminosity produced per unit mass accreted depends on the
gravitational potential. This depends on spin, as the last stable
orbit is dragged inwards for a rotating black hole, but it also
depends on the nature of the accretion flow at this point. If
gravity is the only force acting then the gravitational energy
gained in the region between the last stable orbit and the horizon
simply remains with the rapidly infalling material (stress-free
boundary assumption: Novikov \& Thorne 1973). However, there can be
pressure forces and/or magnetic forces which might violate this in a
hot and/or magnetically dominated flow (Abramowicz et al.\ 1988;
Agol \& Krolik 2000). As well as this uncertainty on the total
energy available, there is also the potentially larger uncertainty
attached to what fraction of this is radiated as opposed to powering
winds and/or a jet and/or being carried along with the material
(advected). The observed X-ray luminosity is only a tracer, and
probably a non--linear tracer, of the total mass accretion rate (e.g
Kording, Fender \& Migliari 2006).

\subsection{Steady State Accretion Discs}
\label{sec:disc}

The structure of the accretion flow is derived by balancing
gravitational heating against cooling so depends at the outset on
which set of heating/cooling processes are assumed to be important.
In the canonical approach of Shakura \& Sunyaev (1973), the
assumption is that viscous stresses convert the gravitational
potential to heat, and that the heat released at a given radius is
radiated locally (so advection, winds and jet cooling are
neglected). The resulting spectrum is rather robust if the energy
thermalizes, so at given radius, $r=R/R_g$ (where $R_g$ is the
gravitational radius $GM/c^2$) the disc emits as a (quasi) blackbody
of temperature $T(r)\propto r^{-3/4}$. At smaller radii there is
more luminosity generated as gravity is stronger, and this is
dissipated over a smaller area, so the temperature increases
inwards.  This solution gives a geometrically thin disc except at
luminosities close to Eddington.

The assumption that the energy thermalizes means that the spectrum
is (to zeroth order) independent of the details of the viscosity
mechanism. Thus the models can predict the observed disc spectra
using only phenomenological descriptions of the stresses, most
notably the Shakura--Sunyaev $\alpha$ prescription, where the shear
stresses, $t_{r\phi}$, are assumed to be directly proportional to
total pressure $P_{tot}=P_{gas}+P_{rad}$, so $t_{r\phi}=\alpha
P_{tot}$. A key breakthrough in the last 10 years has been
identifying the physical origin of the stress as the magnetic
rotational instability (MRI: see e.g. the review by Balbus 2005), a
self--sustaining dynamo process. Numerical simulations including
this self--consistent heating are still in their infancy but already
yield significant insights into the nature of the jet (see
Section~\ref{sec:jet}), and promise much future progress (see e.g.
Balbus 2005).

\subsection{Stability and Time Dependence}
\label{sec:stability}

The Shakura--Sunyaev disc solution described above is a global solution, i.e.
it assumes that the mass accretion rate is {\em constant} with
radius. This is not necessarily true, and in fact Shakura--Sunyaev
accretion discs are subject to two major instabilities, one
connected to the ionization of hydrogen which is triggered at fairly
low luminosities and controls the long term outburst behaviour of
the accreting sources, and the other due to radiation pressure,
which should occur at higher luminosities.

The accretion flow stability at a given radius is a function of the
heating and cooling mechanisms chosen. Generically, the flow is {\em
thermally} unstable if small perturbations in temperature grow i.e.
if a small increase in temperature causes a further rise in
temperature. Similarly, the flow is {\em viscously} (or secularly)
unstable if a small increase in mass accretion rate, $\dot{M}$, leads
to a larger increase in $\dot{M}$ so that the disc is eaten away at
that radius.  The timescale for these instabilities to grow are
$t_{th}\sim \alpha^{-1} t_{dyn}$ and $t_{visc}\sim \alpha^{-1}
(H/R)^{-2} t_{dyn}$, where $H$ is the vertical scale height of the disc, and
the dynamical (orbital) timescale, $t_{dyn}= 4.5\ (m/10)\
(r/6)^{3/2}$~ms for a Schwarzschild black hole of mass $m $M$_\odot$
(see e.g. Frank, King \& Raine 2002; hereafter FKR, and Kato, Fukue \&
Mineshige 1998, hereafter KFM).  For a thin disc, the thermal
timescale is much faster than the viscous one as $H/R\ll 1$. Hence the
thermal instability takes place without there being time to change the
amount of material in the disc at that radius if $H/R \ll
1$. Conversely, both timescales are more or less equal for a
geometrically thick flow so the disc structure can respond to the
changing temperature (FKR, KFM).

\subsubsection{Hydrogen Ionization Instability}
\label{sec:hion}

A Shakura--Sunyaev disc at low mass accretion rates is unstable both
thermally and viscously for temperatures around that associated with
the ionization of hydrogen, i.e.  $10^{4-5}$~K. For low mass
accretion rates, the temperature is low so the material is mostly
neutral. The opacity is also low, much lower than predicted by
free--free processes as the electrons are bound (see e.g. Cannizzo
\& Reiff 1982). However, the opacity rises extremely steeply between
$10^4$--$10^5$~K, so that a small increase in temperature of the
material produces a huge increase in opacity as the highest energy
photons on the Wien tail of the thermal distribution are able to
ionize some of the hydrogen in the disc. This means that these
photons are absorbed, so their energy stays in the disc, increasing
its temperature, rather than escaping to cool the material.  More of
the photons can ionize H, so more are trapped in the disc,
increasing the temperature still further. This thermal runaway only
stops when H is almost completely ionized, so that there is no
longer a rapid jump in opacity of the disc to give the rapid
decrease of disc cooling and hence the rapid increase in disc
temperature (KFM; FKR).

The thermal instability then triggers the viscous instability, as the
increased temperature means an increased mass accretion rate through
the annulus (as the viscous stresses are proportional to pressure in
the Shakura--Sunyaev $\alpha$ prescription).  This is now larger than
the input mass accretion rate (which was just enough to take the
temperature of the annulus to $10^4$~K to trigger the instability), so
the material in the disc is eaten away at this radius. The pressure
decreases, so the heating decreases, so the temperature decreases
until hydrogen is able to recombine. This triggers the thermal
instability again, this time with the cooling running away until H is
mostly neutral again at temperatures below $10^4$~K. But this low
temperature gives lower mass accretion rate than input so the disc
builds up again (KFM; FKR).

\begin{figure}
\begin{center}
\begin{tabular}{cc}
\includegraphics[width=0.4\textwidth,bb=500 630 100
150,clip,angle=90]{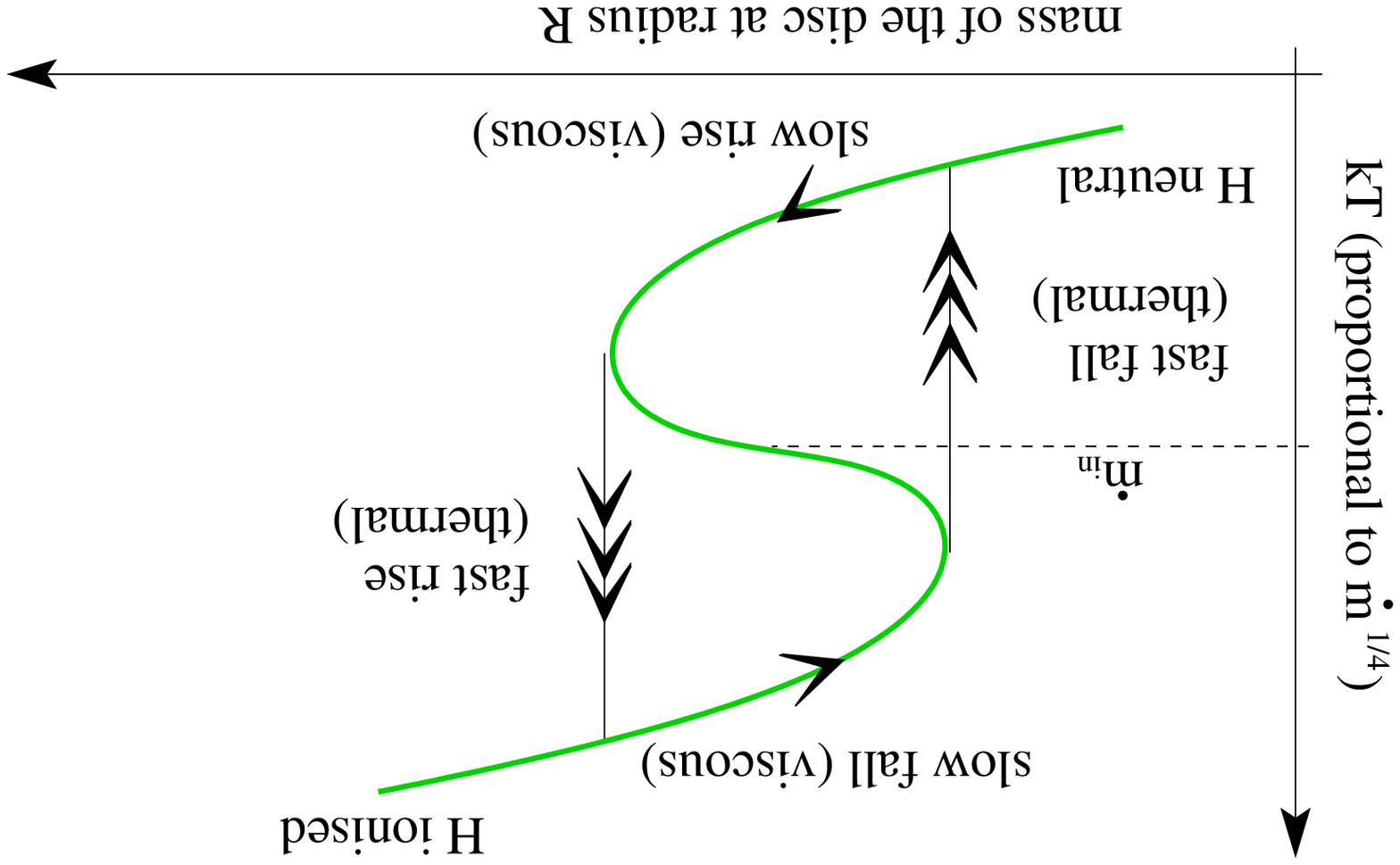} &
\includegraphics[width=0.4\textwidth,angle=0,bb=300 470 550
720,clip]{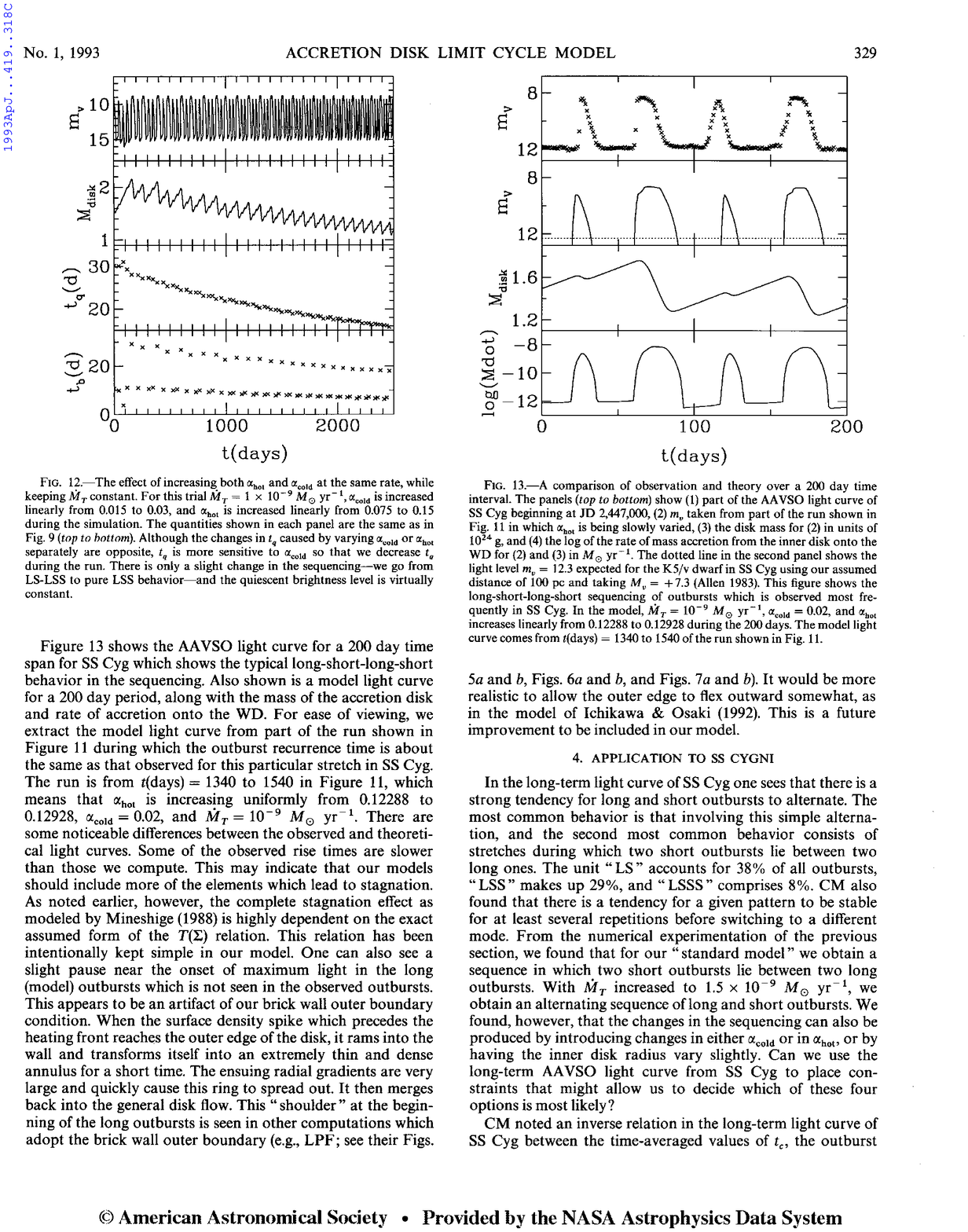}
\end{tabular}
\end{center}
\caption{The hydrogen ionization instability. The figure on the left
shows the local effect of the instability, where the
mass accretion rate through the disc jumps discontinuously
at a given radius. The right hand panel shows results from
theoretical models of the effect of this
local instability on the global disc structure. The entire disc
alternates between periods of outburst and quiescence, matching very
well to the observed behaviour of dwarf novae such as SS Cyg, as shown
in the top panel. From Cannizzo (1993).}
\label{fig:scurve}
\end{figure}

This description of the H ionization limit cycle is a purely local
instability, i.e. at a single radius in the disc. However, the
discontinuous jump in temperature and hence in mass accretion rate
means that it will affect the next annulus of disc material. This
can have a global impact on the disc structure if the difference in
mass accretion rate between the ionized and neutral material is
large enough. The whole disc structure can then do limit cycle
behaviour between a Shakura--Sunyaev disc with larger mass transfer
rate than from the companion, so the whole disc is being eaten away
on the viscous timescale, to a quiescent disc where H is mostly
neutral everywhere.  The quiescent disc structure is very unlike
that predicted by the Shakura--Sunyaev equations as it does not have
constant mass accretion rate at all radii. Instead it has more or
less constant temperature, so has mass accretion rate decreasing at
smaller radii (e.g. the review by Lasota 2001).

The disc instability model (DIM) of the observed outbursts can be
summarized as follows (e.g. Lasota 2001). The quiescent disc builds
up from steady mass transfer from the companion. Eventually this
gets hot enough to trigger the H ionization instability at some
radius. The increased mass accretion rate then increases the mass
accretion rate through the next radius in the disc. This triggers
the H ionization instability, and a heating wave propagates inwards
(and outwards) through the entire disc.  This increased mass
accretion rate is maintained until the outer disc temperature dips
below the H ionization temperature. This propagates a cooling wave,
which switches the entire disc back into the quiescent state after
only a small fraction of the mass in the disc has been accreted.
Such alternating periods of disc outbursts and quiescence are seen
in the dwarf novae subclass of disc accreting white dwarf systems
(Cataclysmic Variables). The characteristic light curve of the disc
is shown in Fig.~\ref{fig:scurve} (KFM, FKR)

However, time-dependent disc codes can only produce this behaviour if
the viscosity is dramatically different above and below the H
ionization instability. The observed outburst behaviour of accreting
white dwarfs (simpler laboratories of disc physics as they are less
extreme than neutron stars and black holes: see e.g.  reviews by Osaki
1996 and Lasota 2001) can only be produced if the scaling of the
stress with total pressure changes by a factor 5--10 (Smak 1984). This
may have a physical origin in the
behaviour of the MRI. H is neutral in quiescence, so
there are very few free electrons to anchor the magnetic fields. The
MRI may be suppressed, and much less efficient hydrodynamic processes
such as spiral waves, probably dominate the stress (Gammie \& Menou
1998). This may only be the case in the Galactic binary
discs. Supermassive black holes have Shakura--Sunyaev discs with much
lower density, so recombination is not so effective at suppressing
free electrons from e.g. potassium, iron etc., so even when H is
mostly neutral then the MRI may still be able to operate. The disc
instability is then purely local, not global, and does not lead to the
same outburst/quiescence behaviour (Menou \& Quaeterat 2001) despite
the disc temperature crossing the H-ionization regime (Siemiginowska,
Czerny, \& Kostyunin 1996; Burderi, King \& Szuszkiewicz 1998).

\begin{figure}
\begin{center}
\begin{tabular}{cc}
\includegraphics[clip=true,width=0.6\textwidth,angle=0]
{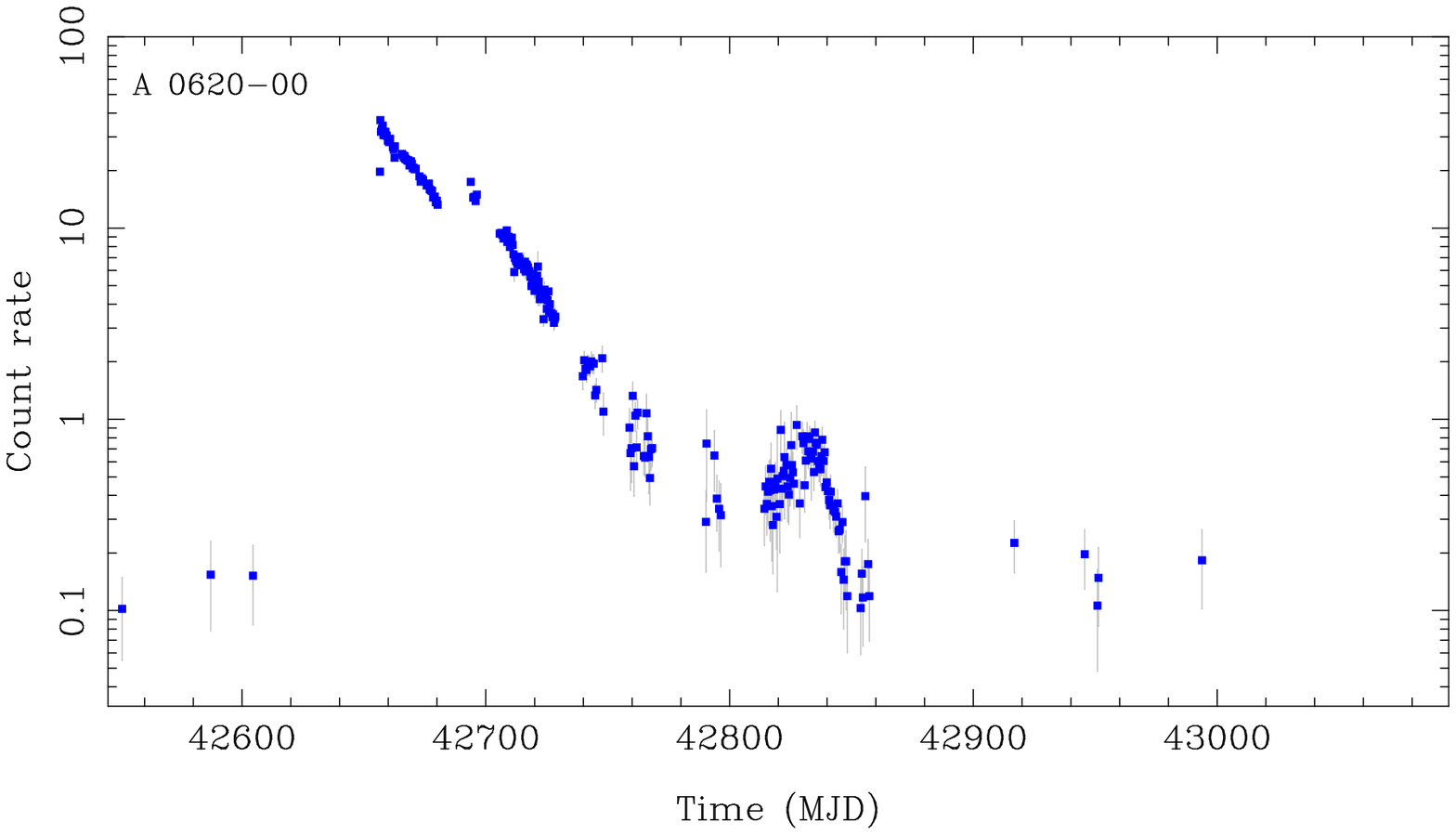} &
\includegraphics[clip=true,width=0.37\textwidth,angle=0]
{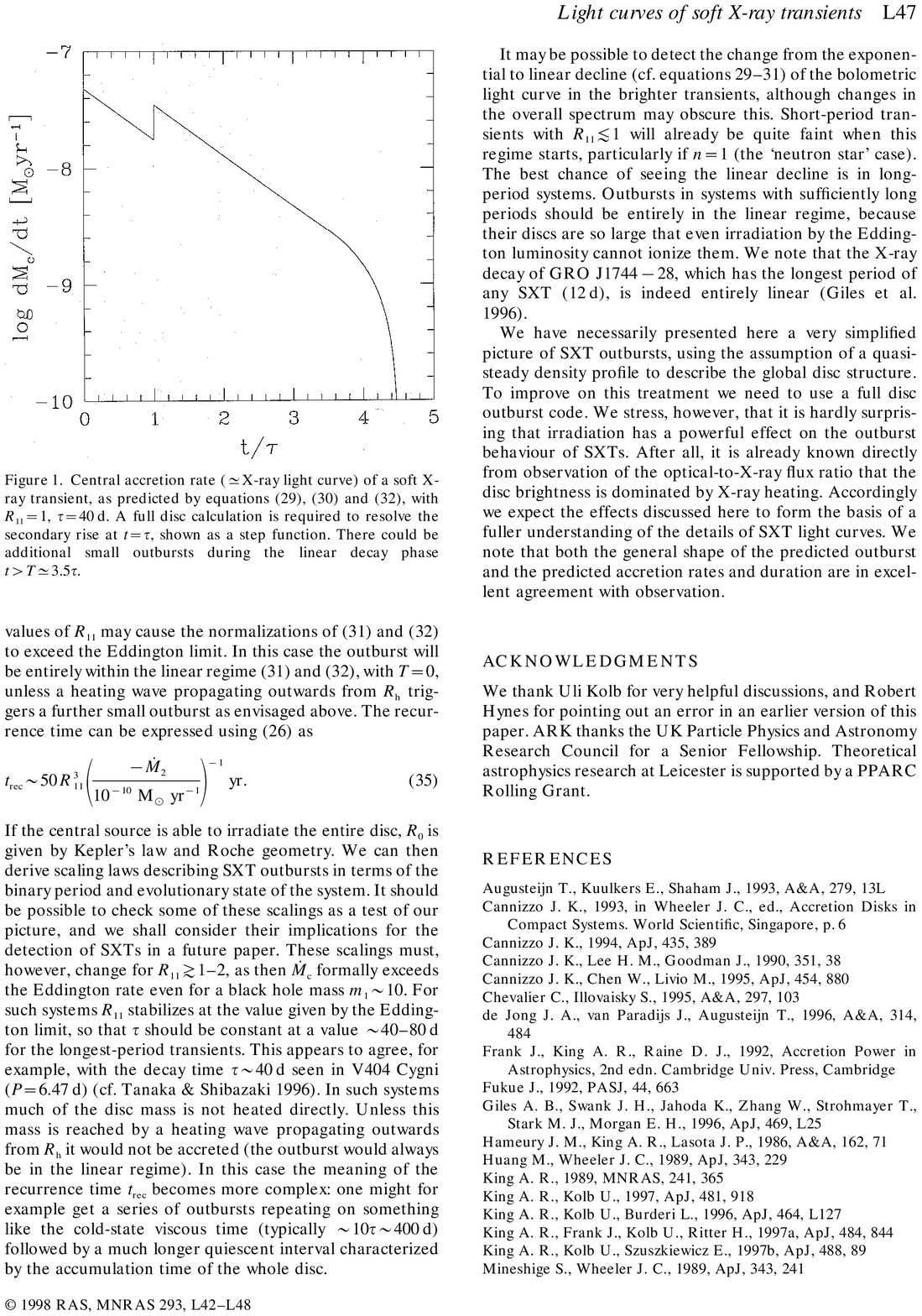}
\end{tabular}
\end{center}
\caption{The left panel shows the
X-ray light curve of the BHB A~0620--00 from the Ariel V all
sky monitor. The fast rise, followed by an exponential decay (with a
reflare) is well matched by theoretical models of the hydrogen
ionization instability controlling the fast rise, with irradiation
preventing the disc switching back into quiescence, leading to
the exponential decay when the irradiation is strong enough to keep
the whole disc above the hydrogen ionization temperature, and then
to a linear decay as this region shrinks in size
(From King \& Ritter 1998). }
\label{fig:fred}
\end{figure}

The disc instability in neutron star (hereafter NS) and black hole binaries
(hereafter BHB) gives very
different behaviour to that of the white dwarfs. Fig.~\ref{fig:fred}
shows an example of this. While the quiescent disc and fast rise to
outburst can be modelled by the same codes as work for the white
dwarf discs, they cannot produce the quasi--exponential decay (KFM;
Lasota 2001). This is because the hugely luminous
inner disc in these X-ray binary systems means that irradiation is
also important, which can keep the disc hot even at large radii (van
Paradijs 1996).  If the irradiation is strong enough to prevent H
recombining then the disc is kept on the hot branch, with a higher
mass accretion rate than supplied by the companion. This eats away
the disc, pulling its temperature down, reducing the mass accretion
rate through the disc which decreases the X-ray irradiation.  This
gives the characteristic exponential decay if the whole disc is
irradiated, but eventually the irradiation becomes weak enough that
the outer disc temperature goes below the hydrogen ionization and
switches to the cool branch. However, the cooling front is prevented
from propagating inwards on the viscous timescale as its innermost
radial extent is set by the radius of the irradiated region. This
leads to linear decays (King \& Ritter 1998; Lasota 2001).

Since we observe such exponential decays (see Fig~\ref{fig:fred}),
the disc must be irradiated. The optical light from discs in X-ray
binaries is also much brighter than predicted by the
Shakura--Sunyaev disc, again pointing to the importance of
irradiation (van Paradijs \& McClintock 1994).  However, the disc
shape is probably convex so that it self-shields the outer disc for
irradiation from the bright, inner regions (Dubus et al.\ 1999). One
potential way around this is if there is tenuous material above the
disc which can intercept some small fraction of the X-ray flux and
scatter it down onto the disc (Dubus et al.\ 2001; Lasota 2001).
Such material can clearly be seen in the small subset of systems
which are seen almost edge on. In these cases the disc and/or
companion star can completely obscure a direct view of the bright
inner disc and intrinsic hard X-ray coronal emission (and the
boundary layer between the disc and surface in the neutron stars),
so that the X-ray source is only seen via scattering in an accretion
disc corona (ADC: e.g. Frank, King \& Lasota 1987; FKR; see
Section~\ref{sec:winds}).  The name is somewhat confusing as this
material is in all probability some sort of large scale wind/outflow
from the outer disc (Begelman, McKee \& Shields 1983) and so is
physically separate from the intrinsic hard X-ray emitting region
which is small, and close to the central object but is also
sometimes termed `the corona'.

\begin{figure}
\begin{center}
\includegraphics[clip=true,width=0.9\textwidth,angle=0]
{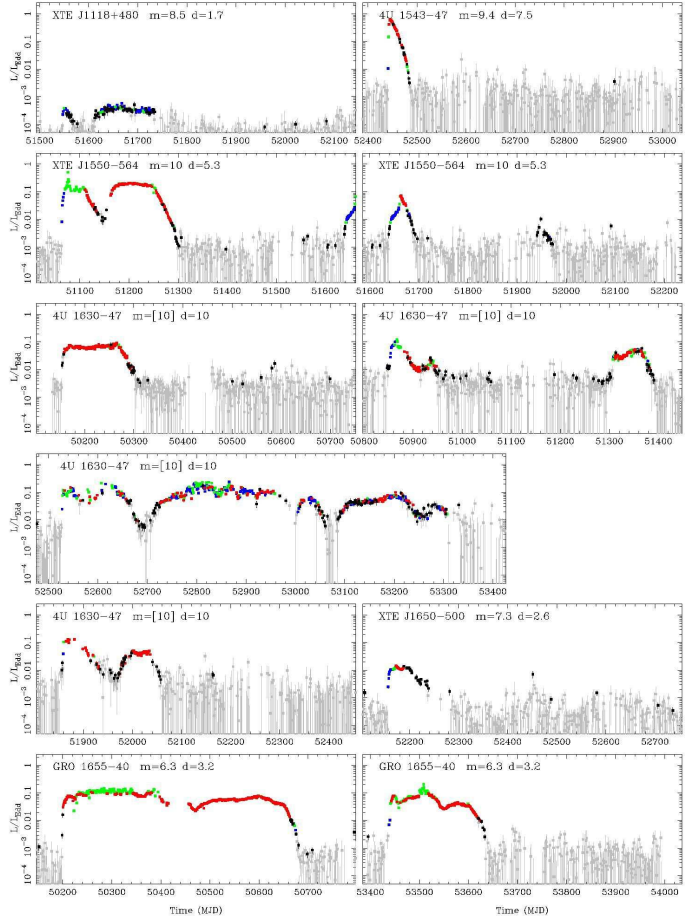}
\end{center}

\caption{{\it RXTE All Sky Monitor} light curves of BHB LMXB showing
the outburst behaviour. Blue, green and red points indicate that the
spectral hardness corresponds to the hard, very high and soft
states, respectively, while black indicates that the uncertainties
on the colour are too large to assign a state. Grey points
correspond to non-detections (3$\sigma$). $m$ is the mass (in
M$_\odot$) and $d$ is the distance (in kpc) used for Eddington
luminosity estimates (see Gierli{\'n}ski \& Newton 2006 for
details).} \label{fig:asm_trans1}
\end{figure}

\begin{figure}
\begin{center}
\includegraphics[clip=true,width=0.9\textwidth,angle=0]
{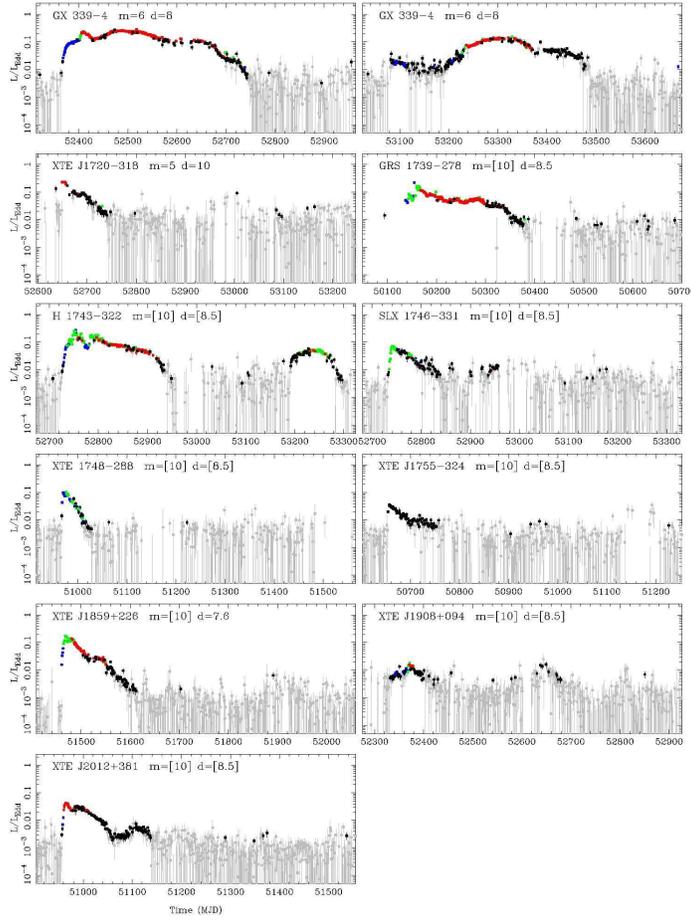}
\end{center}
\caption{As for Fig.~\ref{fig:asm_trans1}}
\label{fig:asm_trans2}
\end{figure}

\begin{figure}
\begin{center}
\includegraphics[clip=true,width=0.9\textwidth,angle=0]
{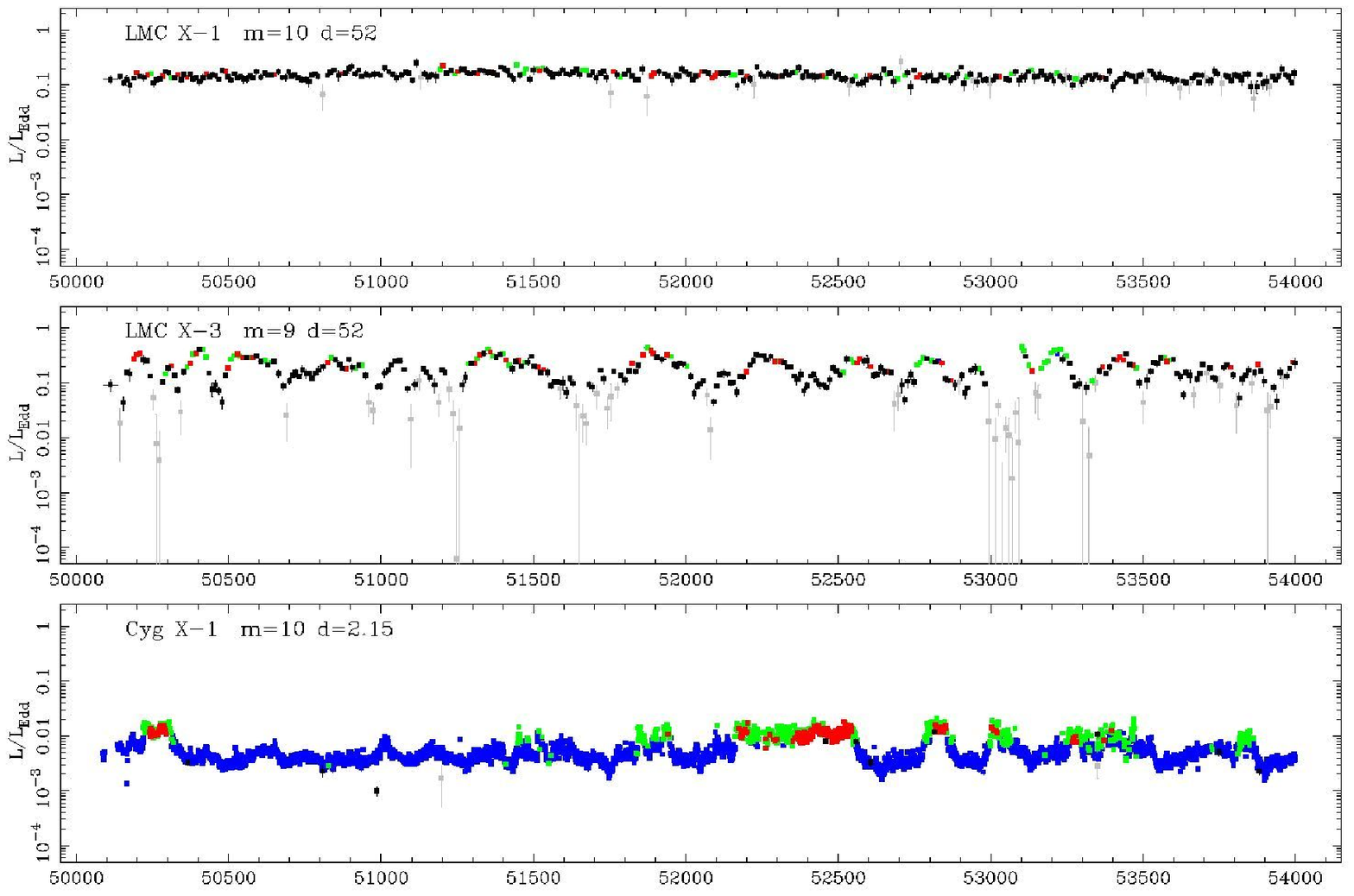}
\end{center}
\caption{As for Fig.~\ref{fig:asm_trans1} but for the BHB HMXB. These
are all persistent sources due to the high mass transfer rate from the
high mass companion star which keeps the whole disc above the hydrogen
ionization instability point. However, although they do not show
dramatic outbursts, they do show variability,
which can lead to spectral transitions.
Note the change in X-axis scale from the previous ASM plots.}
\label{fig:asm_per}
\end{figure}

These models of irradiated discs can also explain the distinction in
disc stability between BHB in Low Mass X-ray Binaries (hereafter
LMXB) and High Mass X-ray Binaries (HMXB) powered (mostly) by Roche
lobe overflow such as Cyg X-1, LMC X-1 and LMC X-3. For a given mass
black hole, a high mass companion star typically has a much higher
mass accretion rate than a low mass one, giving a higher outer disc
temperature. This effect more than compensates for the lower
temperature expected from the larger outer disc radius from the
larger orbit, so the outer edge of the disc is generally above the H
ionization instability in HMXB (van Paradijs 1996). All of the 3
known BHB persistent sources (Fig.~\ref{fig:asm_per}) are HMXB,
though it is also possible for LMXB BHB to be persistent at fairly
low luminosities (Menou, Narayan \& Lasota 1999).

\begin{figure}
\begin{center}
\includegraphics[clip=true,width=0.9\textwidth]
{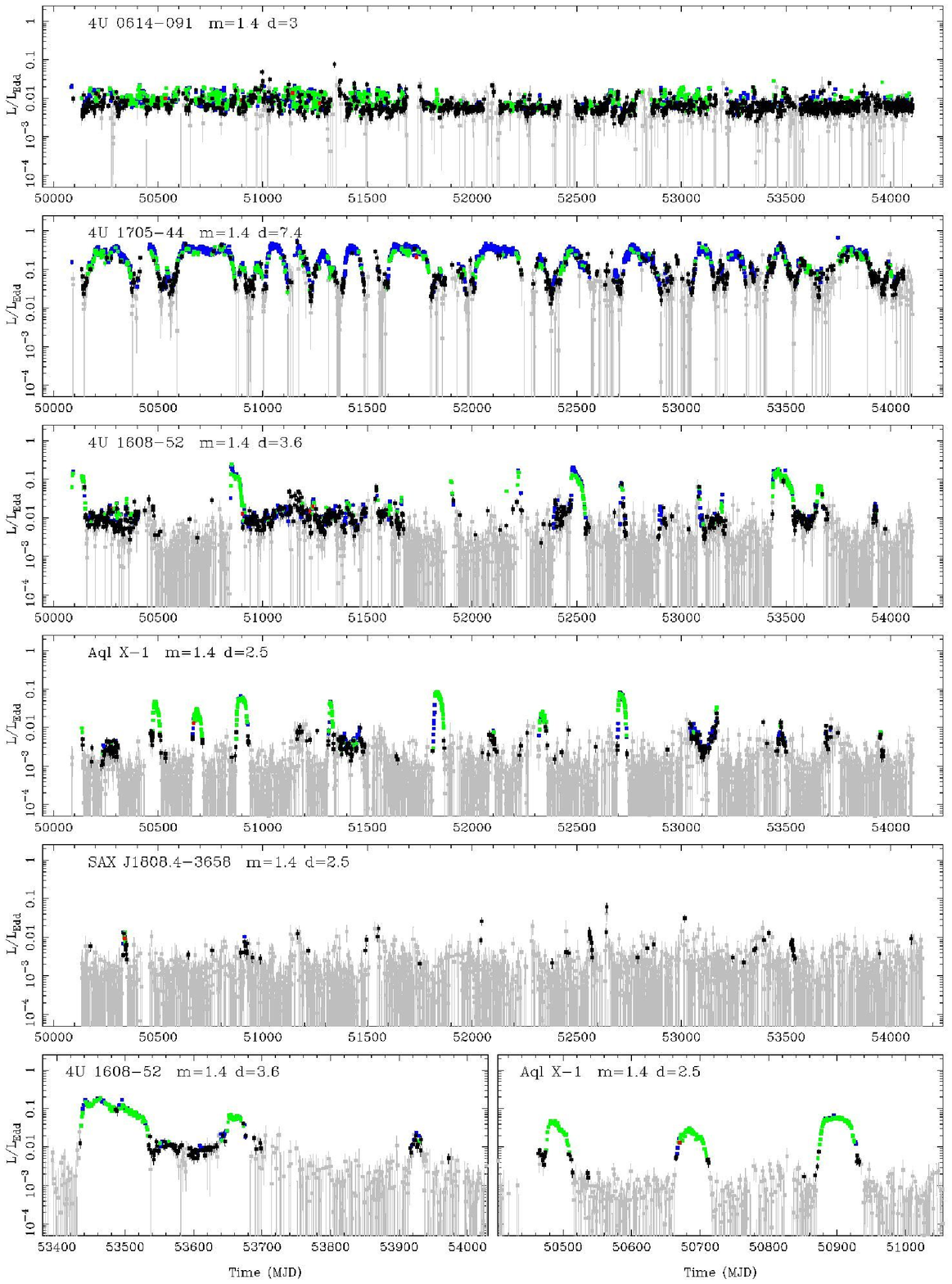}
\end{center}
\caption{As for Fig.~\ref{fig:asm_trans1} but for the neutron star
LMXB systems (atolls). In contrast to the BHB, most of the known
systems are persistent, so are shown on the same X-axis scale as in
Fig.~\ref{fig:asm_per}. Again, they can show smaller scale
variability, and this can trigger state transitions.  However, there
are a few transients, e.g.  4U 1608--52 and Aql X-1, where the
secondary is evolved into a (sub)giant, or the millisecond
pulsars such as SAX J1808.4--3658 where the companion is almost completely
accreted. Selected individual outbursts from  4U 1608--52 and Aql X-1
are shown on the same scale as
the transient LMXB BHB outbursts in Figs.~\ref{fig:asm_trans1} and
\ref{fig:asm_trans2} to highlight the similarities in behaviour.}
\label{fig:ns_lc}
\end{figure}

Similarly, these disc models can also explain the difference in
stability properties between the neutron star and BHB LMXB. A
neutron star primary has lower mass than a black hole, so for a
given mass companion overflowing its Roche lobe requires a smaller
binary orbit.  Smaller orbital separation means a smaller disc due
to tidal truncation so the outer disc does not extend to such low
temperatures, so is more likely to be stable. Many neutron star LMXB
are persistent sources while {\em all} known LMXB BHB are transient
(King, Kolb \& Burderi 1996; Dubus et al.\ 1999). Fig.
\ref{fig:ns_lc} shows light curves of five neutron star systems. Two
of them, 4U 1608--52 and Aql X-1, are recurrent transients, where
the systems are known to have have large orbital periods. This
requires that the companions are somewhat nuclear evolved in order
that they fill their Roche lobes, and the resulting large disc can
go unstable on its outer edge (King et al.\ 1997). This is
presumably also the case for the other known NS transients, apart
from the millisecond pulsars, where the disc is unstable because the
orbit is so small rather than so large. These have companions where
the mass transfer rate is low enough that the outer disc edge
temperature drops below the H-ionization point even though the disc
is tiny (Chakrabarty \& Morgan 1998; Ergma \& Antipova 1999).

While much is now understood, the detailed shape of the light curves
of many of the systems still hold some puzzles. In both NS and BHB,
the accretion rate through the disc can be {\em variable} even in
persistent systems e.g. 4U 1705--44 and Cyg X-1, while the outburst
light curves of many transients, especially the longer period
systems, can be much more complex than simple exponential or linear
decays (see Figs.~\ref{fig:asm_per}, \ref{fig:ns_lc}). The interplay
between the irradiation controlled H-ionization instability and the
tidal instability and/or an enhanced mass accretion rate from the
irradiated companion may go some way to explaining the variety of
light curve behaviour shown in Figs.~\ref{fig:asm_trans1},
\ref{fig:asm_trans2} (e.g. the review by Lasota 2001). Nonetheless,
the match between the disc theory and the observed long timescale
light curves provide compelling evidence for something very like a
Shakura--Sunyaev outer disc in these systems.

\subsection{The radiation pressure instability}
\label{sec:radpress}

The Shakura--Sunyaev disc is also unstable at high mass accretion
rates at small radii due to the rapid increase in heating as the
disc goes from being gas pressure dominated ($P_{\rm gas}= nkT$) to
being radiation pressure dominated ($P_{\rm rad}=\sigma T^4/3c$).  A
small increase in temperature at this point causes a large increase
in pressure, and hence a large increase in heating since the
stresses are assumed to be $\propto P_{\rm tot}$. This gives a large
increase in temperature which is not balanced by any correspondingly
large decrease in opacity to increase cooling, so there is runaway
heating.

In the Shakura--Sunyaev disc equations there is no high mass
accretion rate, high temperature stable solution as in the H
ionization instability, so the runaway predicts complete disruption
of the disc. However, incorporating radial advection into the
equations gives an additional cooling process as some fraction of
the energy is carried along with the flow into the next annulus as
well as there being energy losses through radiation (Abramowicz et
al. 1988). This can balance the increased heating, giving a stable
upper branch and so the possibility of the same sort of
thermal-viscous limit cycle as discussed above for the H ionization
instability. However, there are several key differences.  Firstly,
the disc typically has $H/R\sim 1$ on the advection dominated slim
disc branch. Thus viscous and thermal timescales are not so
different, so the amount of material in the annulus changes with the
temperature. Secondly, there is probably no change in the viscosity
mechanism (MRI for both the gas pressure and radiation pressure
dominated branches). The combination of these two effects mean that
the local instability can propagate, but it is only the radiation
pressure dominated part of the disc which becomes globally unstable
rather than the entire disc.

\begin{figure}
\begin{center}
\begin{tabular}{cc}

\includegraphics[width=0.4\textwidth,bb=500 630 100
150,clip,angle=90]{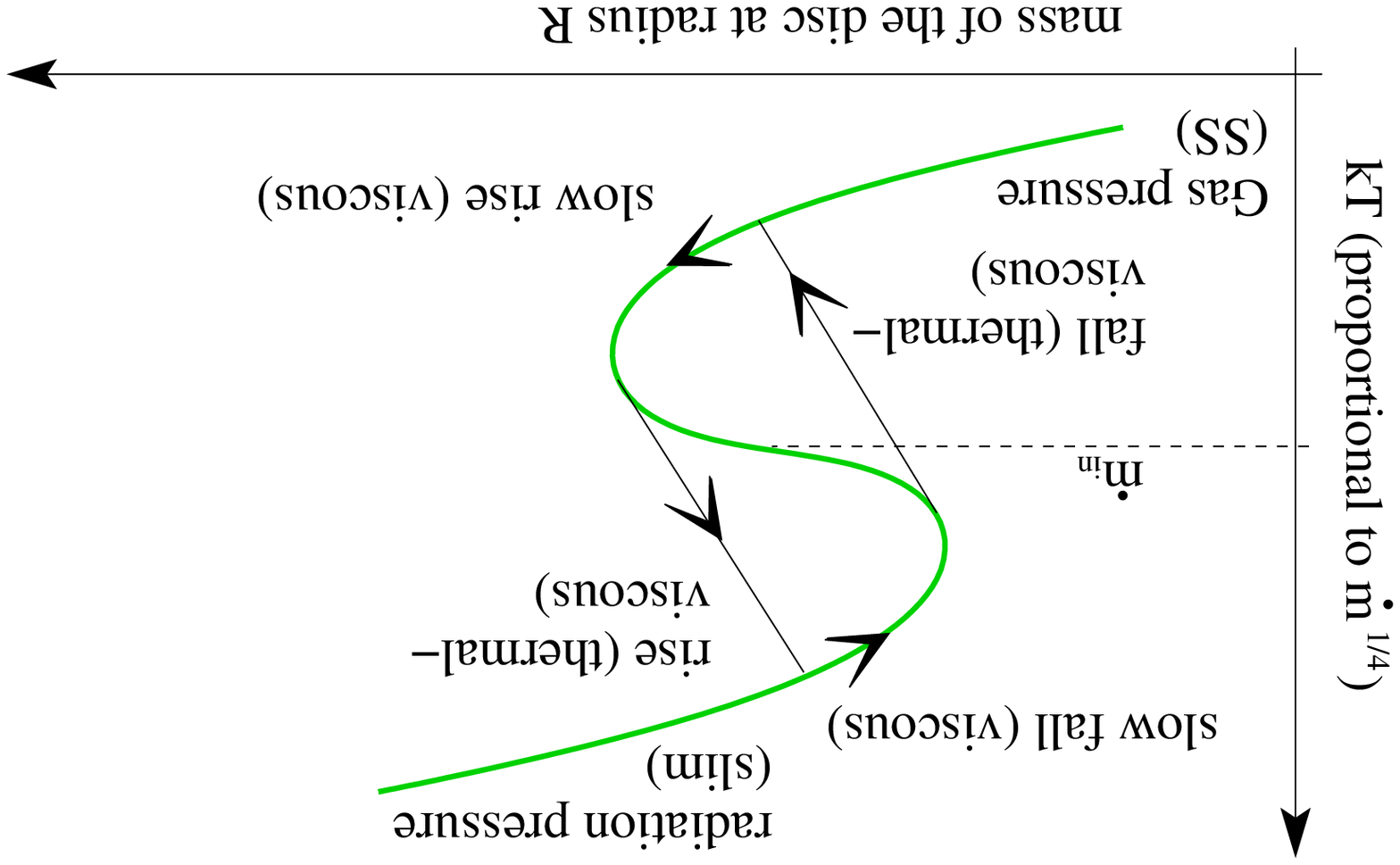} &
\includegraphics[clip=true,width=0.45\textwidth,angle=0]
{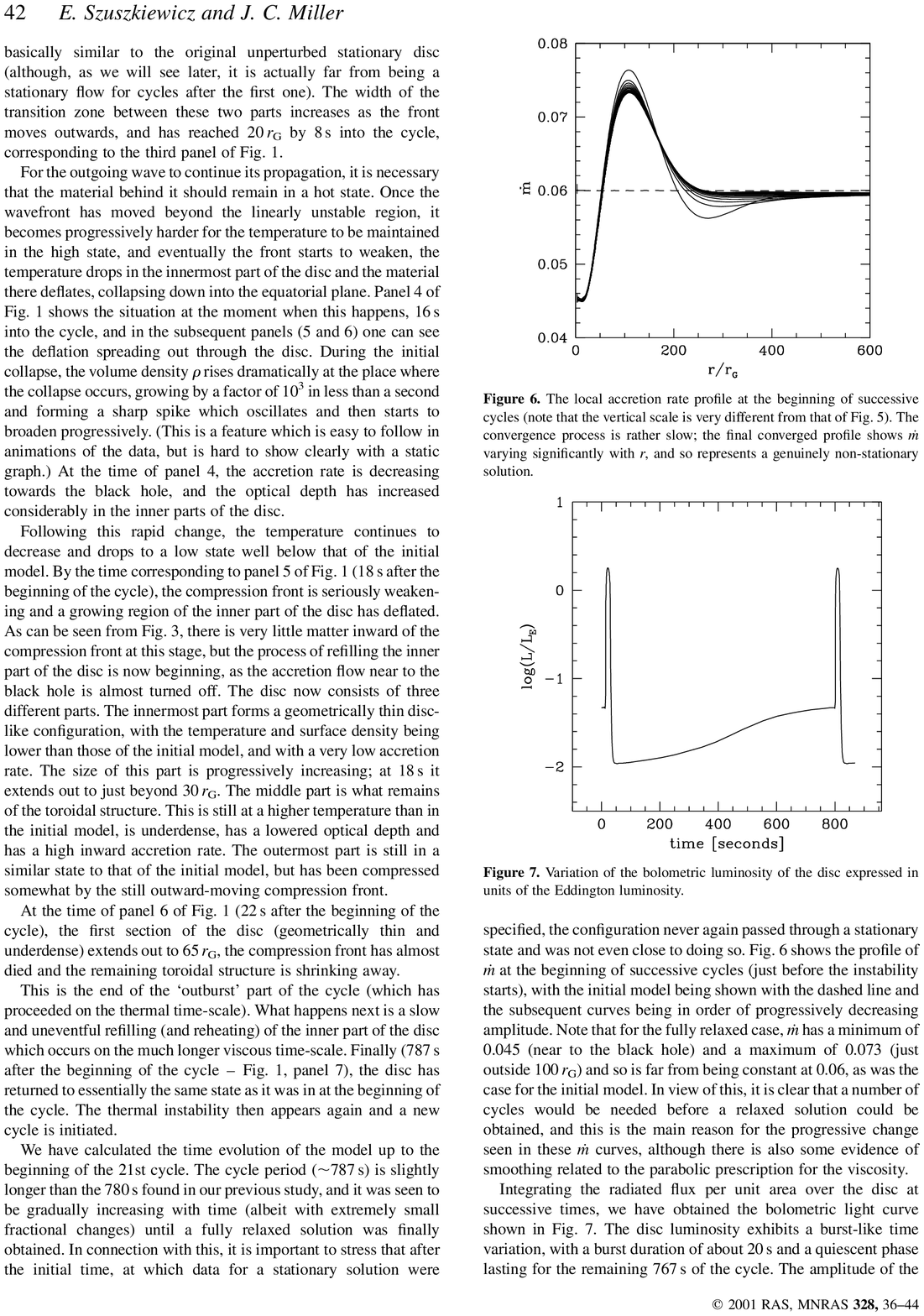}
\end{tabular}
\end{center}
\caption{The radiation pressure instability. The figure on the left
shows the local effect of the instability, where the
mass accretion rate through the disc jumps discontinuously
at a given radius. The right hand panel shows results from
theoretical models of the effect of this
local instability on the global disc structure with
standard Shakura--Sunyaev viscosity ($\alpha=0.1$), for
$L/L_{\rm Edd}=0.06$. No BHB shows anything like this behaviour at this
luminosity (From Szuszkiewicz \& Miller 2001). }
\label{fig:rad_cycle}
\end{figure}

However, unlike the H-ionization instability for the outer disc,
there is very little evidence that the radiation pressure
instability exists as described here. Disc with the classic
Shakura--Sunyaev stress prescription become radiation pressure
unstable at around $\ge 0.06L_{\rm Edd}$, and produce limit cycles
(Honma, Kato \& Matsumoto 1991; Szuszkiewicz \& Miller 2001; Merloni
\& Nayakshin 2006). The inner disc is where most of the
gravitational energy is released, so this instability has a dramatic
effect on the light curve (see Fig.~\ref{fig:rad_cycle}). Yet the
spectra of BHB show {\em
stable} disc spectra  up to at least $0.7L_{\rm Edd}$ (see
Section~\ref{sec:disc}).

This plainly shows that the classic Shakura--Sunyaev stress
prescription is wrong! Alternative stress prescriptions with heating
proportional to gas pressure only are stable everywhere (Stella \&
Rosner 1984). However, the super-Eddington BHB GRS 1915+105 {\em
does} show something which looks very much like a limit cycle in
some of its light curves (see Section~\ref{sec:se}).  Thus it seems
more likely that the effective stress scales somewhat more slowly
with temperature than predicted by radiation pressure, but somewhat
faster than predicted by gas pressure alone.  Analytic estimates of
the effective stress produced by the MRI indicate that it may be
more appropriate to describe the heating as proportional to the
geometric mean of the gas and total pressure (Merloni 2003). Such
discs are locally unstable at $L/L_{\rm Edd}\sim 0.3$, but the
effects of advection from neighbouring annuli mean that this is
damped out, making this prescription stable to somewhat higher
luminosities, around $L/L_{\rm Edd}\sim 0.4$ (Honma, Kato \&
Matsumoto 1991; Merloni \& Nayakshin 2006). This seems very
promising, especially as the exact onset of the instability is quite
sensitive to small changes in the effective $\alpha$ prescription
around this point (KFM). Thus a small tweak in stress scaling from
the geometric mean could probably give the stability limit beyond
$\sim 0.5$ as observed, followed by the onset of the radiation
pressure instability at higher luminosities required to explain the
unique limit cycle like behaviour of GRS~1915+105 (see e.g. the
review by Fender, \& Belloni 2004). This is the only BHB in our
Galaxy which spends significant amounts of time at luminosities
near Eddington, so plausibly its unique variability is connected to
this being the only source with high enough luminosity to trigger
the radiation pressure instability (Done, Wardzi{\'n}ski \&
Gierli{\'n}ski 2004, see Section~\ref{sec:se}).

%% file: spec/spec.tex
\section{The inner accretion flow}
\label{sec:spec}

\subsection{Spectral states in Cyg X-1}
\label{sec:cygx1}

\begin{figure}
\begin{center}
\includegraphics[clip=false,width=0.7\textwidth,angle=0]
{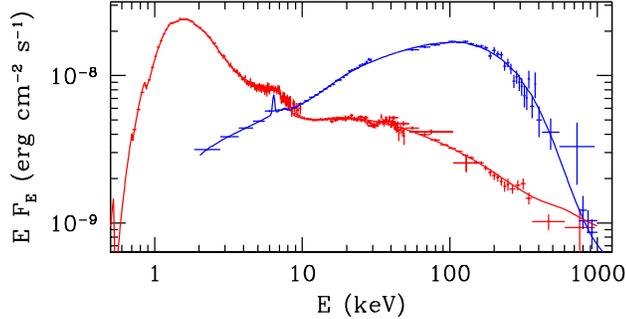}
\end{center}
\caption{X/$\gamma$-ray spectral states in Cyg X-1.
Classical soft (red) and hard (blue) states
taken from Gierli{\'n}ski et al.\ (1999).}
\label{fig:cygx1_states}
\end{figure}

The irradiation controlled H-ionization instability explains much of
the observed long term light curve behaviour in both neutron stars
and black holes.  This is clear evidence for the outer disc having a
structure at least something like the (time dependent)
Shakura-Sunyaev disc models. However, they do {\em not} fully
explain the spectra. Disc spectra simply change in luminosity and
temperature, but maintain their robust, quasi-thermal shape. By
contrast, the observed spectra vary tremendously in shape. This has
been known since the early days of X-ray astronomy, from the
behaviour of the first well studied BHB, Cyg X-1.  This shows two
very different types of spectra, illustrated in
Fig.~\ref{fig:cygx1_states} (Gierli{\'n}ski et al.\ 1999), plotted
in $\nu F_\nu$ so a peak indicates the characteristic photon energy
of the source output. As shown in this figure, the alternatively
named `high', `soft' or `thermal dominant' state is characterized by
a dominant soft component below $\sim10$~keV, accompanied by a
complex non-thermal tail of emission extending to 500~keV and beyond
(Gierli{\'n}ski et al.\ 1999; McConnell et al.\ 2000). The soft
component is interpreted as thermal emission from an optically thick
accretion disc around the black hole, as it can be well reproduced
by a multicolour disc model (Dotani et al.\ 1997 using the {\sc
diskbb} model; Mitsuda et al.\ 1984), which approximates the
spectrum from a Shakura--Sunyaev disc. However, the tail of emission
beyond the disc spectrum which extends to high energies is plainly
not explained as part of the standard disc models.

The discrepancy between observations and the disc predictions are
even more marked in the other spectral shape seen in Cyg X-1, (see
Fig.~\ref{fig:cygx1_states}) where a large fraction of the power is
emitted in a spectrum which looks entirely unlike a disc. These
spectra, alternatively termed `low' or `hard' state, peak instead at
$\sim 100$ keV, though it is accompanied by a low temperature disc
component (see Section ~\ref{sec:lhs}).

The varying (and often confusing) nomenclature reflects the growing
understanding of these spectra. At first, with observations only
covering the classic 2--10~keV bandpass, the difference in count
rate of a factor $\sim 5$ between the two states lead to the `high'
and `low' description (see Fig.~\ref{fig:asm_per}). Increasing
spectral coverage showed that the change in bolometric luminosity
during the transition is much smaller (e.g. Nowak 1995; Zhang et al.
1997; Gierli{\'n}ski et al.\ 1999).  This, together with the
discovery of hysteresis in other BHB (e.g. the review by Nowak 1995;
Section~\ref{sec:hysteresis}) led instead to the `soft' and `hard'
terminology, based on spectral shape rather than intensity. Here we
will adopt `hard state' as denoting the low/hard state and `soft
state' for the high/soft/thermal dominant state.

Thus both spectral states require that there are two components to
the emission from the accretion flow. There is generally some trace
of an optically thick disc, which can be dominant (as in the soft
state) but this is always accompanied by higher energy emission,
which requires that some fraction of the accretion power is
dissipated in optically thin material so that the energy does not
thermalize to the disc temperature (see e.g. the review of radiative
processes in BHB by Zdziarski \& Gierli{\'n}ski 2004).

\subsection{Optically thin accretion flows: structure}
\label{sec:adaf}

One very attractive possibility for the origin of the optically thin
emission component is if the accretion flow itself becomes optically
thin. The key assumption for the disc spectra is that the energy
thermalizes. This requires that there are multiple collisions
between protons and electrons, and multiple collisions between
electrons and photons. This is not necessarily the case, especially
at low mass accretion rates when the density of the flow becomes
low.  Early on it was realized that if the flow is hot then it can
easily become optically thin to electron-photon collisions. This
also implies that the flow is optically thin to electron-proton
collisions (e.g. Stepney 1983) which has very important consequences
as the protons probably acquire most of the gravitational energy (since
gravity acts on mass), yet is it electrons which are by far the more
efficient radiators (Shapiro, Lightman \& Eardley 1976, hereafter
SLE76; Ichimaru 1977; Narayan \& Yi 1995). Incomplete thermalization
of protons with electrons leads to the formation of a two
temperature plasma, where the protons gain most of the gravitational
energy and lose very little of it to the electrons, while the
electrons gain only a small amount of energy via Coulomb collisions
and lose most of it by radiating. Since the flow is optically thin
then this radiation is in the form of Comptonization, bremsstrahlung
and/or cyclo-synchrotron rather than blackbody radiation. The proton
temperature is close to virial, so the flow has a large scale
height, and pressure forces are important as well as centrifugal
forces in balancing gravity (SLE76; Ichimaru 1977; Narayan \& Yi
1995).

The detailed structure of such hot, optically thin, geometrically
thick two temperature flows depends on the conditions assumed.
Without advection the properties of the accreting gas are described
by SLE76. However, advection of gravitational energy by the protons
should always be important for such two temperature flows (Ichimaru
1977; Ion torus: Rees et al.\ 1982; Advection Dominated Accretion
Flow, hereafter ADAF: Narayan \& Yi 1995). The classic ADAF solution
of Narayan \& Yi (1995) is self--similar, imposing a single value
for the advected fraction at all radii, with the further assumption
that advection is only a cooling process. Yuan (2001) relaxes these
assumptions, and shows that the solutions then also include a region
at higher luminosities, where the advected fraction is negative (so
advection is a heating process) and strongly radially dependent
(luminous hot accretion flows, LHAF: Yuan 2001). Both ADAF and the
more general LHAF have proton temperatures which are hot enough for
the material to be formally unbound, i.e. with positive Bernouli
parameter so can form a wind (Narayan \& Yi 1995). The properties of
this wind are not well determined, but it can form an advection
dominated inflow/outflow solution (ADIOS: Blandford \& Begelman
1999) where the mass loss rate probably depends on $\dot{M}$ (Yuan,
Cui \& Narayan 2005). Convection should also be important in these
flows (Blandford \& Begelman 1999), and can dominate in the class of
models termed convection dominated accretion flows (CDAF: Abramowicz
\& Igumenshchev 2001).

The full structure of the flow is complex, even in a hydrodynamic
description with the Shakura--Sunyaev analytic stress prescription.
Yet more complexity should be present with the inclusion of magnetic
fields, giving a Magnetically Dominated Accretion Flow (MDAF: Meier
2005) or jets (Jet Dominated Accretion Flow; JDAF: Falcke,
K{\"o}rding, \& Markoff 2004). This proliferation of properties for
the hot inner flow gave rise to the all inclusive term ?DAF, but
numerical simulations which include the self-consistent heating from
the magnetic dynamo may be a better guide to the properties of the
flow than analytic approximations. These show that in the limit of
no radiative losses the structure is something like that predicted
by the ADIOS models, but these flows also have a relativistic jet
(e.g. Hawley \& Balbus 2002).  This is very important as it shows
that there is no additional physics required to produce the jet: the
MRI in a geometrically thick hot, accretion flow in strong gravity
is sufficient (see Section~\ref{sec:jet}; Meier 2005).

\subsection{Optically thin accretion flows: stability}
\label{sec:adafstability}

All the varieties of hot, optically thin flows are quasi-spherical
with $H/R\sim$ 0.3--0.4. Thus the thermal timescale is still shorter
than the viscous timescale and the stability of the flow at any
radius can be assessed assuming that the surface density remains
constant as for thin discs (see Section~\ref{sec:stability}). In the
SLE76 flows, ion heating by gravity is balanced with ion cooling
through Coulomb collisions heating the electrons. The electrons then
radiate this energy via bremsstrahlung and Compton scattering (both
of which depend only on electron temperature). An increase in ion
temperature causes an initial increase in ion cooling as the ions
have more energy to give to the electrons. The electron cooling is
unaffected so the electrons must heat up. The details of Coulomb
coupling mean that this decreases the efficiency of the energy
exchange, by a factor which more than offsets the original increase
in ion temperature and results in a {\em lower} rate of cooling of
the ions. Thus a small increase in ion temperature leads to a lower
energy loss rate of the ions, so the ion temperature increases still
further and the flow is thermally unstable (Pringle 1976).

However, for all the flows including advection, ion
cooling can be dominated by advective losses rather than Coulomb
heating to the electrons.  The advective losses scale simply with the
ion temperature, so an ion temperature perturbation causes an increase
in ion cooling through advection and the system is stable (Narayan \&
Yi 1995). All the numerical MRI flows are thermally (and viscously) stable
also, as is shown by the fact that time dependent simulations reach an
equilibrium structure with well defined mean properties (e.g. Hawley
\& Balbus 2002)

\subsection{Two types of accretion flow as the origin of spectral states?}
\label{sec:hardsoft}

The two types of spectra seen in Cyg X-1 (hard and soft) are rather
naturally explained by the existence of two very different stable
accretion flow structures, with a hot, optically thin, geometrically
thick flow which can exist only at low luminosities, as well as a
cool, optically thick, geometrically thin disc. Geometrically, these
can be put together into the truncated disc/hot inner flow model
which can explain the observed dichotomy of hard and soft spectra.
At low $L/L_{\rm Edd}$, the inner optically thick disc is replaced
by an optically thin, hot flow, probably through evaporation (Meyer
\& Meyer-Hoffmeister 1994; R{\'o}{\.z}a{\'n}ska \& Czerny 2000;
Mayer \& Pringle 2007). There are few photons from the disc which
illuminate the flow, so Compton cooling of the electrons is not very
efficient compared to heating from collisions with protons. This
ratio of power in the electrons to that in the seed photons
illuminating them, \lhls, is the major parameter (together with
optical depth of the plasma) which determines the shape of a thermal
Comptonization spectrum (e.g. Haardt \& Maraschi 1993). Physically,
\lhls\ sets the energy balance between heating and cooling, and
hence sets the electron temperature. Thus \lhls\ is a more
fundamental parameter to understand thermal Compton scattering than
electron temperature.

In the hard state, the relative lack of seed photons illuminating
the hot inner flow means that \lhls$\gg 1$, producing hard thermal
Comptonized spectra. These can be roughly characterized by a power
law in the 5--20~keV band with photon index $1.5 <\Gamma < 2$ where
the photon spectrum $N(E)\propto E^{- \Gamma}$ [alternatively the
flux $F(E)=EN(E) \propto E^{-\alpha}$ where energy index
$\alpha=\Gamma-1$ and $0.5 < \alpha < 1$]. Conversely, when the mass
accretion rate increases, the flow becomes optically thick, and
collapses into a Shakura--Sunyaev disc (see Section~\ref{sec:disc}).
The dramatic increase in disc flux due to the presence of the inner
disc marks the hard-soft state transition (Esin, McClintock \&
Narayan 1997; Poutanen, Krolik \& Ryde 1997), and also means that
any remaining electrons which gain energy outside of the optically
thick disc material (perhaps from heating in magnetic reconnection
events above the disc) are subject to much stronger Compton cooling,
with \lhls$\le 1$. This results in Comptonized spectra which are
typically much softer, thus the soft state is characterized by a
strong disc and soft tail, roughly characterized by a power law
index of photon index $\Gamma \ge 2$.

\subsection{Yet more Black Hole spectral states}
\label{sec:morestates}

\begin{figure}
\begin{center}
\begin{tabular}{cc}
\includegraphics[clip=true,width=0.38\textwidth,angle=0]
{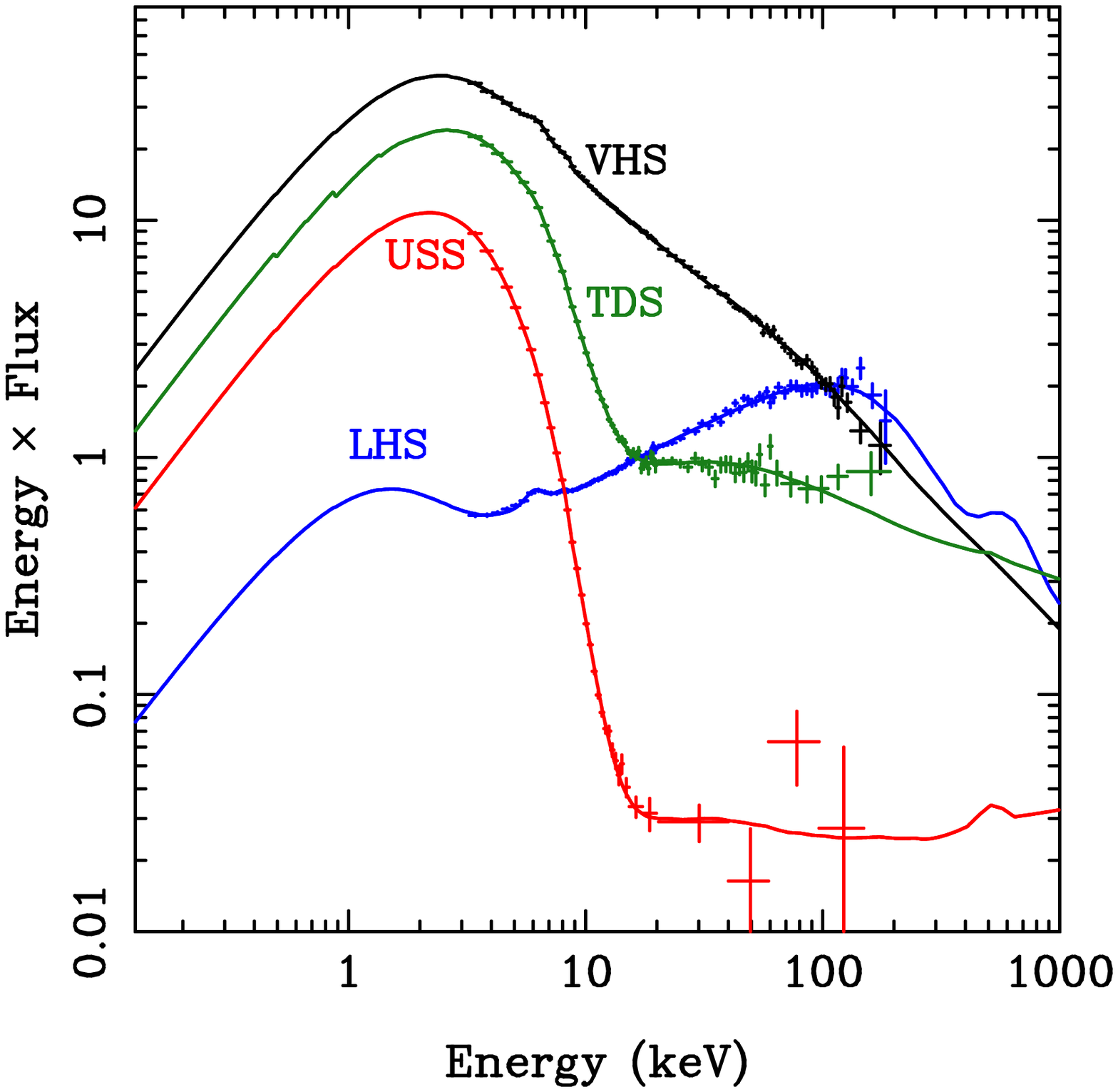} &
\includegraphics[clip=true,width=0.42\textwidth]
{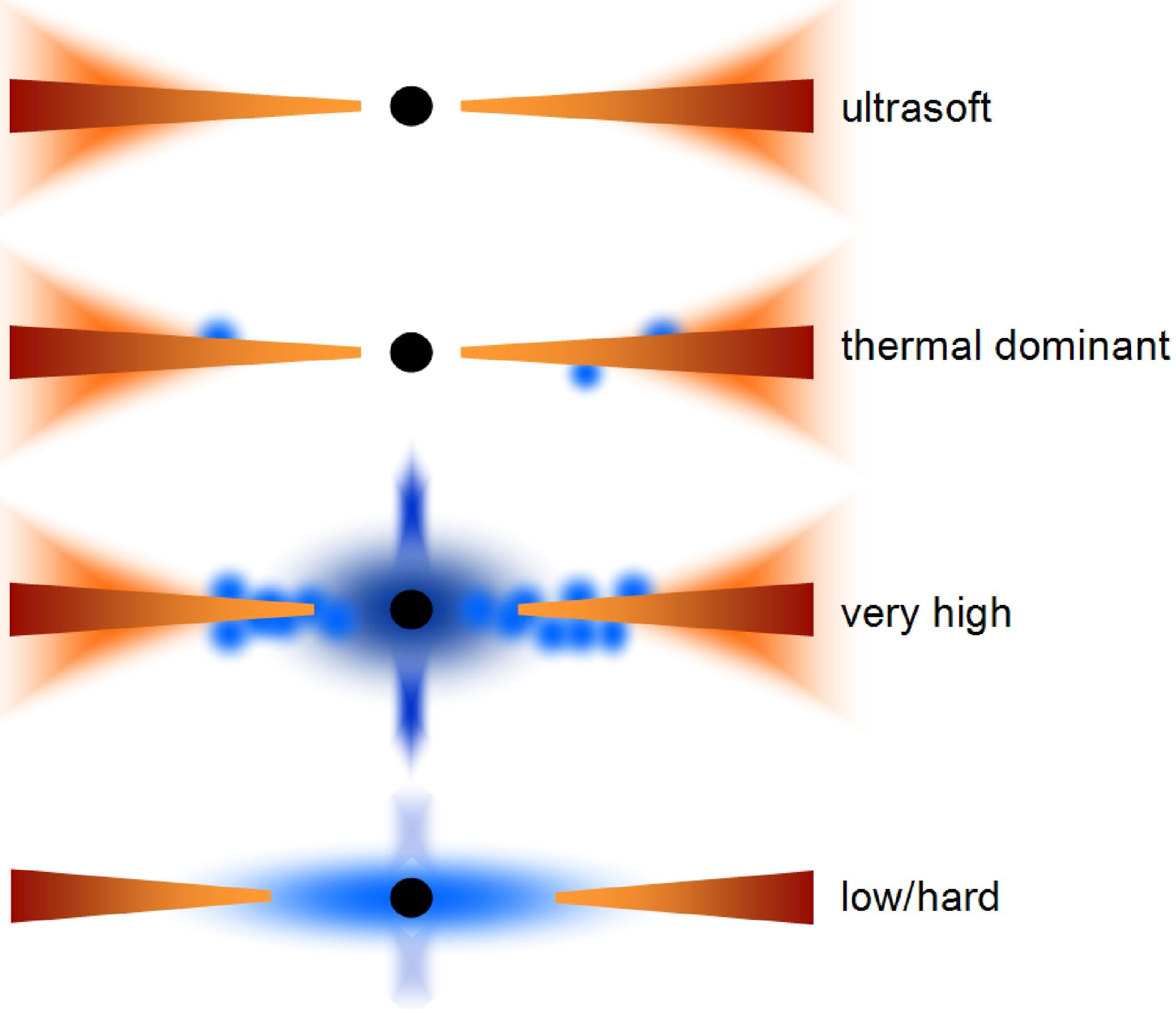}
\end{tabular}
\end{center}
\caption{The left hand panel shows
a  selection of states taken from the
2005 outburst of GRO J1655--40. The right hand panel shows the proposed
accretion flow changes to explain these different spectra, with
differing contributions from the disc, hot inner flow and its
associated jet, active regions above the disc and a wind.}
\label{fig:states}
\end{figure}

Cyg X-1 actually shows only a very small range of variability, with
$L/L_{\rm Edd}$ changing by only a factor $\sim 3$ on long
timescales (e.g. Done \& Gierli{\'n}ski 2003). The increase in data
from transient BHB in the early 1990's from the {\it Ginga}
satellite showed a further variety of spectral shapes could be
observed at high accretion rates. These are similar to the soft
state in that their X-ray emission peaks below 10~keV, as expected
for a disc. However, the high-energy emission accompanying the disc
spectrum can also be very strong, but is much steeper than that seen
in either the hard or soft states. These are termed the very high or
steep power law dominant state (hereafter very high state; see
figure~\ref{fig:states}b), initially discovered during a bright
outburst of GX~339--4 (Miyamoto et al.\ 1991) and GS~1124--68
(Miyamoto et al.\ 1993), and now generally seen in the brightest
phase of many BHBs including GRS~1915+105 (Reig, Belloni \& van der
Klis 2003; Done et al.\ 2004), GRO~J1655--40 (Sobczak et al.\ 1999a;
Kubota, Makashima \& Ebisawa 2001), XTE~J1550--564 (Sobczak et al.
1999b; Kubota \& Makishima~2004), and 4U~1630--472 (Abe et
al.~2005). Similar spectra with strong disc emission together with a
strong steep tail are also seen at lower luminosities, during the
transitions, where they are termed Intermediate state (hereafter IS:
Ebisawa et al.\ 1994; Belloni et al.\ 1996). It has long been
realized that these IS spectra share many characteristics of the
very high state (Belloni et al.\ 1996), so here we will treat them
both together as very high state (but see Gierli{\'n}ski \& Newton
2006).

Figure~\ref{fig:states} shows a selection of spectra seen from the
BHB, GRO J1655--40, at different mass accretion rates, showing
examples of these states (hard, soft, soft with extremely weak tail,
sometimes called ultrasoft, and an extreme example of the very high
state). As can be seen from this figure, these states are {\em not}
a unique function of mass accretion rate as measured by bolometric
luminosity. The hard state seen on the rise is as luminous as the
soft state seen later in the outburst (see e.g. XTE J1550--564 in
Fig.~\ref{fig:asm_trans1}). This lack of one--to--one correspondence
between spectral state and $L/L_{\rm Edd}$ is termed hysteresis and
is probably due to non-stationary behaviour of the accretion flow
associated with the dramatic rise in transient outbursts (see
Section~\ref{sec:hysteresis}). Since the IS is seen during the
transitions, then this likewise can occur over a wide range of
luminosities.

However, despite this severe complication, the general picture is
now clear that soft and very high states are typically high
luminosity states, while the hard state is seen at lower
luminosities. Comprehensive reviews of the observational properties
of these spectral states are given by e.g. Tanaka \& Lewin (1995)
and Remillard \& McClintock (2006).

Thus while we have two theoretical stable accretion flow models, a
disc and an optically thin, hot (messy) flow, there are (at least)
three different types of spectra to explain. As outlined in
Section~\ref{sec:hardsoft}, the hot flows plus a truncated disc can
generically match the hard state properties (see also
Section~\ref{sec:lhs}), while the spectra seen at high $L/L_{\rm
Edd}$ show clear signs of being dominated by the disc. At these high
luminosities the disc is likely to extend down to the last stable
orbit (see Section~\ref{sec:tds}), but even the soft-state spectra
are always accompanied by a high-energy tail. This shows that there
must be some sort of optically thin dissipation which can co-exist
with the majority of the accretion flow being in the form of a disc.
This could be due to some small fraction of the flow in a state
analogous to the hot, optically thin (messy) flow seen in the hard
state, but with properties modified by the strong Compton cooling
(Esin 1997; Janiuk, {\.Z}ycki \& Czerny 2000) and thermal conduction
(R{\'o}{\.z}a{\'n}ska \& Czerny 2000; Liu, Meyer \& Meyer-Hofmeister
2005). There are also alternatives to these smooth flows in models
where the energy dissipation is instead inherently very
inhomogeneous, perhaps due to magnetic reconnection of flux tubes
rising to the surface of the disc, as was first suggested by Galeev,
Rosner \& Vaiana (1979), and finds some support in the inherently
variable (in both space and time) dissipation produced by the MRI
(e.g. Hawley \& Balbus 2002).

One way to put all these mechanisms together into a plausible model
for all the spectral states is sketched in Fig.~\ref{fig:states}b,
similar to that first proposed by Esin et al.\ (1997). In the
sections below we will outline how this model works to explain the
observed spectra of each state. We discuss alternatives to the
truncated disc in Section~\ref{sec:alternatives}.

\section{The low/hard state}

\subsection{Hot inner flow and truncated disc}
\label{sec:lhs}

\begin{figure}
\begin{center}
\begin{tabular}{cc}
\includegraphics[clip=true,width=0.35\textwidth,angle=0]{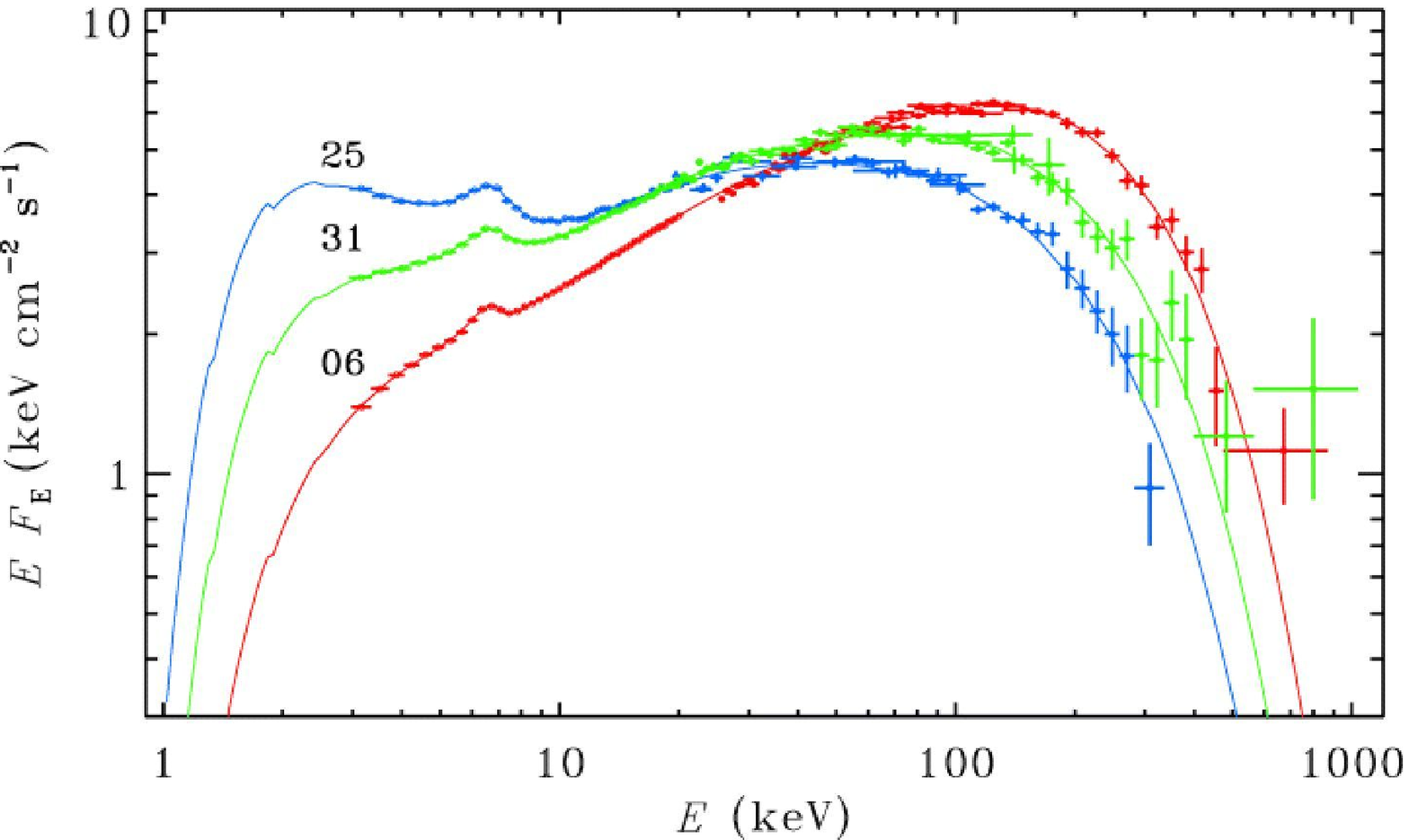}
&
\includegraphics[clip=true,width=0.6\textwidth,angle=0]{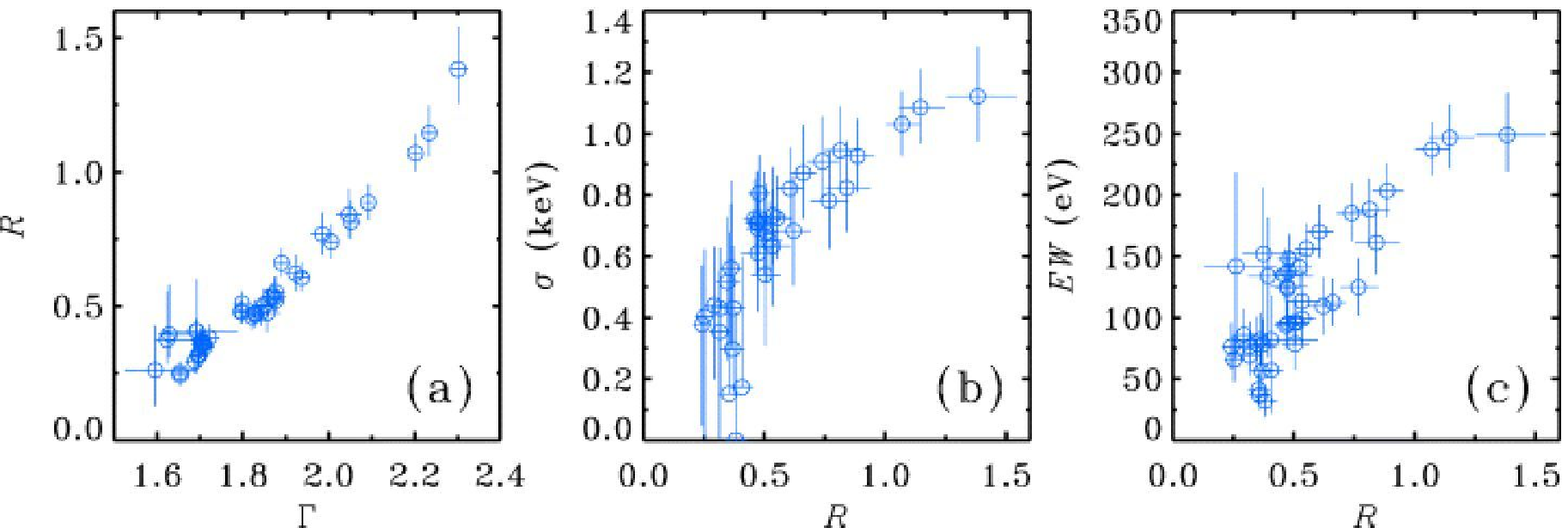}
\end{tabular}
\end{center}
\caption{Range of (absorbed) low/hard state in Cyg X-1, together with
  the correlated changes in spectral index and (neutral) reflected
  fraction, reflected fraction and amount of smearing of the
  associated iron line, and reflected fraction and equivalent width of
  the iron line. These all change in a manner qualitatively
  consistent with the disc extending down further into the hot inner
  flow (from Ibragimov et al.\ 2005).}
\label{fig:lowhard}
\end{figure}

At low luminosities the spectra peak at 100~keV rather than the
expected disc temperature of $<$~0.5~keV (see
Fig.~\ref{fig:states}).  These hard-state spectra clearly show that
the structure of emitting region is very different to that expected
from a disc. The spectra are broadly fit by thermal Comptonization
models, with electron temperatures $kT_e=$~75--110~keV for the best
studied case of Cyg X-1 (Ibragimov et al.\ 2005).
Fig.~\ref{fig:lowhard} shows representative spectra for this state.
The range of spectral slopes indicate a range in \lhls\ from $\sim$
5--15 (Ibragimov et al.\ 2005). This is fairly naturally produced in
the context of the truncated disc model. At very low luminosities
the disc is truncated far from the hole and there is little overlap
between the hot flow and cool disc so few seed photons illuminate
the flow, giving \lhls $\gg$ 1. As the disc moves progressively
inwards it increasingly extends underneath the hot inner flow so
that there are more seed photons intercepted by the flow, decreasing
\lhls\ (as in the geometries sketched in Fig.~\ref{fig:lowhard}).

The maximum luminosity which can be carried by this flow marks the
transition luminosity to the soft state. However, this point is
rather unclear both theoretically and observationally! Theoretically
the problem is that the structure of the hot flow is not yet well
understood. Models of ADAFs show they can only exist up to $L/L_{\rm
Edd} \sim 1.3\alpha^2\sim 0.01$ for $\alpha=0.1$ (Esin et al.\
1997), though this can be extended up to $L/L_{\rm Edd}\sim 0.1$ in
the LHAFs (Yuan et al.\ 2007), and potentially further still since
these analytic models do not include the magnetic fields produced by
the MRI which may give enhanced energy release in the rapid infall
region of the flow close to the black hole (Agol \& Krolik 2000).
Enhanced magnetic pressure due to the collapse of the hot flow may
also be important in increasing the maximum luminosity (Machida,
Nakamura \& Matsumoto 2007). Observationally, the problem is that
the transition happens at a range of luminosities, even in a single
object! The best studied case is that of XTE J1550--564, where the
transition occurs from $L/L_{\rm Edd}\sim 0.2$ to $L/L_{\rm
Edd}=0.003$ (Done \& Gierli{\'n}ski 2003). This range can be
suppressed by considering only the soft-to-hard state transition on
the decline, ignoring the hard-to-soft transitions on the rapid rise
to outburst which are typically at higher $L/L_{\rm Edd}$. The
transitions on the slow decline pick out a fairly constant value of
$L/L_{\rm Edd}\sim 0.02$ (Maccarone 2003) but there are exceptions
even here, with the 1998 outburst of XTE J1550-564 staying in the
soft state down to $L/L_{\rm Edd}\sim 0.003$ (Gierli{\'n}ski \& Done
2003). This variety of luminosities for the transition is one
manifestation of hysteresis, where the same source can show
different spectra at the same luminosity (see
Section~\ref{sec:hysteresis}). Nonetheless, the properties of the
hard state are well matched by a model of a hot flow with a
truncated disc which progressively moves inwards as the mass
accretion rate increases.

\begin{figure}
\begin{center}
\begin{tabular}{cc}
\includegraphics[clip=true,width=0.35\textwidth,angle=90]
{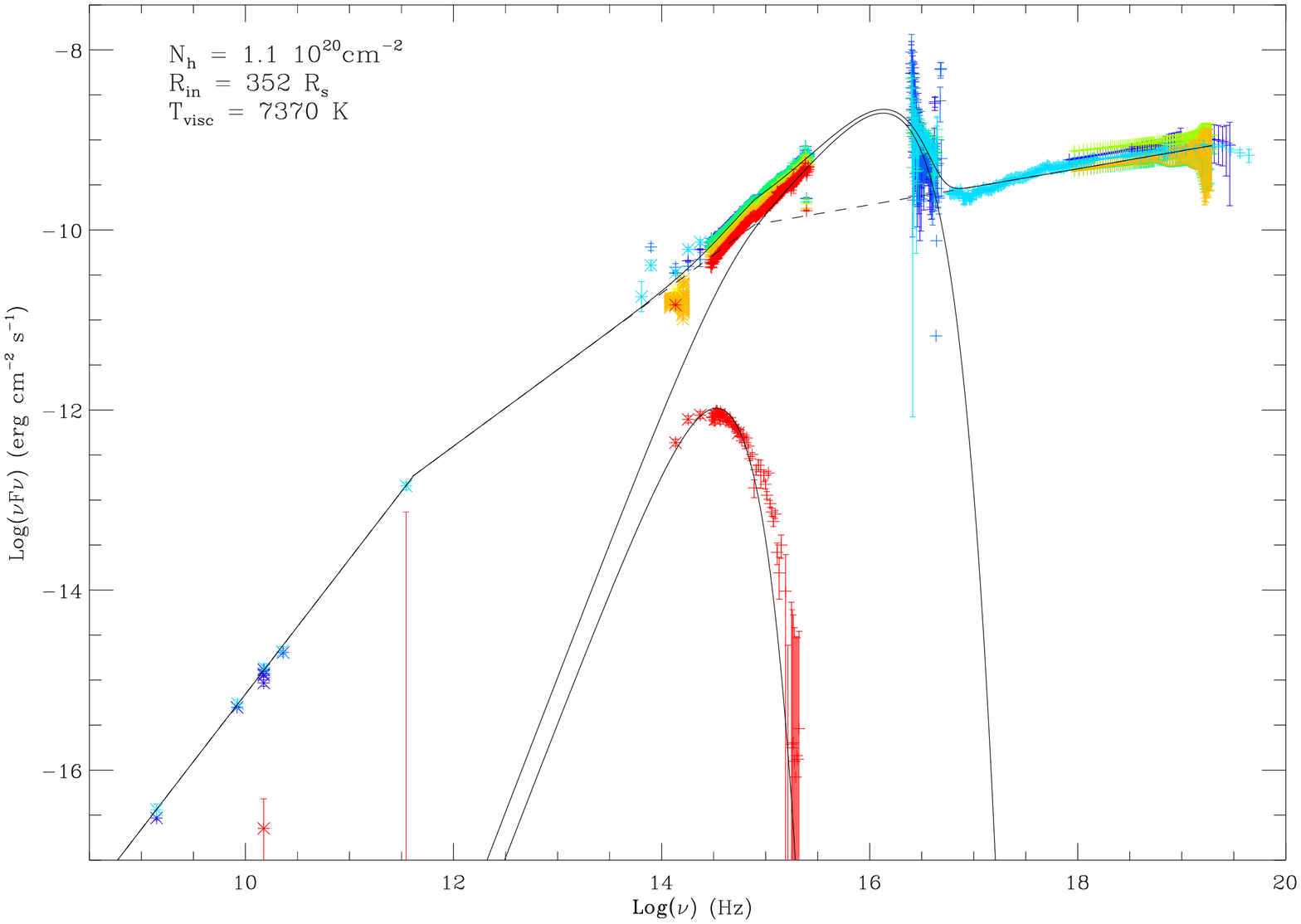} &
\includegraphics[clip=true,width=0.4\textwidth,angle=0]
{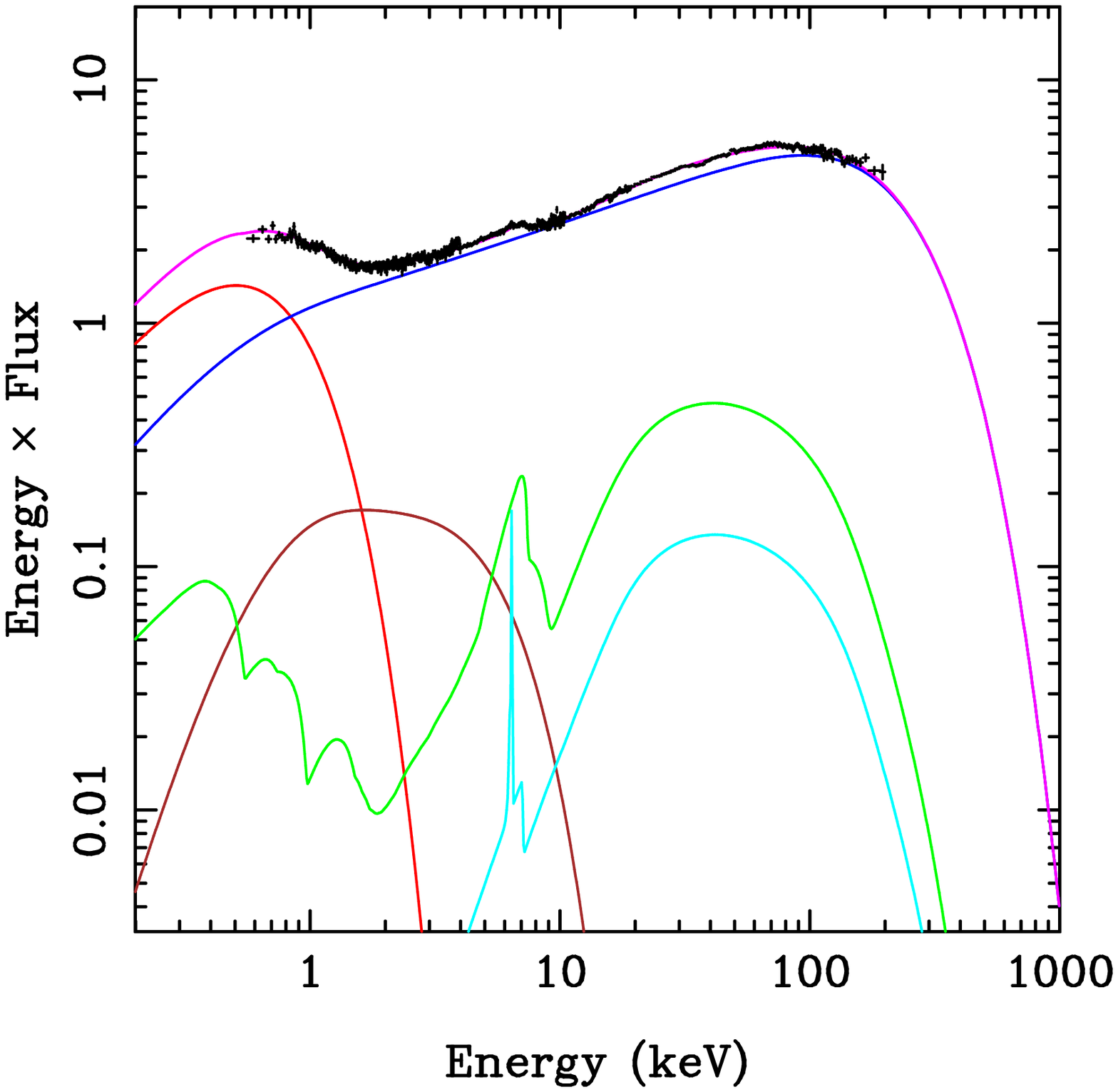}
\end{tabular}
\end{center}
\caption{The soft component in the hard state. left panel: the disc in XTE
J1118+480 at $L/L_{\rm Edd}\sim 10^{-3}$ (Chaty et al.\ 2003).  Right panel:
the disc in Cyg X-1 at $L/L_{\rm Edd}=0.02$ (after di Salvo et al.\ 2001).
The disc spectrum (red) is accompanied by a small tail to higher energies (brown),
producing a soft excess which is around 10 per cent of the total soft
emission and at somewhat higher temperature. The other components of the model are
thermal Comptonization (dark blue), smeared (green) and narrow (light blue) reflection.}
\label{fig:ls_disc}
\end{figure}

The behaviour of the intrinsic emission from the disc itself is
harder to track as the disc temperature is below 0.5~keV. This means
that it cannot be easily observed with the {\it RXTE} PCA instrument
(3--20~keV), which has by far the most data on BHBs. Instead of
multiple monitoring observations there are only occasional snapshots
with satellites covering the softer X-ray bandpass ({\it ASCA}, {\it
BeppoSAX}, {\it Chandra} and {\it XMM-Newton}) and these data are
also often heavily absorbed by the considerable interstellar column
seen towards many BHB as they are in the plane of our Galaxy.
Nonetheless there are some indications that the disc behaves as
predicted by these truncated disc models. The most convincing of
these is the BHB XTE J1118+480, one of the few sources at high
Galactic latitude. The very low galactic column to this object gives
the best view of the EUV/soft X-ray region where the disc should
peak.  This is also a transient where the outburst peak luminosity
was very low, at $L/L_{\rm Edd}\sim 10^{-3}$, so it remains clearly
in the hard state during the entire outburst (see
Fig.~\ref{fig:asm_trans1}). Multiwavelength campaigns
(IR--optical--UV--X-ray) show that the disc has very low maximum
temperature of $\sim$~10--40~eV, well fit with a truncation radius
of $\sim$100--300 $R_g$ (e.g.  Esin et al.\ 2001; Frontera et al.
2001; Chaty et al.\ 2003, see Fig.~\ref{fig:ls_disc}). The total
spectrum also points to the geometry being inhomogeneous. The hard
X-ray spectrum clearly shows that the X-ray source is illuminated by
rather few seed photons i.e. that \lhls$\gg$~1. However the total
spectrum has the soft and hard luminosities about equal
(Fig.~\ref{fig:ls_disc}). This is consistent with the disc
subtending rather a small solid angle at the source of energetic
electrons, also as predicted by the truncated disc models (see
Fig.~\ref{fig:states}).

No other BHB in the hard state has comparable data, but there are
snapshot observations of the soft component in Cyg X-1 at $L/L_{\rm
Edd}\sim$ 0.01--0.02. Here the disc temperature is of order
0.1--0.2~keV (Ba{\l}uci{\'n}ska-Church et al.\ 1995; Ebisawa et al.
1996; Di Salvo et al.\ 2001), consistent with a truncated disc where
the inner radius is more like 50~$R_g$ (di Salvo et al.\ 2001).
Again, there is roughly as much power in this disc spectrum as there
is in the hard X-ray component, yet the hard X-ray spectrum implies
that the source is photon starved, with \lhls$\gg$~1. However, these
broadband data also show that the soft spectrum is more complex than
expected from simple disc plus Comptonization and its (mainly neutral)
reflection. There is spectral curvature which can be modelled as an
additional, higher temperature soft component or as a slightly steeper
power law which contains 5--10 per cent of the total X-ray power
(Ebisawa et al.\ 1996; di Salvo et al.\ 2001; Ibragimov et al.\ 2005
see Fig.~\ref{fig:ls_disc}; Makishima et al. 2007). GX~339--4 also
shows a similarly complex soft component in the hard state (Wilms et
al. 1999), as does GRO~J1655-40 (Takahashi et al 2007).  The origin of
this is not well understood. It could be connected to a warm layer or
hot spots on the disc (di Salvo et al.  2001), or to uncertainties in
reflection modelling (Done \& Nayakshin 2001), or could be an artifact
of intrinsic curvature of the Comptonized continuum (such as can be
produced by anisotropic Comptonization from a sphere rather than
planar geometry: Haardt \& Maraschi 1993) or to time averaging over a
variable shape Comptonization component (Poutanen \& Fabian 1999;
Revnivtsev, Gilfanov \& Churazov 1999). Whatever its origin, this
component means that extracting the true disc temperature from limited
bandpass data is difficult. In general, a small, hot `disc' spectrum
is more likely to be indicating this unknown component than the true
disc spectrum. This may be the origin of the apparently untruncated
disc seen in the hard state in GX~339--4 (Miller et al.\ 2006, see
Section \ref{sec:challenges} below).

The hard X-ray continuum is seen together with some reflected
emission from the accretion disc which is generally not highly
ionized. Both the solid angle subtended by this reflecting material
and the amount of relativistic smearing increase as the hard state
spectrum steepens (e.g. Gilfanov, Churazov, \& Revnivtsev 1999)
although there is some systematic uncertainty in the amount of
reflection which is produced due to the poorly understood spectral
curvature of the complex soft excess (Ibragimov et al.\ 2005).
However, the trend for an increasing reflected fraction with steeper
spectra is plainly qualitatively consistent with the geometries
sketched in Fig.~\ref{fig:lowhard} where the increasing penetration
of the cool disc into the inner hot flow gives more seed photons to
cool the flow. This decrease in \lhls\ results in steeper continua
spectra, together with a larger solid angle subtended by the disc
and larger relativistic effects. This can give both a good
qualitative {\em and} quantitative description of the spectral
index--reflection correlation seen in the data, (Zdziarski,
Lubi{\'n}ski \& Smith 1999; Zdziarski et al.\ 2003) especially given
the uncertainties in determining the amount of reflection. As well
as being dependent on details of how the continuum is modelled
(Wilson \& Done 2001; Ibragimov et al.\ 2005) there are also
theoretical uncertainties on the shape of the reflected spectrum
when the material is ionized. Low energy line emission contributes
to the spectrum as well as simple electron scattering of the
incident continuum as the material is heated and ionized by the
X-ray illumination (Ross \& Fabian 1993; {\.Z}ycki \& Czerny 1994).
Compton upscattering in the upper, X-ray heated layers of the disc
can give additional broadening to the spectral features (Ross,
Fabian \& Young 1999) and there should also be a range of ionization
states present, from both radial and vertical stratification, and
this can be highly complex if the disc is in hydrostatic balance
(Nayakshin, Kazanas \& Kallman 2000).

\begin{figure}
\begin{center}
\begin{tabular}{cc}
\includegraphics[clip=true,width=0.42\textwidth,angle=0]{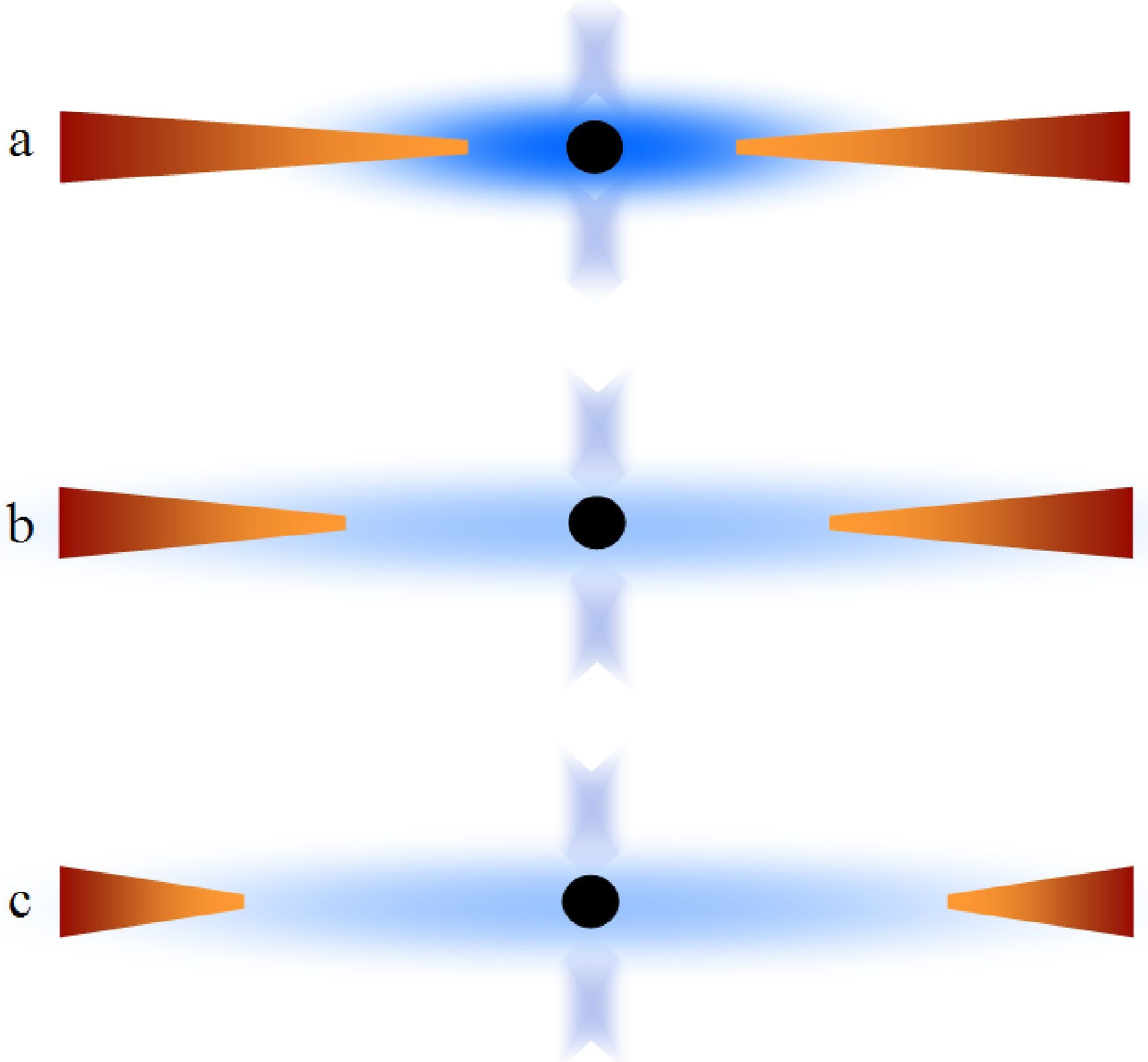}
&
\includegraphics[clip=true,width=0.38\textwidth,angle=0]{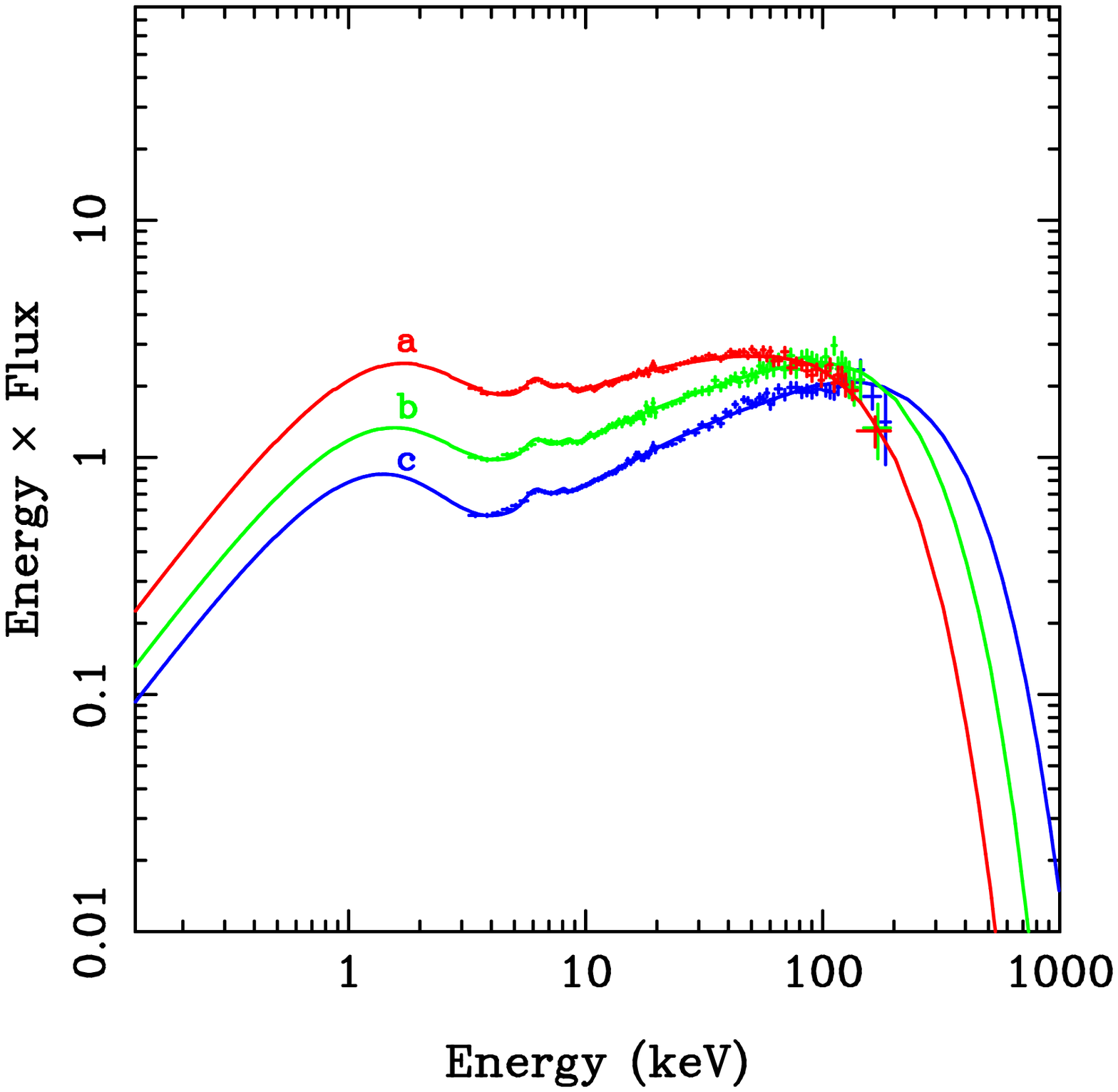}
\end{tabular}
\end{center}
\caption{Range of low/hard state geometries in the truncated disc
  model, together with their predicted spectra. When the disc is
  truncated far from the black hole, few disc photons are intercepted
  by the hot flow. Thus the Comptonized spectrum is hard, while a
  large fraction of disc photons are seen directly. As the disc
  extends further underneath the hot flow the larger fraction of disc
  photons intercepted means the spectrum becomes softer as the
  electron temperature is cooler, while the disc is hotter but less
  distinct.}
\label{fig:lhs_eqpair}
\end{figure}

By contrast, the properties of the thermal Comptonizing region are
more robust.  Fig.~\ref{fig:lhs_eqpair} shows a series of models for
the hard state which quantify the effect of the geometry changes,
where the disc extends progressively further inwards as a function
of $\dot{M}$ and the fraction of disc flux intercepted by the hot
flow also progressively increases (based on the {\sc eqpair} code of
Coppi 1999).  These models have the disc inner radius decreasing by
a factor 2, while the covering fraction of the hot flow increases
from 0.2 to 0.6. The optical depth in the hot flow is fixed at
unity, and \lhls\ decreases from 15 to 2.5. This model incorporates
Compton (and Coulomb) cooling self consistently, and predicts electron
temperatures dropping from 110~keV to 70~keV. This bears a strong
resemblance to the observed hard-state spectra shown in
Fig~\ref{fig:lowhard}.

\subsection{Alternative Geometries for the hard state}
\label{sec:alternatives}

Alternative geometries which include an untruncated disc and
(mostly) isotropic source emission have significant problems in
matching the observed features of the hard-state spectra.  The
continuum spectral shape rules out slab corona models as the spectra
all peak at high energies. Such hard spectra from thermal
Comptonization are only possible if the luminosity in seed photons
within the X-ray region is less than that in the hot electrons i.e.
\lhls~$\gg$~5. A disc extending underneath an isotropically
radiating hot electron region would intercept around half the
Comptonized emission in a slab corona geometry. Part of this
emission can be reflected, but reflection cannot be efficient at
100~keV (where the spectrum peaks) due to electron recoil. The
maximum albedo of the disc even for completely ionized reflection is
around 0.3 for the hardest spectra, so the rest of the illuminating
flux is thermalized in the disc. In a slab geometry all this
thermalized emission is seen by the hot electrons, giving
\lhls~$\le$~2/0.7 = 2.8 (corresponding to photon indices of $\Gamma>
2$), much smaller than that required to fit the hardest spectra
(with $\Gamma\sim 1.5$) of \lhls~$>$~10 (Haardt \& Maraschi 1993;
Stern et al.\ 1995; Malzac, Dumont \& Mouchet 2005).

A patchy corona, perhaps made up of individual magnetic flares,
allows part of the reprocessed flux to escape without
re--illuminating the hot electron region, so it can produce the
required hard spectra. However, the one advantage of a slab corona
is that the reflected flux is also intercepted by the disc, and
hence Comptonized into continuum flux. This suppresses the amount of
recognizable reflected emission, so that $\Omega/2\pi < 1$ as
observed in the hard state (Haardt et al.\ 1994; Petrucci et al.\
2001). A patchy corona allows the reprocessed flux to escape, but
this also allows the reflected flux to escape, so $\Omega/2\pi\to 1$
for the hardest spectra, in direct conflict with the observations
(Malzac, Beloborodov \& Poutanen 2001; Malzac et al.\ 2005) A patchy
corona can only be retrieved if the disc has an extremely ionized
skin, making a completely reflective layer, while the observed
near--neutral reflection is produced deeper in the disc. Such
complex ionization structure, with a rapid transition between
ionized and neutral material, can be produced by hard X-ray
illumination (Nayakshin et al.\ 2000). However, the extreme
ionization reflection produced by the skin is only indistinguishable
from the continuum at low energies. The reflection albedo at high
energies is independent of the ionization state of the material, and
electron recoil in the disc means the reflected emission must drop
at $\sim$~100~keV. Thus a spectrum with a large contribution from
extremely ionized material should be steeper in the 50--100~keV band
than the 5--20~keV band. This is not observed (Maccarone \& Coppi
2002; Barrio, Done \& Nayakshin 2003).

A disc extending down to the last stable orbit cannot be ruled out
completely though. Firstly, the predicted steepening of the
50--200~keV spectrum in ionized reflection models could be
counterbalanced by a flattening in the underlying high energy
continuum, though it seems contrived that this should cancel out to
look like an unbroken power law.  Secondly and perhaps more plausibly,
the continuum source could be connected to the jet, and be in motion
along the jet axis, expanding away from the disc at moderately
relativistic velocities. Mild beaming of the X-ray emission could
results in a reflected fraction of $\sim 0.3$ as observed, even if an
untruncated disc were present.  Similarly the beaming suppresses the
soft seed photon flux from the disc so that the resulting continuum
spectrum is hard.  The observed increase in reflected fraction with
spectral slope shown in Fig.~\ref{fig:lowhard} could be explained if
the expansion velocity decreases with luminosity (Beloborodov 1999;
Malzac et al.\ 2001), and perhaps this change in beaming pattern also
produces increased illumination of the inner disc, giving the observed
increase in smearing (Fig.~\ref{fig:lowhard}).  However, this is {\em
opposite} to the increase in jet velocity at the transition inferred
from radio data (Fender, Belloni \& Gallo 2004; see
Section~\ref{sec:jet}). An increase in jet velocity is also required
for the internal shock models to explain the dramatic radio flares
seen on the hard--to--soft (but {\em not} on the soft--to--hard)
transitions, as these require that faster jet from the brighter,
softer hard state collides with the previous, slower jet material
(Fender, Belloni \& Gallo 2004; see Section~\ref{sec:jet}).

Models where the X-rays are produced directly in the jet were
proposed by Markoff et al.\ (2001). These produce the hard X-rays by
synchrotron emission from the high-energy extension of the same
non-thermal electron distribution which gives rise to the radio
emission. However, the observed shape of the high-energy cutoff in
the hard state is very sharp, and cannot be easily reproduced by
synchrotron models (Zdziarski et al.\ 2003). These jets are also
generally radiatively inefficient, yet the observed luminosity
change at the hard--soft state transition where the flow changes to
a radiatively efficient disc is not large (less than a factor of a
few, see Fig.~\ref{fig:cygx1_states}: Maccarone 2005). Instead,
there are now composite models where the X-rays are from
Comptonization by thermal electrons at the base of the jet, while
the radio is from non-thermal electrons accelerated up the jet
(Markoff et al.\ 2005). The base of the jet expands at weakly
relativistic speeds, and the resulting weak beaming away from the
disc is similar to that proposed by Beloborodov (1999), though the
model of Markoff et al.\ (2005) has this in the centre of a
truncated disc. Given that the hot inner flow probably relates
physically to the base of the jet (see Section~\ref{sec:jet}), then
this model converges on the truncated disc/inner hard X-ray source
geometry shown in Fig.~\ref{fig:lhs_eqpair}, though with some
additional weak beaming of the hard X-rays as in the Beloborodov
(1999) model.

The magnetized accretion--ejection models of Ferreira et al.\ (2006)
have a similar outer disc--inner jet structure, though here the
inner, optically thick disc is still present down to the last stable
orbit. However, its properties are very different from the standard
Shakura--Sunyaev disc in that the angular momentum of the infalling
material is transported self consistently by the jet. The transition
radius between this jet dominated disc and the standard accretion
disc is variable, producing the range of behaviour seen in the hard
state spectra in a similar way to the truncated disc/hot inner flow
models.

However, all these alternative models require that the X-ray source is
mildly beamed away from the disc. This is challenged by the
observation that the hard state BHB show no trend in their properties
as a function of inclination (Fender et al. 2004; Narayan \&
McClintock 2005). Systematic studies of hard state properties over a
sample of BHB at various inclinations should constrain the amount of
beaming (e.g. amount of reflection or spectral slope versus low
frequency QPO) and show whether these models are allowed by the data.

Thus all currently viable models for the hard state converge on a
geometry where the standard disc extends down only to some radius
larger than the last stable orbit, with the properties of the flow
abruptly changing at this point. The only alternative to a hot inner
flow for the origin of the X-ray emission is a mildly relativistic
outflow, where the velocity decreases as the source brightens to
produce the observed softer spectra and higher reflected
fraction. This is probably inconsistent with the very attractive
internal shock models for the bright radio flares at the
hard--to--soft transitions, and may also be at odds with the observed
uniformity of hard state properties across BHB with different
inclinations.

\subsection{Challenges to the truncated disc geometry for the hard state}
\label{sec:challenges}

While the alternative flows discussed above face some observational
challenges, there is also current controversy over the truncated
disc/hot inner flow models over claims that the data show evidence for
an {\em untruncated disc} in the hard state. Firstly, there are
occasional observations of extremely smeared iron line and reflection
features in this state in XTE J1650--500 (Miller et al.  2002; Minutti
et al.\ 2004) and in GX 339--4 (Miller et al.\ 2006).  These are
fairly bright hard states (XTE J1650--500 is close to an intermediate
state) so while the truncated disc models do predict that the inner
radius is not at the last stable orbit, there is no requirement for it
to be very recessed. However, the claimed broadening is such that the
disc extends down to the innermost stable orbit for a fast rotating
black hole, in conflict with the truncated disc models. For XTE
J1650--500, Done \& Gierli{\'n}ski (2006) refit the data to show that
the smearing can be significantly reduced to a level compatible with a
truncated disc if there is also resonance iron K line {\em absorption}
from an outflowing disc wind (see Section~\ref{sec:winds} for more on
outflows and winds). While this has yet to be explicitly demonstrated
for the GX~339--4 data, the same spectral degeneracy probably occurs
in all moderate resolution CCD data. Higher resolution grating or
bolometer data is required to unambiguously show the presence or
absence of such absorption features, and hence reveal the shape of the
underlying continuum and reflection components. Further ambiguity
comes from the shape of the reflected emission, as it is plainly
ionized in these data. All current ionized reflection models are
calculated for conditions appropriate to AGN discs (e.g. Ballantyne,
Ross \& Fabian 2001), where the intrinsic thermal emission is in the
UV/EUV, so that ionization of heavy elements is determined only by
photoionization. Yet for BHB the disc is at soft X-ray temperatures,
so collisional processes are also important in setting the ion
populations, reducing the irradiation required. This changes the shape
of the reflected spectrum due to changing the vertical temperature
structure and hence the amount of internal Comptonization (Ross,
Fabian \& Young 1999). The vertical structure also depends on the
illumination, especially if the disc is in hydrostatic equilibrium
(Nayakshin et al 2000; Done \& Nayakshin 2007). Thus both better data
and better reflection models are required before the broad iron line
results can be claimed to clearly rule out the truncated disc geometry
(Done \& Gierli{\'n}ski 2006). Nonetheless, this is an important goal,
as it is possibly the only way to detect a non--radiative, weakly
illuminated, optically thick inner disc (e.g.  Beloborodov 1999;
Malzac et al.\ 2001; Ferreira et al.\ 2006) and hence understand the
nature of the inner accretion flow.

\begin{figure}
\begin{center}
\begin{tabular}{cc}
\includegraphics[clip=true,width=0.42\textwidth,angle=0]{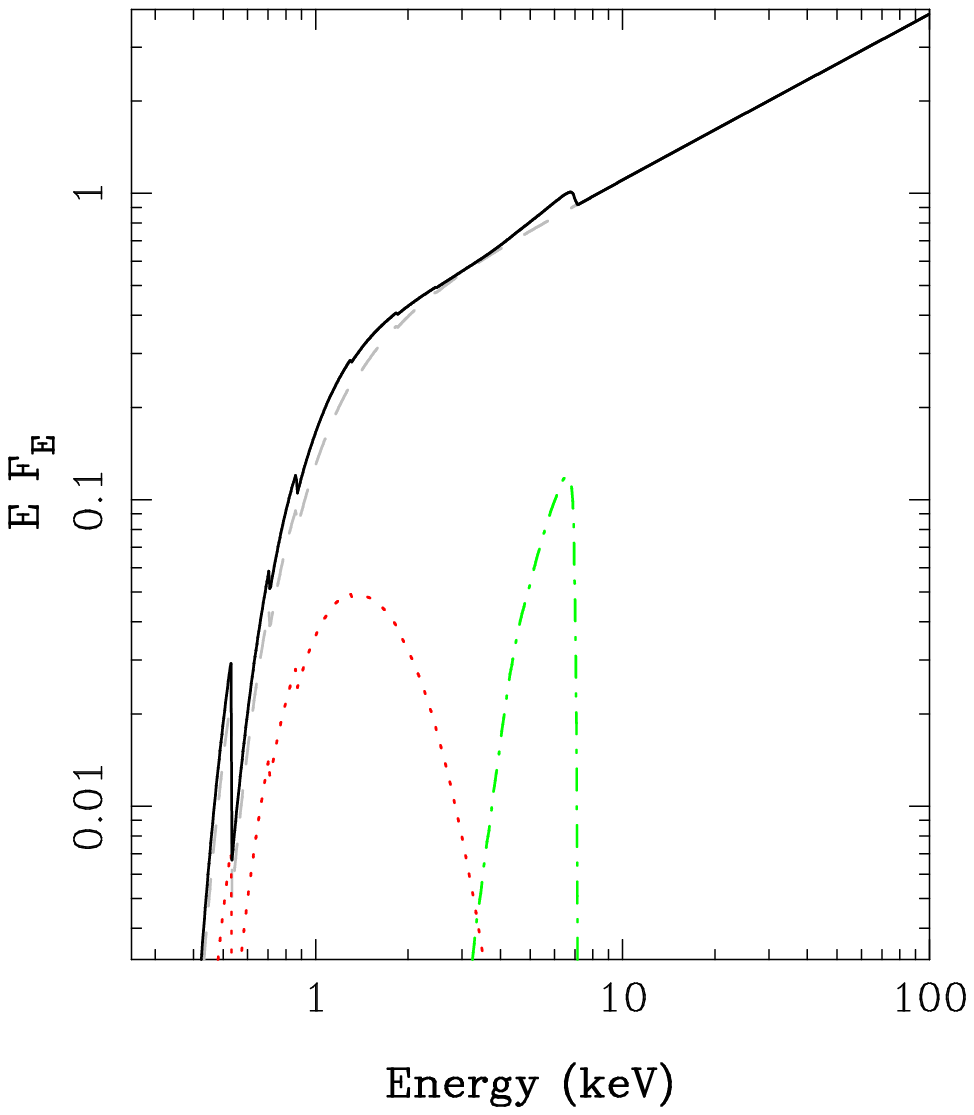}
&
\includegraphics[clip=true,width=0.48\textwidth,angle=0]{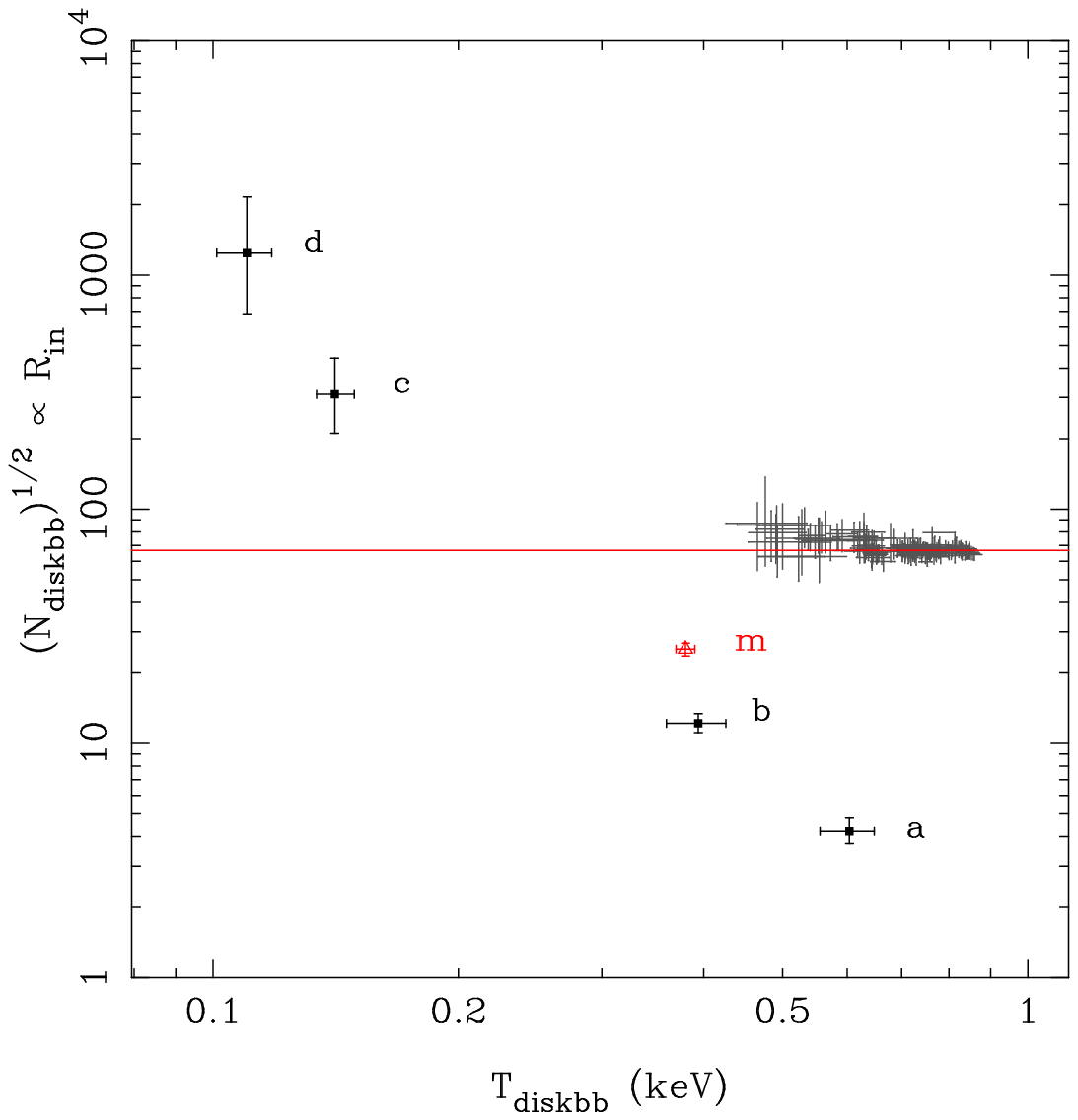}

\end{tabular}
\end{center}
\caption{The left hand panel shows a bright hard state ($L_{\rm
bol}/L_{\rm Edd}\sim 0.02$, assuming mass of 6~M$_\odot$, distance
of 4~kpc) model spectrum of GX~339--4 from simultaneous {\it
XMM-Newton} and {\it RXTE} data (first model in table 1 of Miller et
al.\ 2006). The disc is shown in dotted red curve, the Gaussian line
in dash-dotted green curve and the power law in dashed grey curve.
The disc contribution to the 0.1--100 keV flux is only 2.5 per cent
($L_{\rm disc}/L_{\rm Edd}\sim 6\times 10^{-4}$), so its parameters
are dependent on the continuum model. The right panel shows the
square root of the {\sc diskbb} model normalization, which is
proportional to its inner radius. The red point (marked $m$)
corresponds to the observation in the left panel. The derived inner
radius is smaller than the approximately constant radius seen in
disc-dominated spectra from {\it RXTE} monitoring (grey points, red
line shows the mean; after Gierli{\'n}ski \& Done 2004). This is
also seen in archival {\it ASCA} data in the hard state when the
disc is determined against a simple power law ($a$), or Comptonized
continuum ($b$). However, a more complex spectral form, modelled
with an additional power law ($c$) or Comptonization component ($d$)
gives the opposite result. The disc is then larger than that seen in
the high/soft state, consistent with a truncated geometry. }
\label{fig:millergx339}
\end{figure}

There is a further challenge to the hot inner flow models, from the
residual disc emission in the sources with only moderate galactic
absorption ($<$0.5$\times 10^{22}$ cm$^{-2}$). These show spectra
with a rise at low energies, which can be fit by disc models with
luminosity and temperature indicating that the disc extends down
close to the black hole (GX~339--4: Miller et al.\ 2006; SWIFT
J1753.5--0127: Miller, Homan \& Miniutti 2006; XTE J1817--330:
Rykoff et al.\ 2007).  However, disc luminosity and temperature are
very difficult to uniquely determine unless the disc is the dominant
spectral component (see Section~\ref{sec:tds} and
\ref{sec:vhs_disc}). These apparent disc components typically carry
less than 5--10 per cent of the flux in the instrument bandpass, so
details of how the tail is modelled become very important.

\begin{figure}
\begin{center}
\begin{tabular}{c}
\includegraphics[clip=true,width=0.48\textwidth,angle=0]{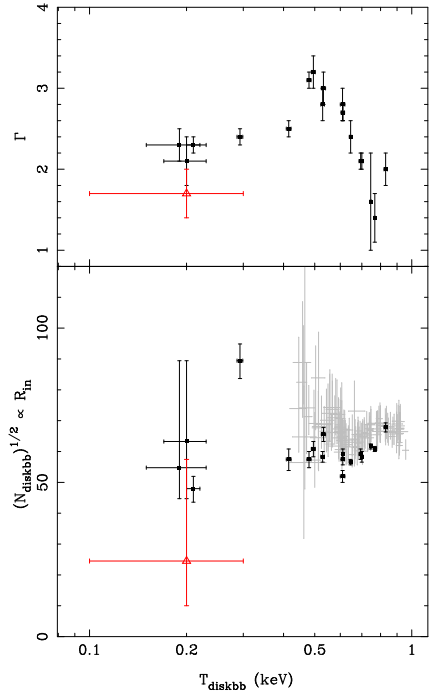}
\end{tabular}
\end{center}

\caption{The outburst of XTE J1817--330, with data fit to a disc
plus power law model. The upper panel shows the power-law spectral
index while the lower panel shows square root of {\sc diskbb}
normalization ($\propto R_{\rm in}$). The black points are the {\it
SWIFT} best-fit results from table 3 in Rykoff et al.\ (2007) while
the grey data show results from {\it RXTE} monitoring of this
outburst (soft state only). Both {\it SWIFT}  and {\it RXTE} data
give similar results for the inner radius of the disc-dominated
state, and show that this is approximately constant. The {\it SWIFT}
data show that the source is always in the disc-dominated or
intermediate state apart from one potential observation in the hard
state (red triangle). This has large uncertainties (so may be an
intermediate state rather than hard state) but is consistent with a
disc which extends down to the same or smaller inner radius than
that seen in the disc-dominated data. This result is similar to the
apparently small radii discs seen in simple fits to hard state data
in GX~339--4 (see Fig.~\ref{fig:millergx339}). Hence, its properties
depend on the detailed model of the hard-state continuum used, and a
soft excess/low-energy continuum curvature will allow a truncated
disc fit to these data.}

\label{fig:rykoff}
\end{figure}

We show this explicitly for the GX~339--4 data of Miller et al.
(2006) in Fig.~\ref{fig:millergx339} (left panel), where one of their
model fits to the data (an absorbed disc, power law and broad line) is
shown as a deconvolved $\nu F_\nu$ spectrum. The contribution of the
disc is extremely weak, less than 20 per cent of the total flux even
at its peak energy, and it carries less than 1 per cent of the
bolometric luminosity. Plainly the derived parameters of this
component will be very dependent on the detailed modelling of the
spectrum at soft energies. We illustrate this using archival data from
GX 339--4 in its 1995 September 12 hard state observation from {\it
  ASCA} (Wilms et al.\ 1999). The right hand panel of
Fig.~\ref{fig:millergx339} shows the square root of {\sc diskbb}
normalization ($\propto R_{\rm in}$) from these data assuming that the
continuum is given by an absorbed disc plus power law (a) or
Comptonized (b) spectrum. Both these give a disc which is {\em
  smaller} than the disc seen in the same object in {\it RXTE}
observations of its high/soft, disc dominated spectra (see Section
\ref{sec:tds}). This is directly opposite to the predictions of the
truncated disc model. However, there are clear indications that the
soft spectrum in the hard state is more complex (see
Fig.~\ref{fig:ls_disc} and Section \ref{sec:lhs}; Ebisawa et al.\
1996; Wilms et al.\ 1999; di Salvo et al. 2001; Ibragimov et al. 2005;
Takahashi et al. 2007; Makishima et al. 2007). Fitting a more complex
form, so that the non-disc continuum is either a broken power law (c)
or low temperature Comptonization plus the harder power law (d) gives
a {\em larger} inner radius than that seen in the disc dominated
states, consistent with the truncated disc models. These derived radii
will all increase if the disc normalisation is corrected for the
photons scattered into the Comptonised spectrum (Kubota et al. 2001;
Kubota \& Makishima 2004; Kubota \& Done 2004; Done \& Kubota 2006;
Makishima et al. 2007).

Thus the derived disc radii in hard-state spectra are sensitive to the
assumed continuum shape. A power law or downwards curving continuum
(soft deficit) at low energies gives an inner disc radius which is
much smaller than that derived from an upwards curving (soft excess)
continuum. The red point in Fig.~\ref{fig:millergx339} (right panel,
marked as $m$) shows {\sc diskbb} normalization from the power-law
continuum fit to {\it XMM-Newton}/{\it RXTE} data of Miller et al.\
(2006). While these data have not yet been fit to more complex
continua, it seems likely that increasing the non--disc continuum at
low energies would force the residual disc emission to be at lower
temperature and larger radius. These uncertainties in the underlying
spectral shape clearly preclude any definitive measure of the disc in
these data, and there are also further uncertainties in the spectral
modelling of the disc spectrum in the low/hard state as the
stress--free inner boundary condition is probably not appropriate and
the disc may also be significantly heated by irradiation and/or
conduction. 

Rykoff et al. (2007) also claim that the evolution of the disc
component in an outburst of XTE J1817--330 is incompatible with the
truncated disc models. We replot their results from multiple {\it
SWIFT} observations in Fig.~\ref{fig:rykoff} for the disc plus power
law models, where the upper panel shows the power-law spectral
index, and the lower panel the derived disc radius. This outburst
was also followed by the {\it RXTE}\/ satellite, and we include {\sc
diskbb} normalization as derived from the PCA data as the grey
points in Fig.~\ref{fig:rykoff}. The two instruments give results
which are consistent, and the combined datasets show clearly that
the disc-dominated state has a well defined, approximately constant
radius. However, the {\it SWIFT} spectral indices show that the
Rykoff et al. (2007) data contain only one possible spectrum in the
hard state ($\Gamma<2$), and that its signal--to--noise is limited.
Thus their data do not put any strong constraints on the disc in the
hard state, contrary to their conclusions. The point lies below the
constant radius inferred from the disc dominated data (though its
large uncertainty means it is also consistent with this value),
similar to the simple continuum fits to the hard state GX~339--4
data shown in Fig.~\ref{fig:millergx339}b. We have re--extracted these
data and find that a more complex continuum form, with upwards 
curvature, allows a much larger disc, consistent with the truncation models.

Thus there are no observations which unambiguously conflict with the
truncated disc models, while there are a tremendous amount of data
which can be fit within this geometry, including spectra, rapid
variability characteristics (see Section~\ref{sec:timing}) and jet
properties (see Section~\ref{sec:jet}). While the truncated disc
model is indeed currently a very simplified version of what must be
a more complex reality, nonetheless the range of data it can
qualitatively (and sometimes quantitatively) explain gives
confidence that it captures the essence of the hard state.

\section{High mass accretion rates: thermal dominant and very high states}
\label{sec:tds}

\subsection{Disc spectra in the soft state}

The spectra in Fig.~\ref{fig:states} which are dominated by a
thermal component and where the tail is only a small fraction of the
total bolometric luminosity (soft state) show convincing evidence
for a disc. However, this is actually surprising as the classic
Shakura--Sunyaev disc is unstable at high luminosities when
radiation pressure starts to dominate the total pressure (see
section~\ref{sec:radpress}), pointing to the heating being somewhat
different than assumed (Section~\ref{sec:radpress}). Nonetheless,
this uncertainty in stress prescription has very little impact on
the zeroth order predicted disc spectra. The fact that the observed
soft-state spectra are so close to the expectation of an optically
thick, geometrically thin disc means that the accretion structure is
understood at some level, so the disc spectra can be used to probe
the dramatically curved space-time in the vicinity of the black
hole. In particular, in the Shakura--Sunyaev models the emitting
disc only extends down to the last stable orbit round the black
hole, and then free falls rapidly to the event horizon. The lack of
stress at the last stable orbit gives a clear inner edge to the
disc, and so provides an observational diagnostic of the spin from
the observed temperature and luminosity of the disc spectrum (see
Section~\ref{sec:disc}).

Simple spectral fitting for the soft component is based on the {\sc
diskbb} model which ignores any inner disc boundary condition,
describing the disc local temperature as $T(r)=T_{\rm
in}\cdot(r/r_{\rm in})^{-3/4}$ (Mitsuda et al.\ 1984).  The {\sc
diskbb} model is parameterized by $T_{\rm in}$ and $r_{\rm in}$, the
maximum observed disc temperature and the apparent disc inner
radius, respectively, where the disc bolometric luminosity $L_{\rm
disc}$ can be related to these two spectral parameters as $L_{\rm
disc}=4\pi r_{\rm in}^2 \sigma T_{\rm in}^4$.  However, several
corrections are required to derive the true inner disc radius from
these measured parameters. In order of importance these include the
stress--free inner boundary condition (which means the temperature
drops to zero at the innermost stable orbit, so the peak temperature
is from larger radii and hence is lower: Gierli{\'n}ski et al.\
1999; Kubota et al.\ 2001; Gierli{\'n}ski et al.\ 2001), spectral
hardening due to incomplete thermalization of the escaping radiation
(colour temperature correction: Shimura \& Takahara 1995; Merloni et
al. 2000; Davis et al.\ 2005), and relativistic corrections (which
can either increase or decrease the observed temperature depending
on inclination: Cunningham 1975; Zhang, Cui, Chen 1997).

\begin{figure}
\begin{center}
\begin{tabular}{cc}
\includegraphics[clip=true,width=0.54\textwidth,angle=0]
{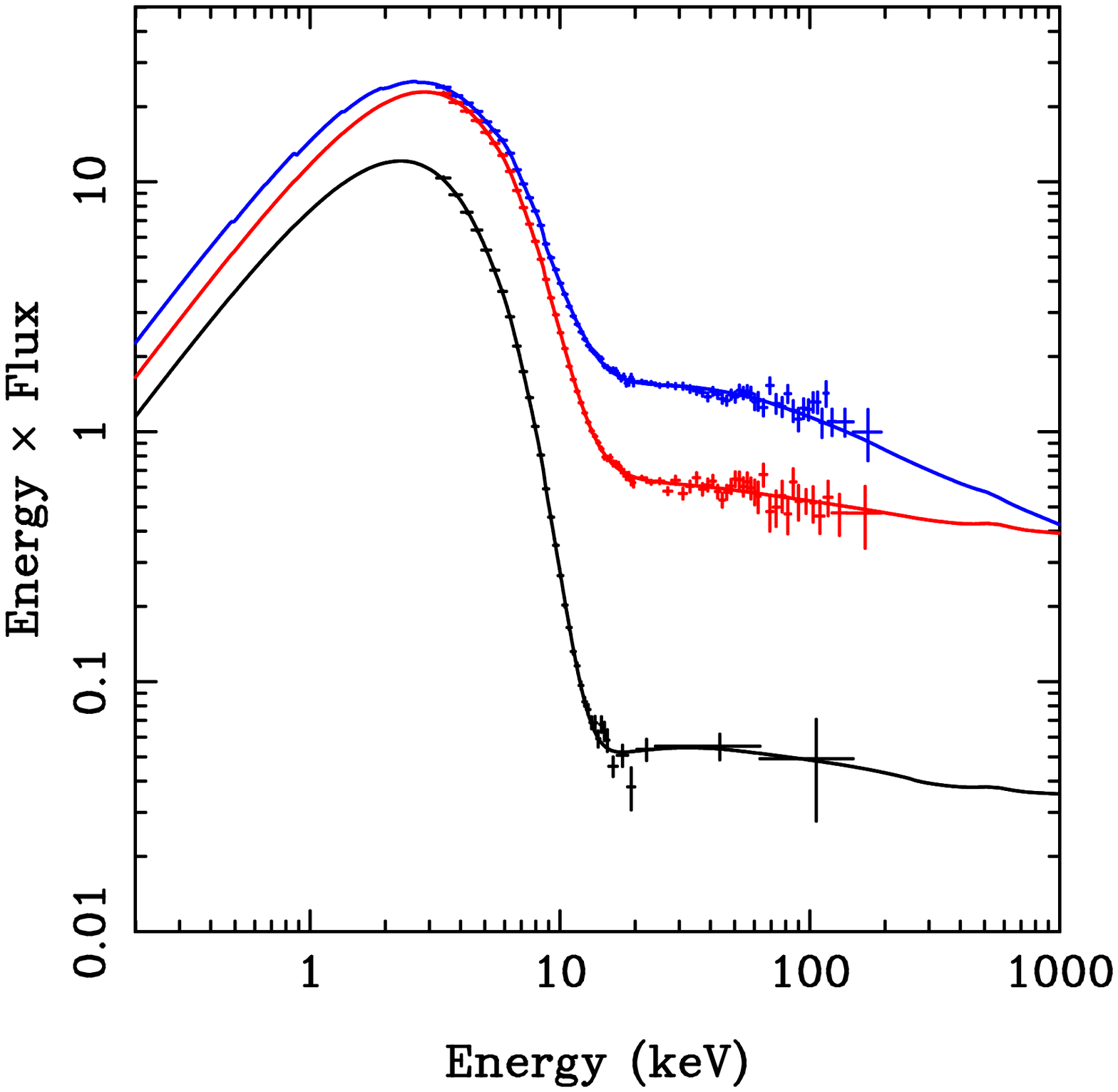} &
\includegraphics[clip=true,width=0.39\textwidth,angle=0]
{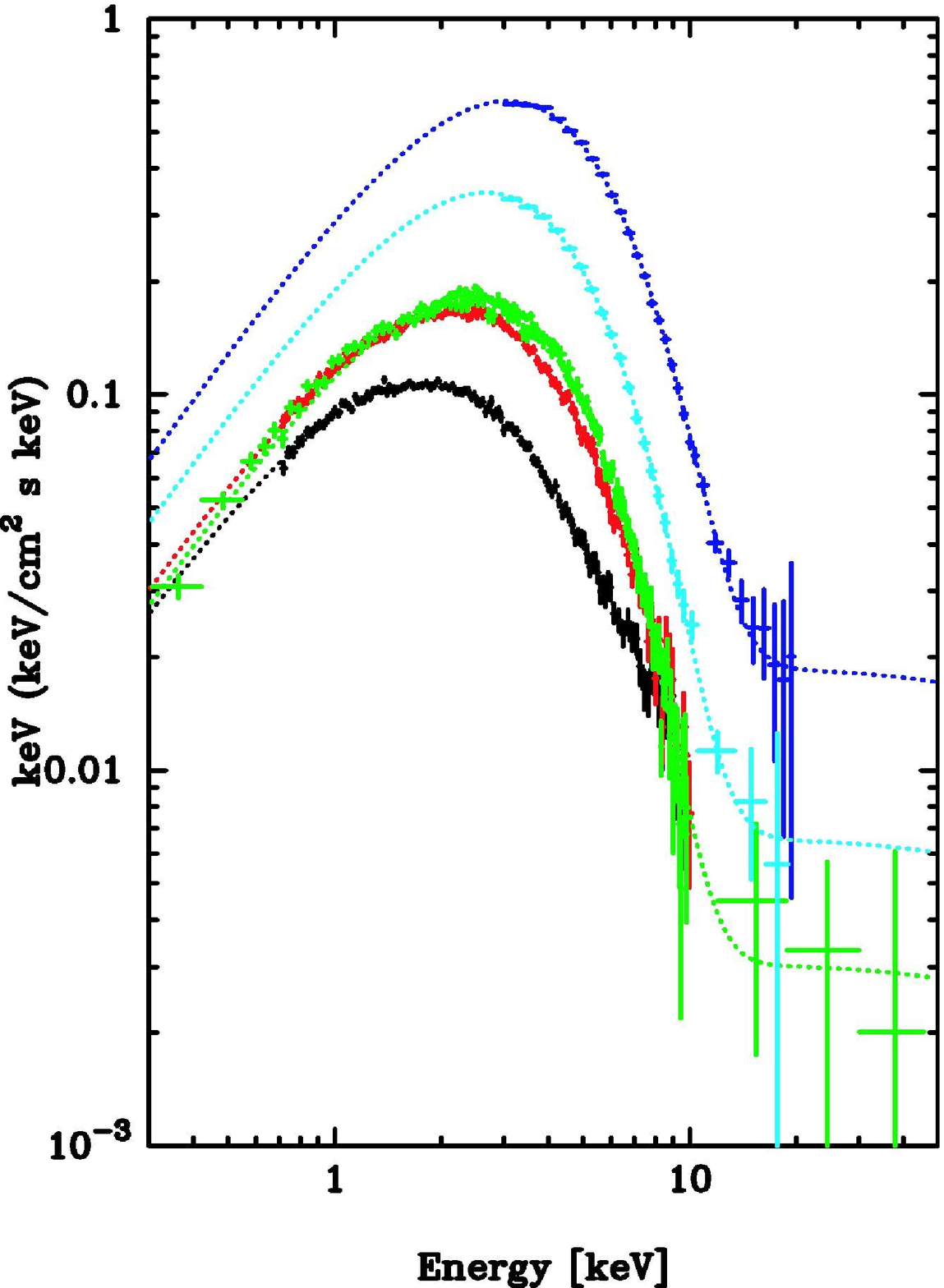}
\end{tabular}
\end{center}
\caption{A range of soft-state spectra at different luminosities
from the BHB 4U~1655--40 from RXTE PCA and HEXTE data showing that the
hard tail is indeed negligible in this state (left). The right panel shows
LMC X--3 from ASCA (black and red), BeppoSAX (green) and RXTE PCA 
(cyan and blue) data. The disc peak is clearly well covered by the
instruments with softer response, and these give results which are
consistent with those derived from RXTE data (Davis et al. 2006).}
\label{fig:4u1543}
\end{figure}

\begin{figure}
\begin{center}
\includegraphics[clip=true,width=0.9\textwidth,angle=0]
{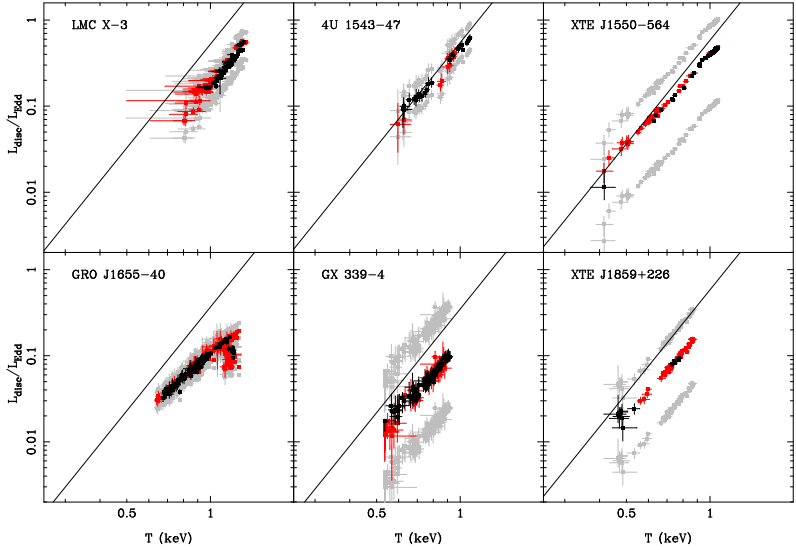}
\end{center}
\caption{Luminosity--temperature relations for the black holes with
disc dominated spectra fit with {\sc diskbb} for those which
span the largest range in luminosity. Grey points show the range of uncertainty
due to mass and distance estimates.}
\label{fig:tlum}
\end{figure}

Nonetheless, irrespective of the true disc radius, compelling
evidence for the standard disc formalism is given by the {\em
observation} that the value of $r_{\rm in}$ is usually observed to
remain constant in the soft state as $L_{\rm disc}$ changes
significantly (Ebisawa et al.\ 1991; 1993; 1994). An alternative,
even more direct way to present the same result is to plot the
observed disc luminosity and temperature against each other for
multiple observations (Kubota et al.\ 2001; Kubota \& Makishima
2004; Kubota \& Done 2004; Gierli\'{n}ski \& Done 2004; Davis et
al.\ 2006; Shaffee et al.\ 2006, see Fig.~\ref{fig:4u1543}).
Fig.~\ref{fig:tlum} shows this for the BHB with soft-state spectra
which span the largest range in luminosity. Clearly these sources
are consistent with an approximate $L\propto T^4$ relation over
factors of 10--50 change in disc luminosity, as predicted for a
constant inner radius for the accretion disc. This is exactly the
behaviour predicted by General Relativity at the last stable orbit,
so these data confirm that gravity is consistent with Einstein's
predictions even in the strong field limit. This is fantastic--but
also highlights the fact that the point at which corrections due to
GR become large is below the last stable orbit! Given that the
closeness of this to the event horizon, this shows that
observational tests of newer theories of gravity will be very
challenging (Gregory et al.\ 2004).

\subsubsection{Black hole spin}
\label{sec:spin}

Assuming GR is indeed the correct theory, these data also give a way
to measure the spin of the black hole assuming there are some
constraints on distance, mass and inclination from the binary
parameters. However, then the details of the corrections discussed
above are a key issue. Several attempts have been made to assess
each of these separately. Gierli{\'n}ski et al.\ (2001) and Kubota
et al.\ (2001) address the stress free inner boundary correction,
while Zhang et al.\ (1997) tabulated the general relativistic
correction factors affecting the escaping radiation from the disc
around a Schwarzschild and maximal Kerr black hole. Both these
effects are calculable from first principles. However, this is not
the case for the colour temperature correction, as this depends on
the disc opacity as a function of both frequency and vertical depth
in the disc. This means that the calculations are sensitive to the
(unknown) stress prescription. In general, opacity is a strongly
decreasing function of frequency, so the lower energy radiation can
thermalize while the emission at higher energies does not. Thus the
emission is a blackbody only up to the frequency at which the true
opacity becomes small. To get the same amount of energy out with
incomplete thermalization requires that the emission extends out to
higher energies as a blackbody is the most efficient emission
process at a given temperature. This modifies the blackbody
radiation, in a way which depends on the vertical density and
temperature structure of the disc (Shakura \& Sunyaev 1973).

The difference between the temperature inferred from the high-energy
rollover of the thermal emission, and that expected from a true
blackbody is termed a colour temperature correction, $f_{\rm col}$.
Calculations using the vertical structure predicted by a
Shakura--Sunyaev $\alpha$ disc show that this is remarkably constant
with both radius and $L/L_{\rm Edd}$, with $f_{\rm col}\sim 1.8$
predicted over the range $0.005\le L/L_{\rm Edd}\le 0.5$ (Shimura \&
Takahara 1995). This robustness was challenged by Merloni et al.\
(2000), who found much larger $f_{\rm col}$ at small mass accretion
rates. However, the origin of this discrepancy may be that Merloni
et al.\ (2000) assumed a constant vertical density profile. This is
only appropriate for a radiation pressure dominated disc and breaks
down at low $\dot{M}$ where the disc is gas pressure dominated and
has stronger density/temperature gradients (Gierli{\'n}ski \& Done
2004).

These results are now superseded by calculations which include full
metal opacities. The Shakura--Sunyaev models assume that the only
sources of opacity are free--free and electron scattering, yet
bound--free (photoelectric absorption edge) opacity is important
even at fairly high temperatures (Davis et al.\ 2005). These new
models also directly incorporate the stress free inner boundary
condition, and self--consistently propagate the escaping radiation
through the curved spacetime (Davis et al.\ 2005; Davis \& Hubeny
2006). These confirm that the colour temperature correction can
remain fairly constant over large changes in luminosity (Davis et al
2005), and fits to soft-state spectra give estimates for spin
ranging from 0.1--0.8 for the 5 BHB with the best data (Davis, Done
\& Blaes 2006; Shaffee et al.\ 2006; Middleton et al.\ 2006).  Such
moderate (as opposed to maximal) spins also match with the
theoretical predictions for the birth spin distribution of black
holes, as the pre-supernovae core before stellar collapse is slowly
rotating, and spin up from captured fallback of material is
countered by angular momentum loss in gravitational waves during the
formation process (see e.g. the review by Gammie, Shapiro \&
McKinney 2004). Spin-up through accretion during the lifetime of the
binary is limited in most systems as the companion mass in LMXB is
smaller than the black hole mass (King \& Kolb 1999).

\subsubsection{Deviations of soft-state spectra from $L\propto T^4$}
\label{sec:appstandard}

Fig.~\ref{fig:tlum} shows overwhelmingly that the disc dominated,
soft-state spectra do follow a $L\propto T^4$ relation, as predicted
from simple models of a constant radius, constant colour temperature
correction disc. However, mild deviations from this sometimes occur
at high luminosities, for example GRO J1655--40 and to a lesser
extent in XTE J1550--564 (Fig.~\ref{fig:tlum}) all bend away from
the $L\propto T^4$ relation at temperatures of 0.9--1.2~keV. These
`apparently standard' spectra are similar to the simple soft state,
in that they are still dominated by a strong thermal component, and
the tail to higher energies is weak. However they vary slightly
differently in that the disc luminosity increases rather more slowly
with temperature, more like $T^2$ than $T^4$ (Kubota \& Makishima
2004). Thus either the disc colour temperature correction is
increasing, and/or its inner radius is decreasing, and/or the disc
temperature structure is different from the Shakura-Sunyaev
prediction that $T(r)\propto r^{-3/4}$.

\begin{figure}
\begin{center}
\includegraphics[clip=true,width=0.9\textwidth,angle=0]
{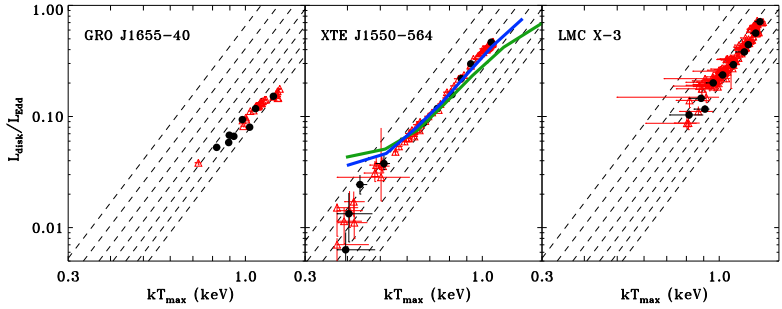}
\end{center}
\caption{Luminosity--temperature relations for the black holes with
disc dominated spectra spanning the largest range in luminosity
fit to the newest models of disc spectra. The blue and green lines
superimposed on XTE J1550--564 show the disc model solutions of Davis
et al.\ (2006) for viscosity of $\alpha=0.01$ and $0.1$,
respectively. A larger viscosity scaling gives a disc which is less
massive, so it becomes optically thin at lower
luminosity/temperature. This bends the relation away from the purely
thermal $L\propto T^4$.
}
\label{fig:shane}
\end{figure}

The new models of disc spectra do predict an increase in colour
temperature correction at high luminosities (Davis et al.\ 2006).
Beyond temperatures of 0.9--1~keV even metals become completely
ionized, and there is very little true opacity, giving an increasing
$f_{\rm col}$. The size of this effect depends on the optical depth
of the disc i.e. on $\alpha$ (Fig~\ref{fig:shane}) as large stresses
mean that the disc is very efficient in transporting material, so
the underlying disc is less dense. Recombination is less effective
so the photosphere is more highly ionized and runs out of true
opacity at lower $L/L_{\rm Edd}$ than for smaller effective
viscosity.

While the models with $\alpha\sim 0.01$ do match the bend in XTE
J1550--564, this is not a particularly convincing explanation
Firstly, the rapid rise to outburst requires $\alpha\sim 0.1$
(Lasota 2001), which is inconsistent with the observed bend (
Fig~\ref{fig:shane}). Secondly, the bend seen in GRO J1655--40 has a
somewhat different shape, and LMC X-3 (and to a lesser extent 4U
1543--47 see Fig.~\ref{fig:tlum}) have no bend at all but cover the
same temperature--luminosity range (Davis et al.\ 2006).

Optically thick advection (radiation trapping) can also cause a
deviation in the $L$--$T$ relation. This is neglected in the Davis
et al.\ (2005) models but should be important in a Shakura--Sunyaev
disc at high $L/L_{\rm Edd}$ (see Section~\ref{sec:radpress}). The
immense energy released close to the plane of the disc does not have
time to escape to the photosphere before being swept along with the
flow to smaller radii. This is more important at smaller radii so
preferentially suppresses the luminosity of the hottest parts of the
disc. For a disc temperature distribution parameterized as a power
law, with $T(r)\propto r^{-p}$, these models predict that $p$ goes
from 0.75 in the standard regime to 0.5 when advection dominates the
inner regions (e.g. Watarai et al.\ 2000). This matches very well to
the behaviour seen (Kubota \& Makishima 2004; Kubota et al.\ 2001),
making these models attractive. Nonetheless, this is a global
physical mechanism so should apply to all BHB, so again this fails
to adequately address why there is no such effect seen in LMC X-3. A
further argument against this bend being from the onset of advection
is that this is predicted to change the observed behaviour only at
$L/L_{\rm Edd}\sim 1$, rather higher than the luminosities inferred
here.

\begin{figure}
\begin{center}
\begin{tabular}{cc}
\includegraphics[clip=true,width=0.45\textwidth,angle=0]{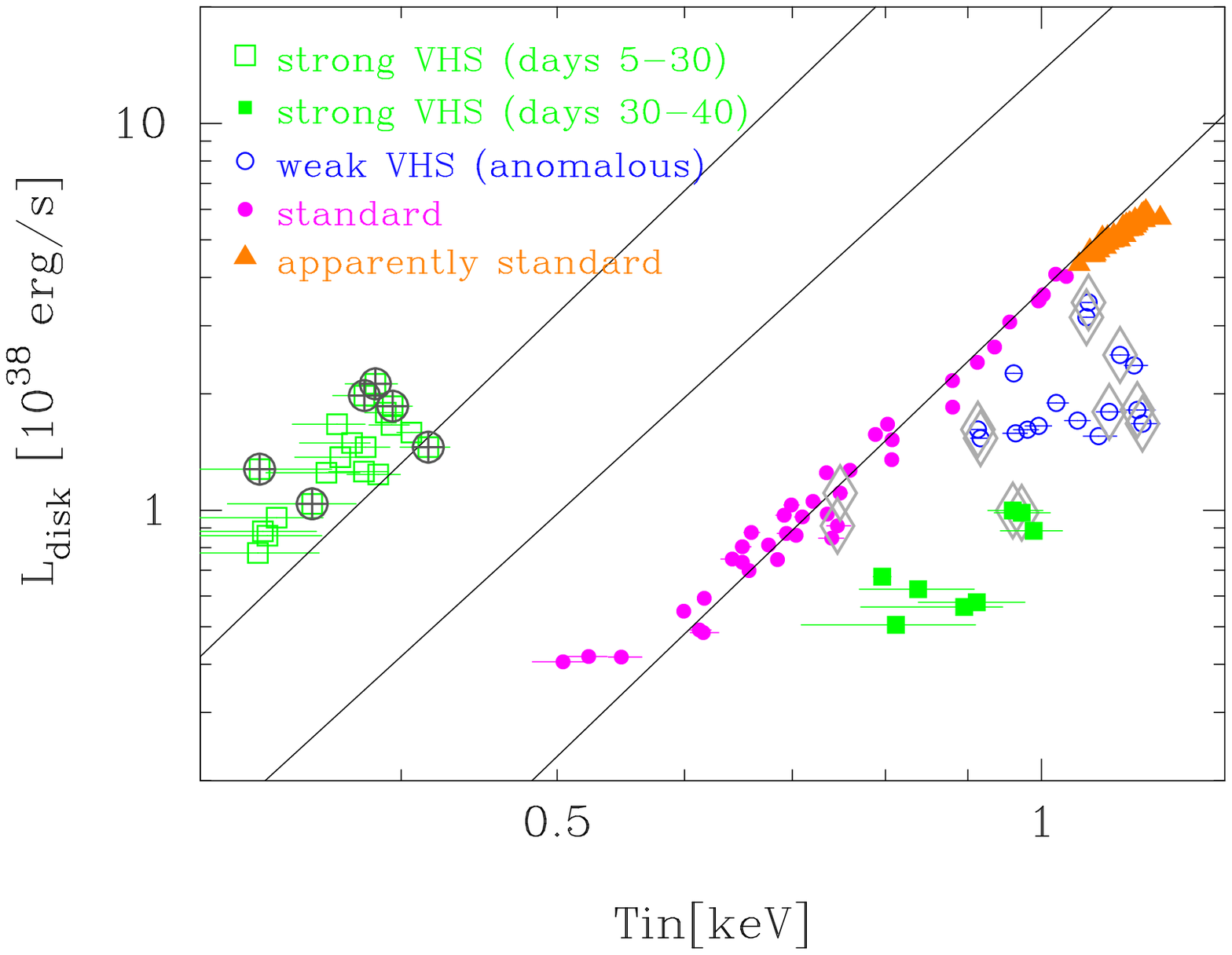}
&
\includegraphics[clip=true,width=0.45\textwidth,angle=0]{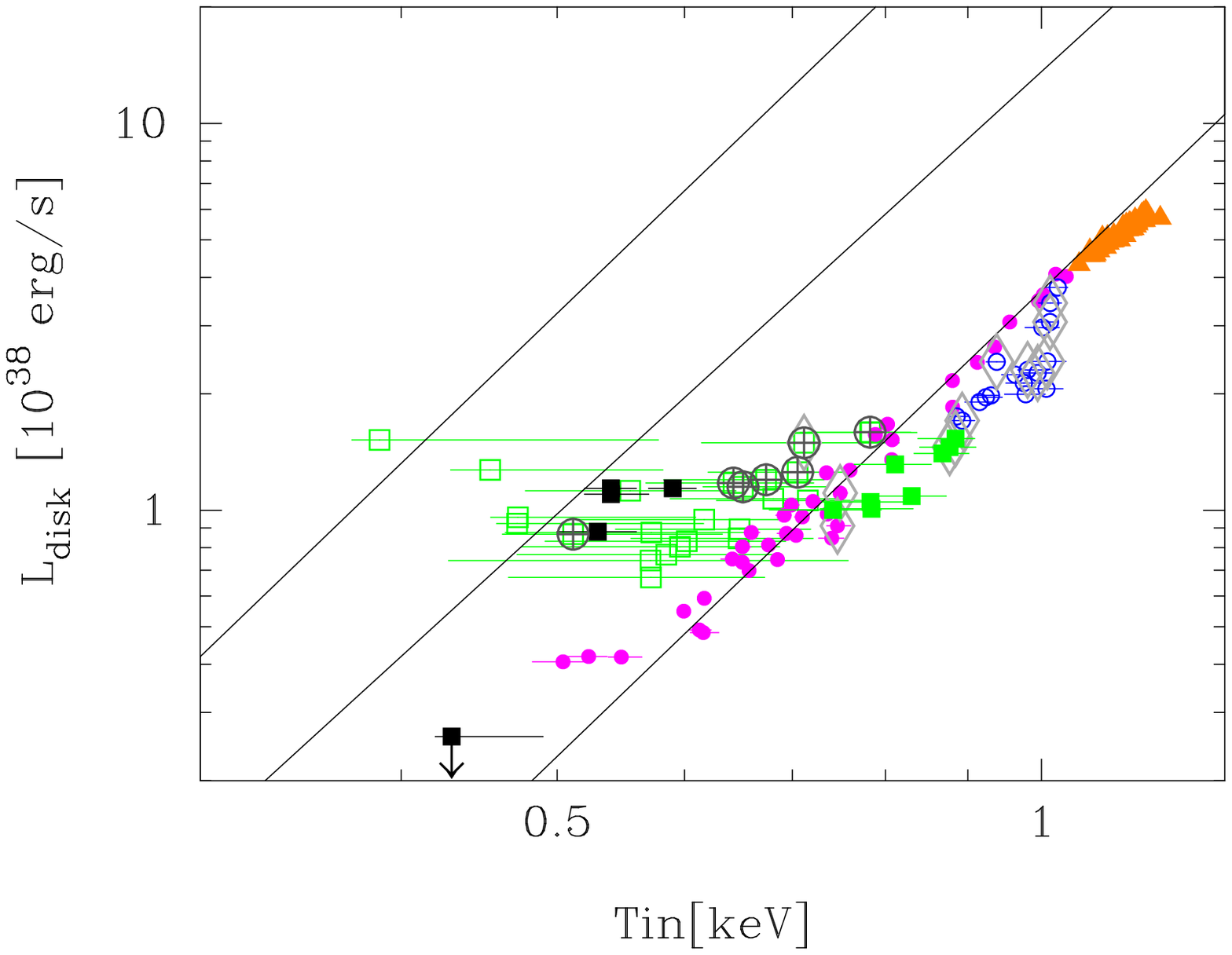}\\
\includegraphics[clip=true,width=0.45\textwidth,angle=0]{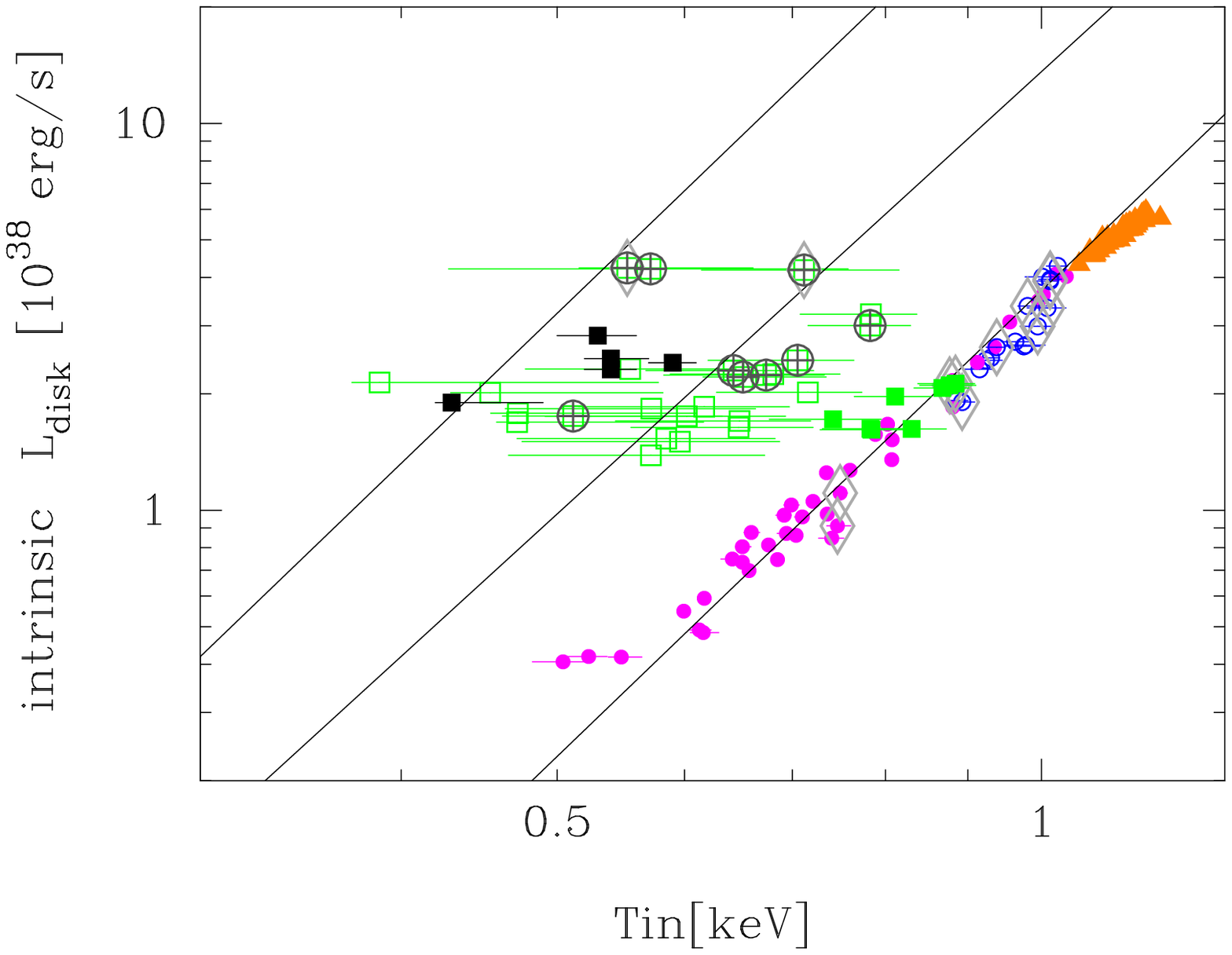}
&
\includegraphics[clip=true,width=0.45\textwidth,angle=0]{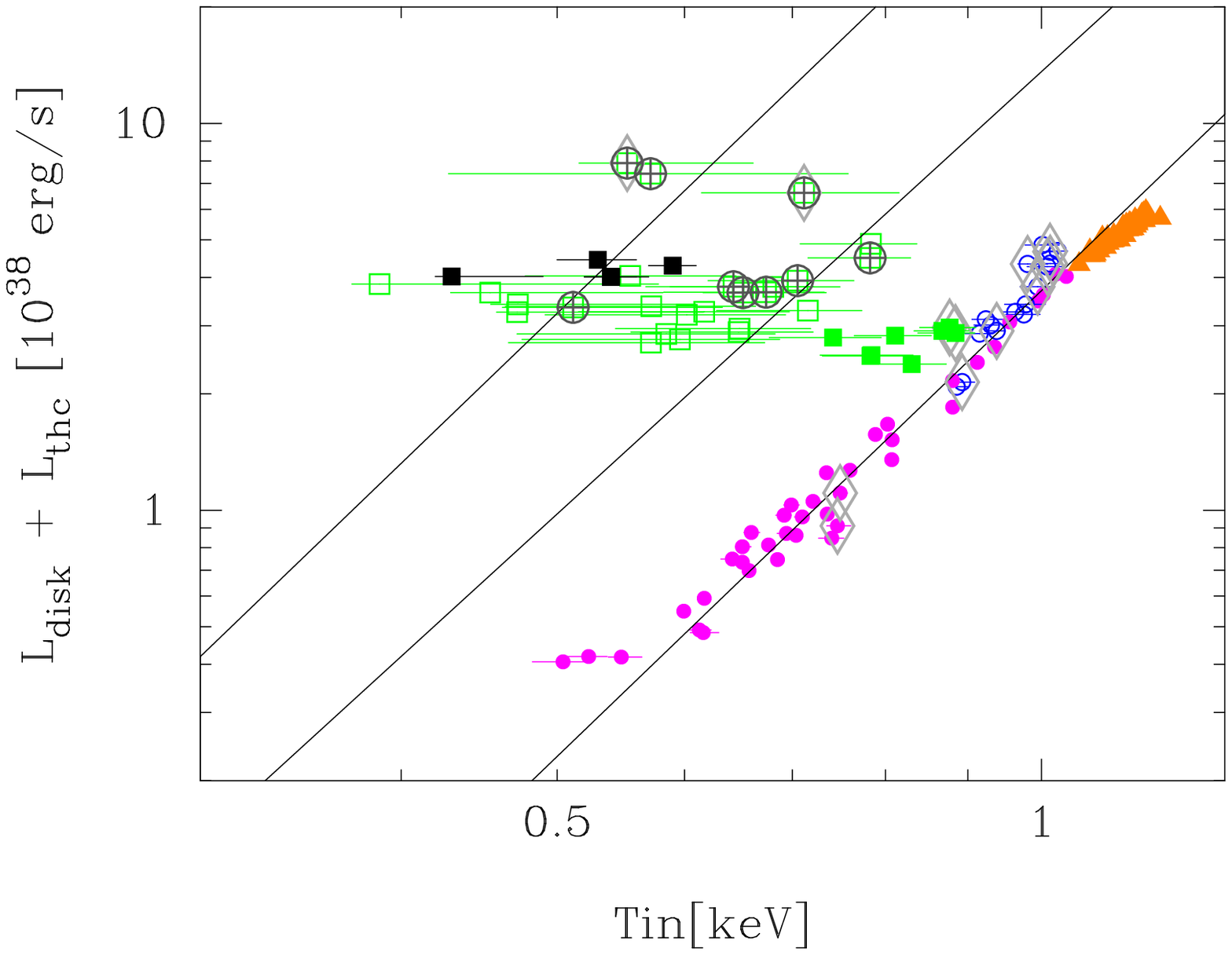}
\end{tabular}
\end{center}
\caption{Luminosity -- temperature relations for XTE J1550--564
as derived from model fits of ($a$) a disc plus power law tail ($b$) an
additional steep thermal Comptonization component. This illustrates
the importance of fitting physical models for the tail when it is
energetically dominant (very high state) though when the tail is weak (soft state) this
has little effect. ($c$) shows the reconstructed disc luminosity
assuming that all the energy in the Comptonized tail was derived from
the disc, while ($d$) shows the reconstructed disc luminosity assuming
that all the photons in the tail came from the disc. Plainly, there
are some extremely Comptonized very high-state spectra which cannot be matched
easily to an untruncated disc geometry (Kubota \& Done 2004). }
\label{fig:vhs_tlum}
\end{figure}

Instead, the difference in behaviour of different sources at the
same inferred temperature and luminosity might be explained by
different inclinations as these deviations are most noticeable for
highly inclined sources. Both GRO~J1655--40 and XTE J1550--564 are
at $\sim70^\circ$, and a further source 4U~1630--47 where the bend
is also seen (Abe et al.\ 2005, not included here as the
interstellar absorption is very high) is also likely to be at
similarly high inclinations (Tomsick, Lapshov \& Kaart 1998).
Conversely, 4U~1543--47 is at fairly low inclination (Orosz et al.
1998), and LMC X-3, while not well determined, must be less than
$\sim 70^\circ$ (Cowley et al.\ 1983). Inclination could affect the
observed spectrum in several ways, perhaps by the disc starting to
become geometrically thick and self--shielding part of its inner
regions, and/or that an equatorial wind from the disc (see
Section~\ref{sec:winds}) starts to become optically thick, with
$\tau_T>1$ so that the luminosity seen at high inclinations is only
$\exp(-\tau)$ (plus some geometry dependent contribution from
scattering into the line of sight) of the intrinsic flux seen at low
inclinations.

\subsection{Disc spectra in the very high state}
\label{sec:vhs_disc}

Much larger deviations from the standard $L\propto T^4$ relation are
seen in soft states where the spectra are no longer disc dominated,
and the tail carries more than $\sim$ 30 per cent of the luminosity.
The importance of the tail underneath the disc emission in these
very high/intermediate-state spectra means that details of how this
is modelled can affect the derived parameters of the disc. A simple
and often used approximation for the shape of the tail is a power
law. However, if the tail is produced by Compton scattering of disc
photons then there is a low energy break at energies close to the
seed photons. This gives a much lower continuum flux underneath the
disc spectrum, so leads to a larger inferred disc luminosity
(Poutanen et al.\ 1997; Done, {\.Z}ycki \& Smith 2002). This effect
recovers the expected untruncated disc properties in spectra where
the Comptonized component is less than 50 per cent of the bolometric
flux (Kubota et al.\ 2001; Kubota \& Done 2004). Additional effects
then become important, as Comptonization conserves photon number.
When the number of photons in the Comptonized spectrum is not
negligible compared to those in the disc then the intrinsic disc
spectrum is brighter than observed (Kubota \& Makishima 2004; Kubota
\& Done 2004). This helps recover the expected untruncated disc in
even more strongly Comptonized spectra, but for the most extreme
very high state the disc still appears distorted (Done \& Kubota
2006; see Fig.~\ref{fig:vhs_tlum} and Section~\ref{sec:vhs_corona}).

This supports the idea that there are two types of very high state
geometry, one where Comptonization is stronger than in the soft
state, but where the disc extends down to the last stable orbit, and
one where the Comptonization is so strong that the underlying disc
structure starts to change (Done \& Kubota 2006). This is similar to
the suggested distinction between the intermediate and very high
states of Esin et al.\ (1997).

\subsection{The high-energy tail in the soft and very high states}
\label{sec:hybrid}

The soft state is always accompanied by a small fraction of emission
in a tail extending to much higher energies. The shape of this tail
can be roughly modelled by a power law of photon index $\Gamma\sim
2-2.2$ extending out beyond 500~keV (Gierli{\'n}ski et al.\ 1999).
This index remains remarkably constant in the soft state
irrespective of whether the tail carries 1 or 10 per cent of the
total power.  If the tail was produced by thermal Comptonization
then this would imply that \lhls\ is fairly constant at $\sim$2 in
the energetic electron region, which is easy to arrange if the seed
photons are produced predominantly by reprocessing (Haardt \&
Maraschi 1993). Since the intrinsic disc flux is high then this
implies that the X-ray regions cover only a very small section of
the disc, motivating the magnetic flare geometry sketched in
Fig.~\ref{fig:states} (Poutanen et al.\ 1997).

However, the high-energy tail is {\em not} produced by thermal
Comptonization.  To extend out to 500~keV and beyond requires a very
high temperature, so the X-ray region must have a rather small
optical depth in order to give $\Gamma\sim 2$. This leads to a bumpy
spectrum, with individual Compton scattering orders separated rather
than merging into an approximate power law (Gierli{\'n}ski et al.
1999). The observed smooth spectrum can instead be produced by
Compton scattering on a non-thermal electron population, where the
index is set predominantly by the shape of the electron distribution
rather than \lhls. This removes the constraints on geometry, since
there is now no requirement for the seed photons to be reprocessed.
Instead, the observed constancy of the spectral index implies some
mechanism for fine tuning the electron acceleration process which is
not yet understood.

While the tail is clearly non-thermal, it is not well fit in detail by
a power law spectrum and its reflection (Gierli{\'n}ski et al.\ 1999;
Zdziarski et al.\ 2001). While some of this may be due to
uncertainties in modelling ionized reflection (see Section
\ref{sec:lhs}) the continuum shape should also be complex. The self
consistent electron spectrum is set by a balance between the
accelerative heating with cooling. The acceleration process may
produce a power law electron distribuition, but the cooling is more
complex. High-energy electrons will cool via Compton scattering (which
does produce a power law distribution) but electron--electron
(Coulomb) collisions are more important than Comptonization for the
low energy electrons. This is a collisional process so produces a
thermal distribution. Thus the self consistent electron distribution
is hybrid, being thermal at low energies with a non-thermal tail to
high energies (Coppi 1999).  This produces complex curvature in the
Compton scattered spectrum, with a mixture of thermal and non-thermal
features which fit very well to the observed soft-state spectra, with
the inclusion of ionized reflection (Gierli{\'n}ski et al.\ 1999;
Ibragimov et al.\ 2005).

\begin{figure}
\begin{center}
\includegraphics[clip=true,width=0.7\textwidth,angle=0]
{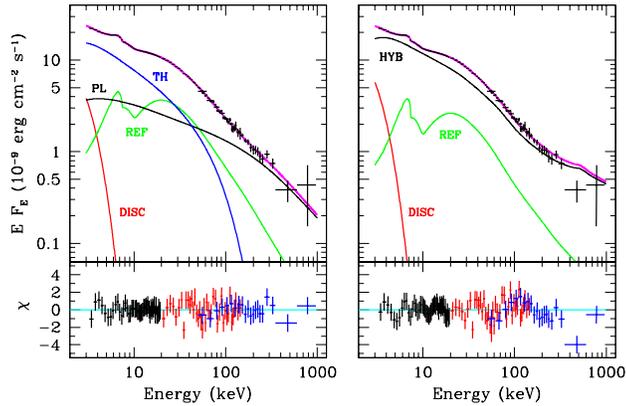}
\end{center}
\caption{Two possible spectral decompositions of the
extreme very high state seen in XTE~J1550--564. The left hand panel shows the fit
assuming that there are two regions of energetic electrons,
one thermal, connected
to the hot inner flow (blue: TH), and one non-thermal (black:
PL). The power-law distribution of electrons does not necessarily give
a power-law photon spectrum. The right panel assumes there is only one
region, where the electron distribution is a hybrid, being thermal at
low Lorentz factors and non-thermal at higher ones. The reflection of
the intrinsic spectrum (green: REF) is always required to highly
ionized and  hence is somewhat uncertain (from Gierli{\'n}ski \& Done 2003).}
\label{fig:vhs_spec}
\end{figure}

The very high state tail is qualitatively similar, showing a mix of
thermal and non-thermal features, but this dual nature can be seen
more clearly than for the soft state.  Simple power law fits do not
give even a fair approximation to the shape of the tail as shown
e.g. Sobczak et al. 1999b, who used different power law indices to
fit the 3--20~keV and 20--200~keV data from this state in XTE
J1550--564. The shape of the tail is somewhat different from the
soft state also: the non-thermal high energy tail is steeper, and
the thermal low energy Comptonization carries a larger fraction of
the total power (Zdziarski et al.\ 2001; Gierli{\'n}ski \& Done
2003). Ionized reflection again adds to this continuum complexity.

Fig.~\ref{fig:vhs_spec} shows a range of possible models which can
fit the tail seen in an extreme very high-state spectrum. It can be
described either by two separate regions, one where there is low
temperature, moderate optical depth ($\tau\sim 2$), thermal Compton
scattering, together with a separate non-thermal electron region
(which is again a hybrid rather than simple power law).
Alternatively the thermal and non-thermal electrons can be
co-spatial, but where not all the intrinsic acceleration is
non-thermal. Some fraction of the power has to directly heat the
electrons in order for the thermal Comptonization to be as important
as observed (Gierli{\'n}ski \& Done 2003).

\subsubsection{Geometry of electron acceleration region in soft and very high states}
\label{sec:vhs_corona}

\begin{figure}
\begin{center}
\begin{tabular}{cc}
\includegraphics[clip=true,width=0.45\textwidth,angle=0]
{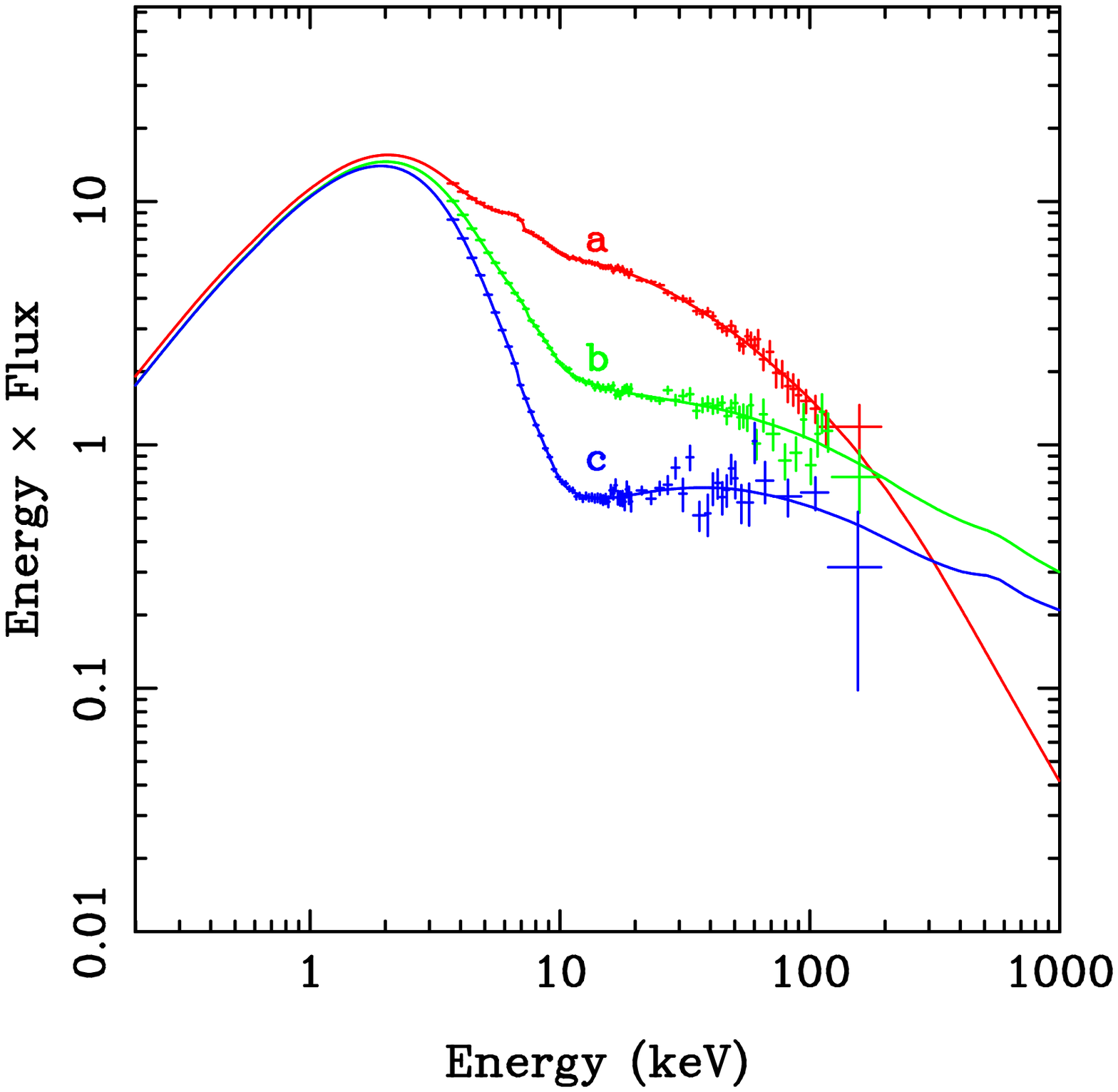} &
\includegraphics[clip=true,width=0.45\textwidth,angle=0]
{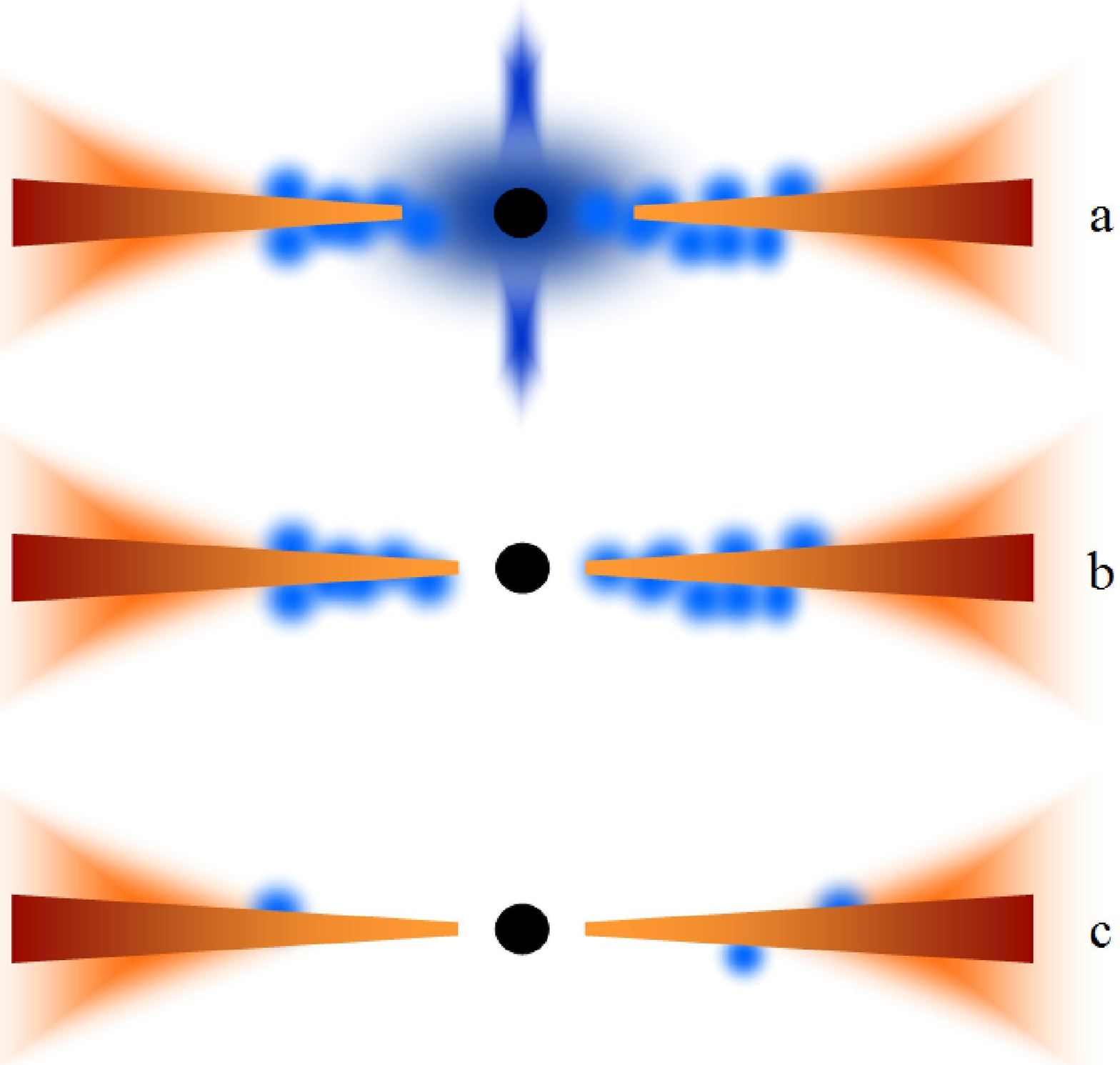}
\end{tabular}
\end{center}
\caption{Left: Extreme very high-state spectrum ($a$) merging smoothly onto
less Comptonized very high ($b$) and then into a soft-state spectrum
($c$). All data taken from XTE~J1550--564. Right: A range
of geometries which can smoothly connect between
extreme very high and soft states.}
\label{fig:trans_vhstds}
\end{figure}

The geometry of the electron region can be constrained from the
spectra, and from the spectral evolution.
Fig.~\ref{fig:trans_vhstds} shows that the very high state can
smoothly connect onto the soft state. The shape of the soft state
spectra, with the disc being dominant, clearly shows that few of
these seed photons are intercepted by the hot electrons. This can be
made more quantitative. The electrons scatter only a fraction $C_f
[1- \exp(-\tau) ]$ where $C_f$ is the covering fraction of the
energetic electron regions over the disc and $\tau$ is their optical
depth. Thus either $C_f$ or $\tau$ must be small in the soft state.

This is plainly not true in the extremely Comptonized very high
state spectra, where the disc spectrum merges smoothly into the
Comptonized emission. Both $C_f$ {\em and} $\tau$ {\em must} be
large in these spectra, though for the more weakly Comptonized very
high state the disc is more visible, requiring that at least one of
$C_f$ or $\tau$ decreases from the extreme very high state through
to the soft state (Fig.~\ref{fig:trans_vhstds}).

\begin{figure}
\begin{center}
\begin{tabular}{cc}
\includegraphics[clip=true,width=0.45\textwidth,angle=0]
{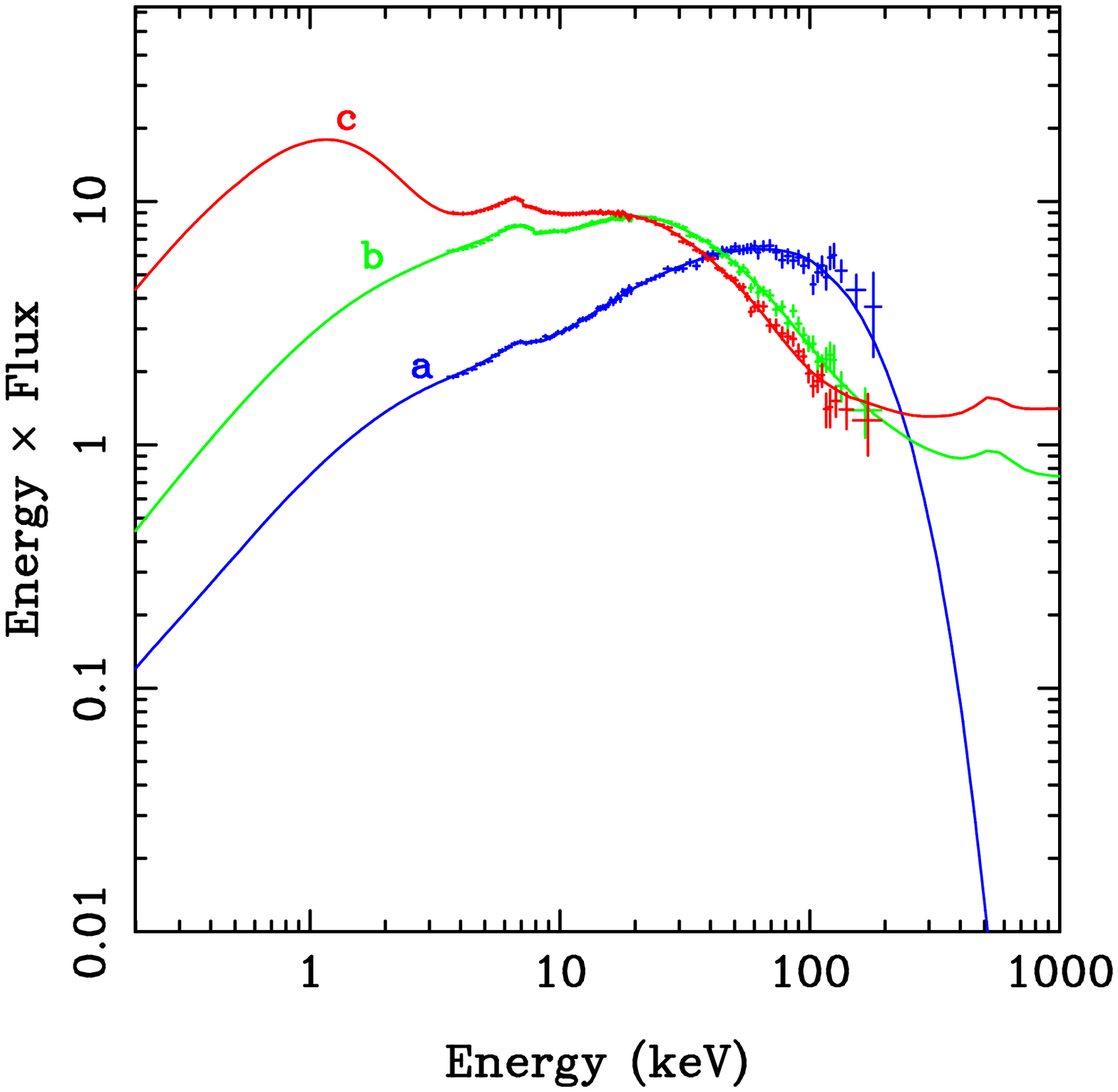} &
\includegraphics[clip=true,width=0.45\textwidth,angle=0]
{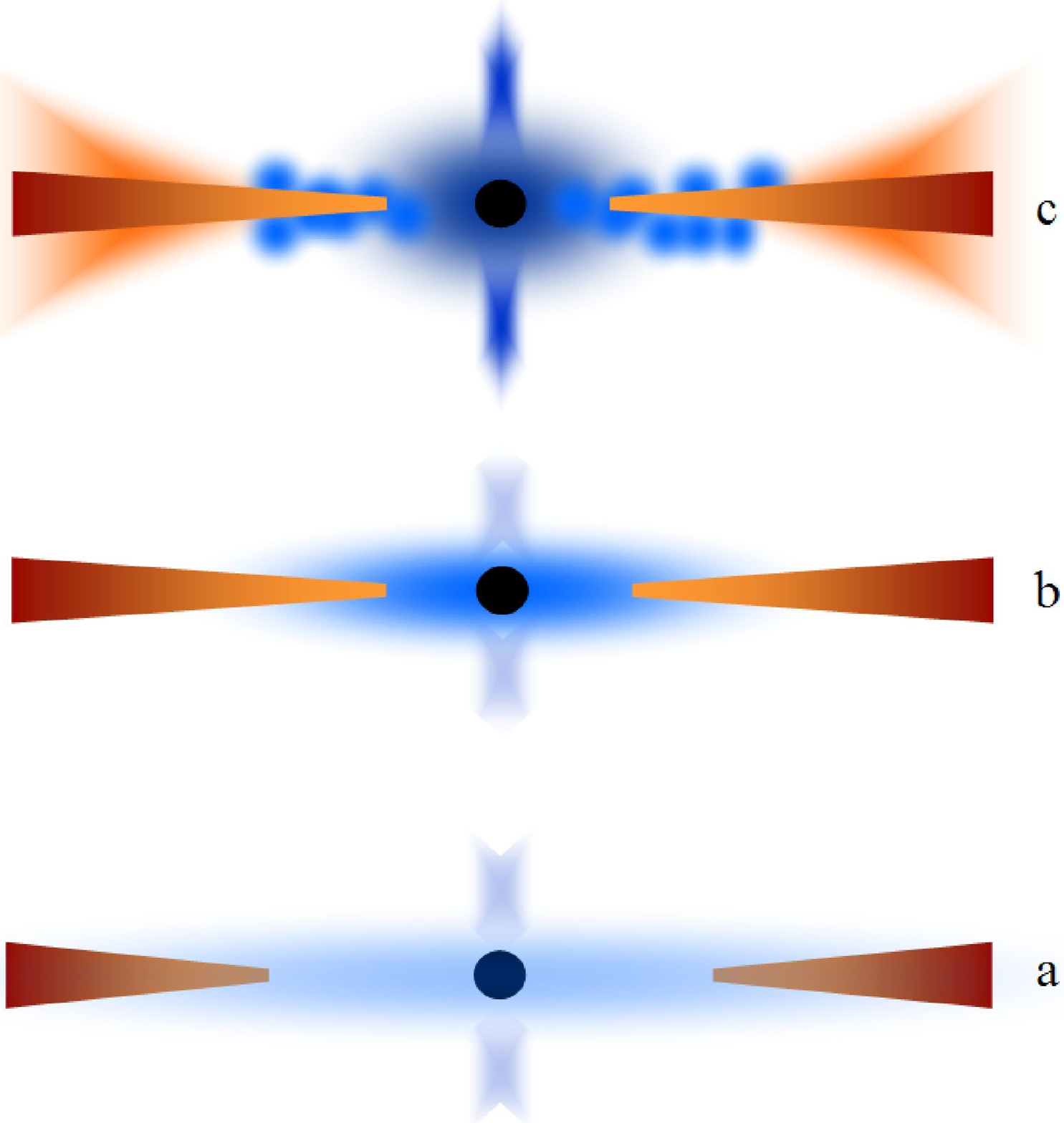}
\end{tabular}
\end{center}
\caption{Left: hard-state spectrum ($a$) merging smoothly onto
less Comptonized very high state ($b$) and then into an extreme very high-state spectrum ($c$)
during the rapid rise to outburst. All data taken from XTE J1550--564.
Right: A range
of geometries which can smoothly connect between the hard state
and extreme very high-state spectra.}
\label{fig:trans_lhsvhs}
\end{figure}

A further constraint comes from the fact that the very high state is
seen to merge smoothly into the hard state state on the rise to
outburst (see Fig.~\ref{fig:trans_lhsvhs}). Given that the extreme
very high-state spectra indicate both thermal and non--thermal
electrons then it is plausible that the thermal component is from
the remains of the hot inner flow, while the non--thermal is from
magnetic flares above the disc (see also Fig.~\ref{fig:vhs_spec}). A
detailed analysis of the disc spectrum also supports the idea that
the disc is slightly truncated in these extreme very high state
(Done \& Kubota 2006), though there are substantial uncertainties in
the models of disc spectra at this point. Nonetheless, the range of
very high state geometries shown in Fig.~\ref{fig:trans_vhstds} and
\ref{fig:trans_lhsvhs} match all current constraints, and form a
bridge between the hard and soft states.

\section{Spectral evolution}
\label{sec:colcol}

The previous sections outlined a physically motivated model in which
spectral changes (particularly the hard-soft spectral transition)
are driven by a changing geometry. There now exists an enormous
amount of data from the X-ray binary systems which can be used to
test this. Done \& Gierli{\'n}ski (2003) systematically analyzed all
the available spectra from many black hole systems, using broad band
`colours' to get an overview of the source behaviour. They fit
physically motivated spectral models to the data in order to get
intrinsic fluxes i.e. corrected for absorption and the detector
response. Ratios of these fluxes roughly relate to the mean spectral
slope across the given energy bands.  Many black holes can then be
plotted on the same diagram, as shown in Fig.~\ref{fig:colcol}. {\em
All} the black holes are consistent with the {\em same} spectral
evolution track on these colour--colour diagrams, and this colour
evolution can be well matched by Comptonization models which are
based on the geometry changes described in the sections above (black
lines superposed on the colour--colour data in
Fig.~\ref{fig:colcol}; Done \& Gierli{\'n}ski 2003).

\subsection{Hysteresis}
\label{sec:hysteresis}

\begin{figure}
\begin{center}
\begin{tabular}{cc}
\includegraphics[clip=true,width=0.45\textwidth,angle=0]
{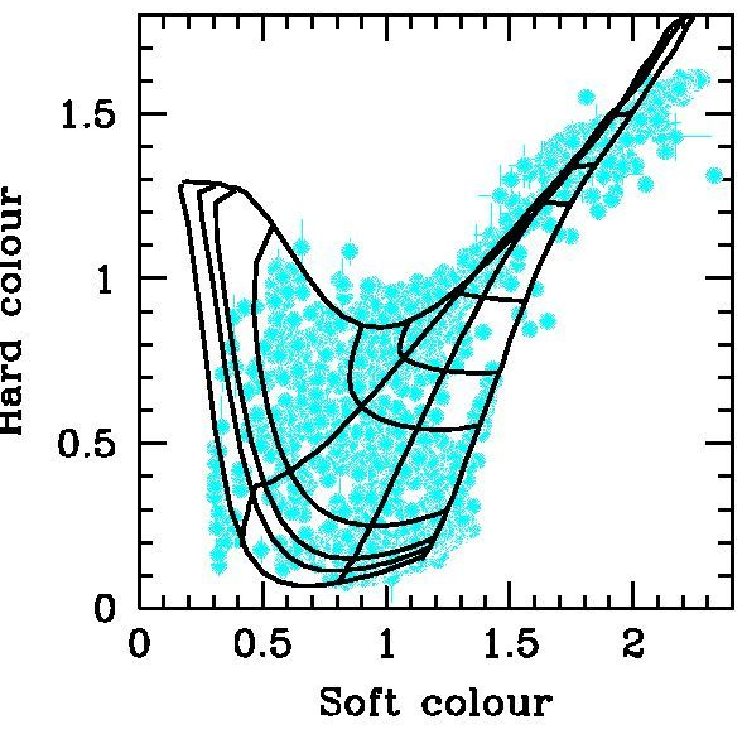} &
\includegraphics[clip=true,width=0.4\textwidth,angle=0]
{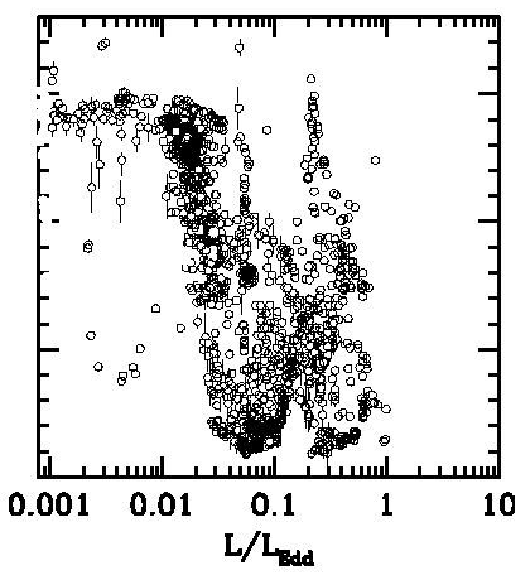}
\end{tabular}
\end{center}
\caption{Left: Colour--colour diagrams of BHBs
(after Done \& Gierli{\'n}ski 2003), overlaid with model tracks
which incorporate the changing geometry of the disc, hot flow and corona.
Right: Colour--luminosity plot of BHB showing how the same spectrum is
seen at very different luminosities. This scatter is seen in
individual objects, where the characteristic hard--to--soft spectral
transition occurs at higher luminosity on the rapid rise to outburst
than on the decline.}
\label{fig:colcol}
\end{figure}

Fig.~\ref{fig:colcol} shows that this well ordered behaviour in
terms of the evolution of spectral shape does {\em not} uniquely
correlate with $L/L_{\rm Edd}$. Many different spectral states can
be seen at a given $L/L_{\rm Edd}$, or conversely, a given spectral
transition e.g. hard-to-soft can be seen at multiple values of
$L/L_{\rm Edd}$.

This is not just due to uncertainties in distance, mass etc. The
{\em same} black hole can show a {\em different} transition
luminosity at different times (see e.g. Remillard \& McClintock
2006).  There is no one--to--one mapping between spectral state and
$L/L_{\rm Edd}$ in most BHB. This is the well known hysteresis
effect (e.g. Nowak 1995, Maccarone \& Coppi 2003), which is most
obvious in terms of the luminosity at which the major hard--to--soft
transition takes place, with hard state having larger luminosities
on the rapid rise to outburst in the hard/very high/soft state
transition than the reverse on the slow decline. This effect occurs
in {\em all} BHB with high enough signal--to--noise to track the
transition except for Cyg X-1 (e.g. Maccarone \& Coppi 2003). Thus
hysteresis is suppressed in some way in this system, and there are
two plausible potential candidate mechanisms for this, and both are
based on the fact that Cyg X-1 is the only bright HMXB BHB source.
The first possibility is that hysteresis is caused by the dramatic
changes in accretion flow during the H ionization instability.  The
high mass companion to Cyg X-1 has such a large mass transfer rate
that the disc is above the H ionization temperature everywhere (see
Section~\ref{sec:disc}) so does not show the violent instability.
Alternatively, since the companion does not completely fill its
Roche lobe, it may be accreting low angular momentum material from
the stellar wind, so forming a rather different accretion flow
(Smith, Heindl \& Swank 2002; Maccarone \& Coppi 2003). A comparison
with the neutron star LMXB favours the former explanation as the
only two systems to show large scale hysteresis are also the only
two in which the disc shows the dramatic H ionization instability
(Gladstone, Done \& Gierli{\'n}ski 2007). The consequent dramatic
change in mass accretion rate takes much longer to propagate through
the thin disc than through the hot inner flow (Smith, Heindl \&
Swank 2002), allowing the system to access non-equilibrium states.
This is a rather different model for the observed behaviour than
those which assume there are two potential steady state solutions at
a given mass accretion rate, where the solution chosen depends on
the past history of the flow (e.g. Zdziarski \& Gierli{\'n}ski 2004)

%% file: ns/ns.tex
\section{Neutron star spectra}
\label{sec:ns}

Black holes and neutron stars have very similar gravitational
potentials as neutron star radii are of the order of three
Schwarzschild radii, the last stable orbit of material around a
black hole. Thus the gravitational potential in which the accretion
flow is embedded is very similar between the two objects, and the
models developed for the black holes should simply carry over to the
neutron star systems.

However, there is a fundamental difference between the two classes
of objects.  Neutron stars have a solid surface, while black holes
do not. Surface processes such as X-ray bursts (from nuclear burning
of the accreted material onto the surface), or coherent pulsations
(from a residual magnetic field) are unique signatures of neutron
stars. These are a sufficient but not necessary condition for a
surface: not all neutron star systems show these (e.g. the review by
Lewin, van Paradijs \& Taam 1993)

By contrast, the boundary layer between the accretion flow and the
surface should {\em always} be present.  In Newtonian gravity an
accretion disc can radiate only half of the gravitational potential
energy with the other half stored as kinetic energy of the rotating
material. This kinetic energy is all radiated at the surface in a
boundary layer if the surface is stationary. In General Relativity
the energy in the boundary layer is even larger, about twice that of
the disc (Sunyaev \& Shakura 1986; Sibgatullin \& Sunyaev 2000).
Neutron stars can of course be rapidly rotating, but even the
fastest confirmed millisecond pulsar PSR J1748--2446ad (at 716 Hz) is
rotating at only approximately half the Keplerian period (Hessels et
al. 2006), where the energy released in the boundary layer should
still be as much as that in the disc (Sibgatullin \& Sunyaev 2000).

Thus the expectation is that accreting neutron stars with low
magnetic fields ($B<10^{8-9}$~G, so that this does not affect the
dynamics of the accretion flow) should have accretion flows which
are similar to black holes at the same $L/L_{\rm Edd}$, but with the
addition of a boundary layer with luminosity comparable to that of
the disc.

Low magnetic fields are found only in the LMXB NS systems. There are
no known NS HMXB with low fields, nor are there any known NS LMXB
with high fields. This can plausibly be explained as dissipation of
the high birth field of $\ge 10^{12}$~G in accretion torques during
the long term evolution of the binary to Roche lobe overflow (see
e.g. the review by Bhattacharya \& Srinivasan 1995). However, even
within the LMXB NS there are a range of B fields. The accreting
millisecond pulsars must have fields around $\sim 10^8$~G in order
for the magnetic pressure to dominate over the ram pressure of the
$L/L_{\rm Edd}\sim 0.05$ flow and hence produce the eponymous X-ray
pulsations (Chakrabarty \& Morgan 1998). The rest of the LMXB NS
systems must have lower surface fields, or they too would produce
similar pulsations during periods of similarly low mass accretion
rate (Vaughan et al 1994; Maccarone \& Coppi 2003). This range of
surface field can be explained from the differences in {\em long
term} mass accretion rates. The millisecond pulsars are unstable to
the disc instability, and are bright only during short outbursts
(see Fig.~\ref{fig:ns_lc}). Their mean mass accretion rate is much
lower than all the other NS LMXB (e.g. Gladstone, Done \&
Gierli{\'n}ski 2007), and this is insufficient to bury their surface
field under the accretion flow, where it can only diffuse out on
timescales of $\sim 100$ years (Cumming, Zweibel \& Arras 2001).

Neutron star LMXBs fall into two categories: atolls (including all the
millisecond pulsars: e.g. Van Straaten, van der Klis \& Wijnands 2005)
and Z sources (Hasinger \& van der Klis 1989). These differ in
luminosity (as well as several other aspects, see
Section~\ref{sec:sens}), with the Z sources being typically brighter
($>$0.5 $L_{\rm Edd}$) while the atolls are seen over the same range
of luminosities as the black holes discussed in the previous sections
(from $<$ 10$^{-3}$ up to $\sim L_{\rm Edd}$). Thus here we consider
only the atolls and millisecond pulsars, as these form a matched
luminosity sample to compare with the BHB.

\begin{figure}
\begin{center}
\includegraphics[clip=true,width=0.8\textwidth]
{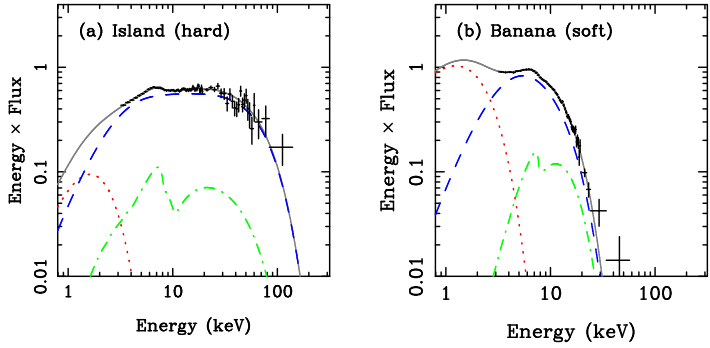}
\end{center}
\caption{Two spectral states of an atoll 4U 1705--44 from {\it
RXTE}. The panels show unfolded and unabsorbed X-ray spectra,
together with the best-fitting models extrapolated below the lower
PCA bandwidth limit (3 keV) in order to demonstrate the soft
component. (a) the hard (island) state (observation id.
40051-03-12-00) fitted by a blackbody (dotted red curve), its
thermal Comptonization (dashed blue) with reflection (dash-dot
green). (b) the soft (banana) state (observation id. 40051-03-14-00)
fitted by a disc blackbody (dotted red), thermal Comptonization of
unseen seed photons hotter than the disc (dashed blue) and
reflection (dash-dot green).}
\label{fig:nsstates}
\end{figure}

The most obvious point of similarity is that the atolls switch
between two distinct spectral states (e.g. Gierli{\'n}ski \& Done
2002; Maccarone \& Coppi 2003): the soft `banana' and the hard
`island', so named from the shapes they make on a colour--colour
diagram. Sources which show transitions between these states trace
out a C (or atoll) shaped path from the hard island state at the top
right of the C down to the lower banana branch at the bottom of the
C and then curving slightly upwards and to the right (bottom right
hand end of the C, termed the upper banana branch) as the source
luminosity increases further (see e.g. Hasinger \& van der Klis
1989). The corresponding spectra are shown in
Fig.~\ref{fig:nsstates}, and the correlation with the BHB soft--hard
transition is clear, with the island state corresponding to the hard
state and the banana branch corresponding to the soft states
(soft/very high).

\subsection{Spectral evolution in atolls}
\label{sec:atolls}

\begin{figure}
\begin{center}
\begin{tabular}{cc}
\includegraphics[clip=true,width=0.6\textwidth]
{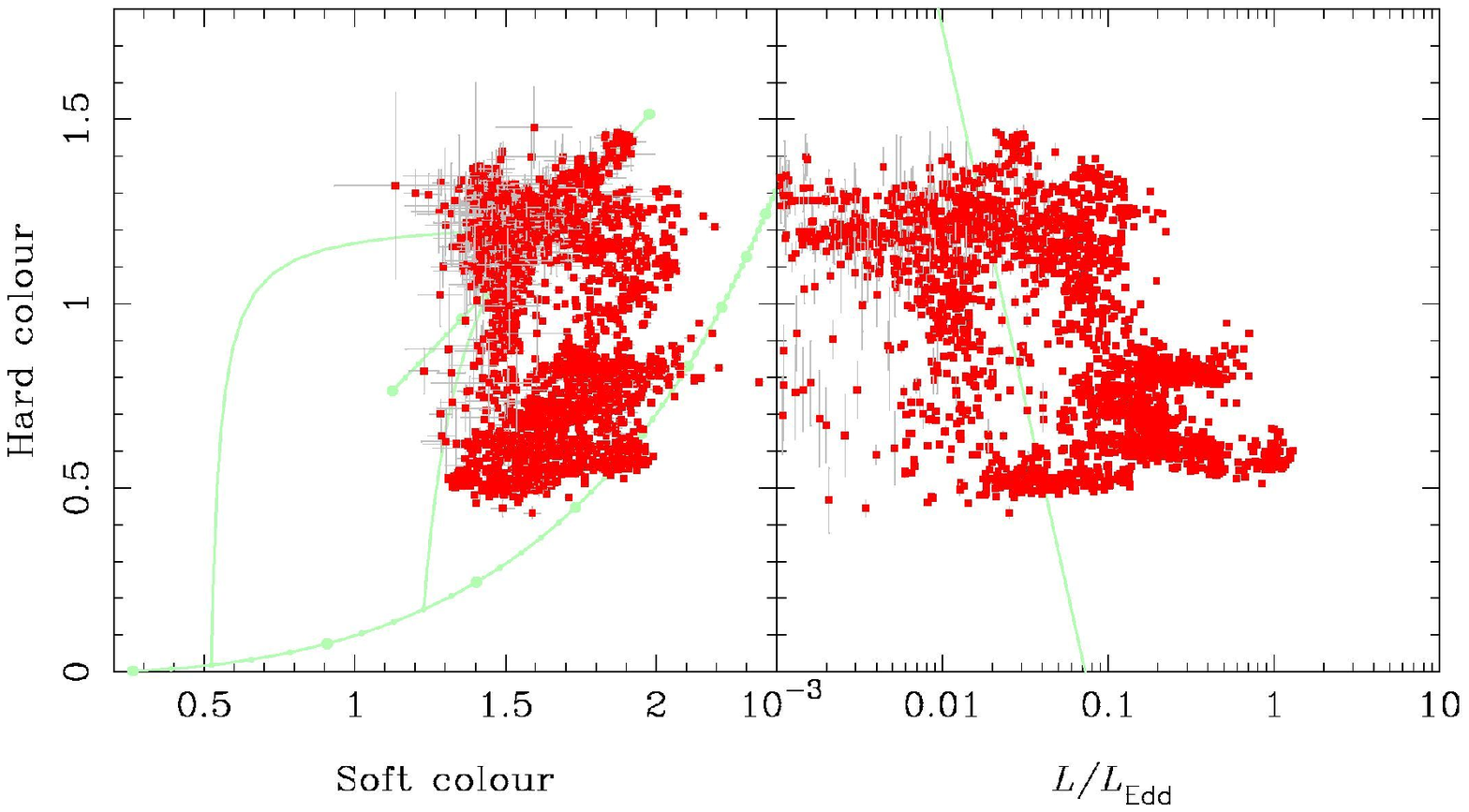} &
\includegraphics[clip=true,width=0.36\textwidth]
{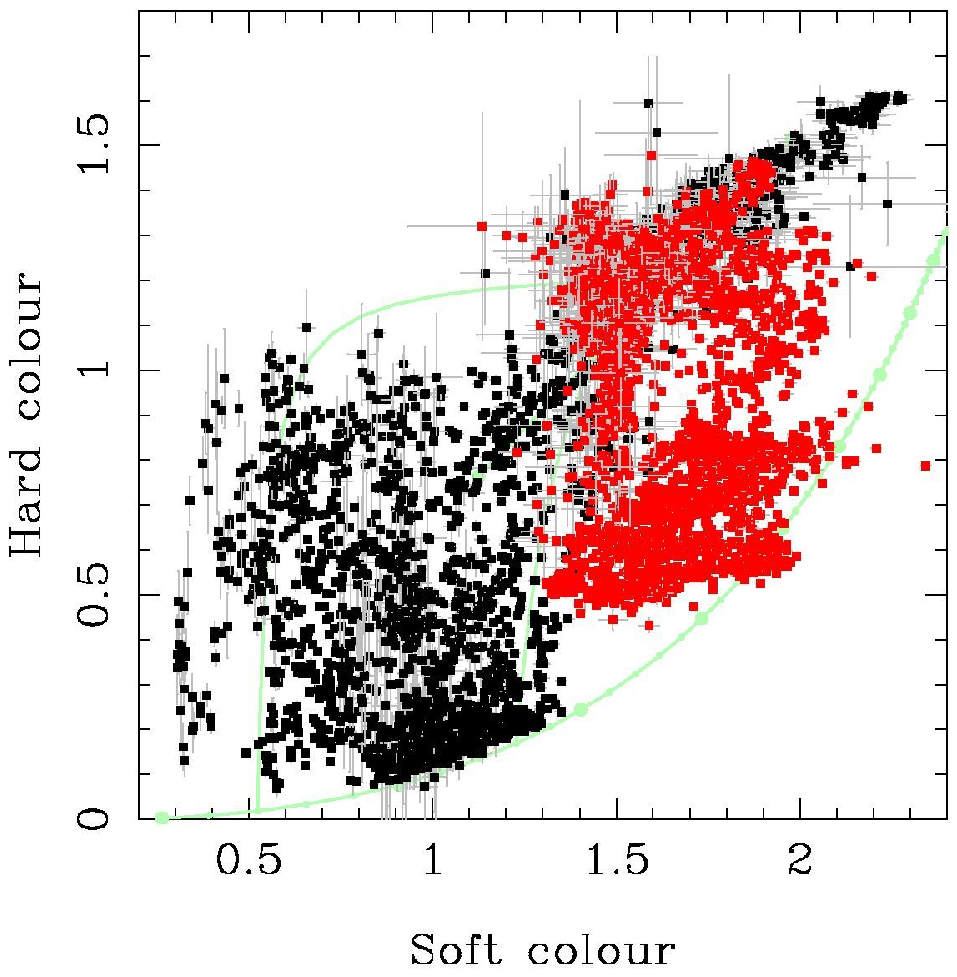}
\end{tabular}
\end{center}
\caption{Left: Colour--colour and colour--luminosity diagrams for all
  low column atolls and millisecond pulsars in Gladstone et al
  (2007). Right: a comparison of
  BHB (black) and neutron star (red) spectral evolution. Plainly there are
  significant differences between these two types of object,
  consistent with the existence of an event horizon in black holes, in
  contrast with the known surface of the neutron star (after Done \&
  Gierli{\'n}ski 2003).}
\label{fig:colcol_ns}
\end{figure}

The left panel of Fig~\ref{fig:colcol_ns} shows this hard--to--soft
spectral transition even more clearly in the colour--colour and
colour--luminosity diagrams for a sample of millisecond pulsars and
atolls (Gladstone et al 2007). These colours are exactly equivalent to
the ones used for the BHB in Fig.~\ref{fig:colcol}, so can be directly
compared (Done \& Gierli{\'n}ski 2003, right panel of
Fig~\ref{fig:colcol_ns}). The very similar gravitational potential and
very similar range in $L/L_{\rm Edd}$ give rise to very different
spectral evolution between BHB and neutron stars.

This difference can be interpreted by using the same truncated disc
model, outlined in Secs.  \ref{sec:hardsoft} and \ref{sec:colcol}, but
taking into account the presence of the solid surface (see
Gierli{\'n}ski \& Done 2002; Done \& Gierli{\'n}ski 2003).  This can
impact the spectrum in two separate ways, firstly through its direct
thermal emission, and secondly through the additional energy released
in the boundary layer.

The truncated disc model for the lowest mass accretion rates in BHB
has few seed photons from the disc illuminating the inner hot flow.
The resulting spectrum has weak disc emission, and a hard
Comptonized spectrum. In the NS at correspondingly low accretion
rates, the disc should be similarly truncated, while the hot inner
flow joins smoothly onto the optically thin boundary layer (Medvedev
\& Narayan 2001; Medvedev 2004). Thus the hot flow is (at least)
twice as luminous as before, so might be expected to be harder than
in BHB. However, the neutron star surface provides an additional
source of seed photons as it is heated by irradiation and/or
conduction. Since these are at the centre of the hot flow then they
form the dominant seed photon flux.  For the expected
quasi--spherical geometry then only a fraction $1-e^{-\tau}$ are
seen uncomptonized. Nonetheless, these residual surface photons can
be the source of the soft component seen in the island state
(Gierli{\'n}ski \& Done 2002b), as the disc is at lower energies due
to truncation and (similarly to the BHB, see Section~\ref{sec:lhs})
it is made more difficult to see by the often high Galactic
absorption column.

The additional source of seed photons from the NS surface means that
NS spectra are never quite as hard as the hardest BHB (Done \&
Gierli{\'n}ski 2003). It also changes the evolution of the spectrum
as a function of luminosity compared to the BHB: NS move
horizontally to the right on a colour--colour diagram, when
brightness increases (Gierli{\'n}ski \& Done 2002a; Muno, Remillard
\& Chakrabarty 2002), whereas the BHB move diagonally (see
Section~\ref{sec:colcol}). This can be explained by the increasing
temperature of the neutron star surface as accretion rate (and
optical depth of the hot flow) increases. With the seed photon
temperature rising, the energy at which the Comptonized spectrum
rolls over at its low-energy end increases, and the contribution of
the uncomptonized soft photons from the NS surface decreases.
Therefore, the soft X-ray part of the spectrum becomes harder and
the soft colour increases. Since the seed photons are from the
neutron star surface rather than from the disc then the disc
truncation radius makes little difference to the spectrum. The seed
photon geometry does not change, nor does the ratio of seed photon
to hot electron luminosity since the seed photon luminosity is tied
to that of the hot flow (via irradiation and/or conduction).  Thus
the Comptonized spectrum remains the same at high energies, away
from the seed photons so the hard colour stays the same, and the
source moves horizontally on the colour--colour diagram
(Gierli{\'n}ski \& Done 2002a, 2002b; Done \& Gierli{\'n}ski 2003).

\begin{figure}
\begin{center}
\includegraphics[clip=true,width=0.45\textwidth]{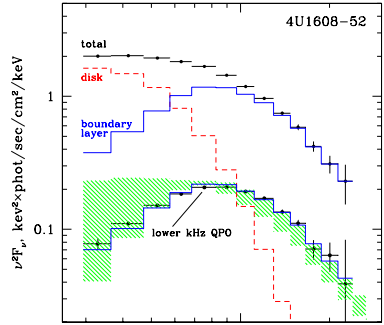}
\end{center}
\caption{Fourier-frequency resolved spectra of the boundary
layer in the atoll 4U~1608 at high $L/L_{\rm Edd}$ (banana branch).
The lower kHz QPO $\nu_l$ has the same spectrum as that inferred from
fitting the `Eastern' model to the total spectrum (black) i.e. where
the disc (red) is at lower energies and the Comptonized boundary layer
(blue) contributes at higher energies.
(from Gilfanov, Revnivtsev \& Molkov 2003).}
\label{fig:fres_bl}
\end{figure}

With further increasing accretion rate the disc eventually moves in
and the hot inner flow collapses into a thin disc. The increased
mass accretion rate is now all concentrated in the equatorial plane
and the boundary layer (or a spreading layer which has similar
properties: Inogamov \& Sunyaev 1999; Suleimanov \& Poutanen 2006)
becomes optically thick to electron scattering, but not yet thick
enough to completely thermalize into a blackbody (Popham \& Sunyaev
2001). Thus the boundary layer spectrum changes dramatically from
hot, optically thin Comptonization to much lower temperature (only
slightly higher than expected for complete thermalization),
optically thick Comptonization, giving a correspondingly dramatic
drop in hard colour from the island state to the banana branch. The
decrease in the inner disc radius is consistent with increasing
frequencies in power spectra as the source makes a transition from
the island state to banana branch (see e.g. the review by van der
Klis 2004; Section~\ref{sec:timing}).

However, detailed X-ray spectral fitting of the soft/banana state
shows that unambiguously decomposing the smooth curved shape into
disc and boundary layer components is difficult. Historically, there
were two different approaches to the spectra. In the Eastern model
(Mitsuda et al.\ 1989) the soft and hard components were identified
with the disc and Comptonized boundary layer, while in the Western
model (White, Stella \& Parmar 1988) they were attributed to
blackbody from the surface/boundary layer and a Comptonized disc,
respectively. However, this ambiguity is now resolved, most
compellingly by variability studies which show that the spectrum of
the variable component is hard. Since the most rapid variability
should be associated with the BL rather that the disc (as is also
the case in BHB: Churazov, Gilfanov \& Revnivtsev 2001), then this
is clear evidence for the `Eastern model' with a lower temperature
disc and hotter, Comptonized BL (Gilfanov, Revnivtsev \& Molkov
2003; Revnivtsev \& Gilfanov 2006; see Fig.~\ref{fig:fres_bl}).
Better spectral models of Comptonization which include the low
energy turndown of the Comptonized emission close to the seed photon
energy also break the spectral degeneracy, and show clearly that
even the spectra alone favour this model (Di Salvo et al.\ 2000;
Gierli{\'n}ski \& Done 2002b)

Thus there is strong evidence for the identification of the cooler
component on the banana branch as the disc, while the hotter one is
the boundary layer (Eastern model).  There is clear, though
indirect, spectral evidence for the NS surface emission as well from
the fact that the seed photon energy is higher than the observed
disc spectrum (Di Salvo et al.\ 2000; Gierli{\'n}ski \& Done 2002b;
Falanga et al.\ 2006). All three components can only be seen
directly around the transition, where the disc has high enough
temperature to contribute to the X-ray bandpass, while the boundary
layer is not yet completely optically thick so as to Comptonize all
the NS surface photons (Fiocchi et al.\ 2007).

\begin{figure}
\begin{center}
\begin{tabular}{cc}
\includegraphics[clip=true,width=0.35\textwidth,angle=0]
{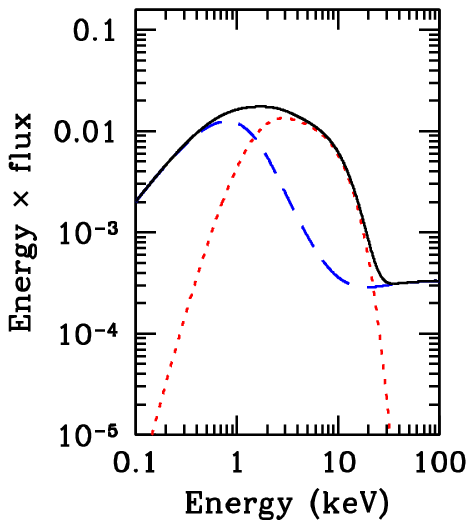} &
\includegraphics[clip=true,width=0.4\textwidth,angle=0]
{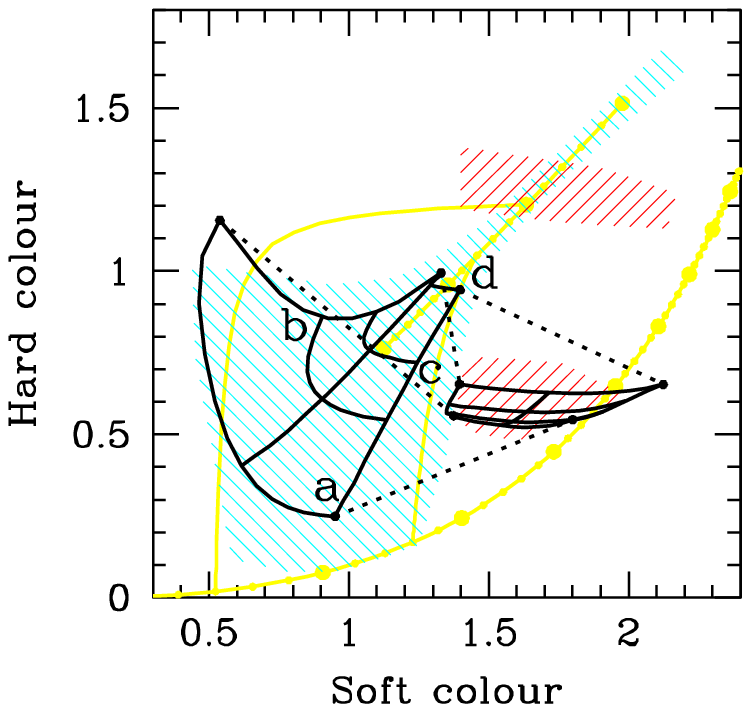}
\end{tabular}
\end{center}
\caption{The effect of the boundary layer.
Left: The dashed blue curve represents a soft-state spectrum of a
black hole with disc emission and a hard power-law tail. The red
dotted curves show additional emission from the boundary layer with luminosity
equal to that of the original spectrum. The total spectrum (solid
curve) represents a typical soft-state spectrum from a neutron LMXB.
A weak tail to higher energies,
observed in many neutron star sources (Paizis et al.\ 2006), may  be a
direct counterpart of the non-thermal Comptonization in black holes.
Right: Colour--colour diagram showing the areas occupied by
black holes (top left to bottom right cyan hatched area) and atolls
(top right to bottom left red hatched area). The grid on the left
represents a range of models (disc + hybrid Comptonization) which
describe the range of soft states from
black holes. Point $b$ on the grid corresponds to
the black hole model on the left panel (dashed blue curve). Adding a
boundary layer transforms the entire grid into the area covered by
the banana state of atolls.
}
\label{fig:ns_bl}
\end{figure}

As the luminosity increases along the banana branch then the disc
becomes more luminous and hotter. This should also follow a
luminosity $\propto T^4$ relation while the disc maintains a
constant inner radius, but decomposing the curvy spectra to this
accuracy is difficult (Done \& Gierli{\'n}ski 2002b; Takahashi \&
Makishima 2006; Lin, Remillard \& Homan 2007). The shape of the
Comptonization also changes, in a way which is consistent with it
becoming more optically thick as the accretion rate increases, and
hence becoming more saturated, approaching a blackbody
(Gierli{\'n}ski \& Done 2002b; Revnivtsev \& Gilfanov 2006). This
competes with the expected increase in temperature of the BL due to
increased luminosity, as more complete thermalization leads to a
lower effective temperature. Thus the BL temperature can remain
constant or even drop as the mass accretion rate increases
(Gierli{\'n}ski \& Done 2002b; Revnivtsev \& Gilfanov 2006), giving
a fairly constant hard colour, while the soft colour (set by the
disc) increases along the banana branch.

Fig.~\ref{fig:ns_bl} shows how the addition of the boundary layer to
the variety of soft states (soft and very high) BHB accretion flow
models reproduces the colour evolution along the banana branch. Thus
the NS LMXB are consistent with the same accretion flow models as
the BHB, where the major spectral evolution is driven by the
decreasing truncation radius of the disc with increasing accretion
rate, but with the addition of the boundary layer expected from the
NS surface. Conversely, the fact that there is a clear difference
between the spectral evolution of BHB and NS LMXB is consistent with
the existence of an event horizon in BHB, the most extreme strong
gravity prediction of Einstein's General Relativity (see e.g. the
review by Narayan 2003 for other evidence for a black hole event
horizon).

%% file: jet/jet.tex
\section{Linking jets to the hot inner flow}
\label{sec:jet}

Another major area of progress in the last few years has been in
terms of understanding the link between accretion flows and jets.
Like everything else, the radio emission strongly correlates with
spectral state in BHB. The low/hard state has a steady jet, with
radio luminosity $L_R\propto L_x^{0.7}$ (e.g. Corbel et al.\ 2003;
Gallo, Fender \& Pooley 2003). This can be quantitatively matched by
an ADAF-like hot inner flow producing the X-ray emission, and acting
as the base of the jet which produces the correlated radio flux
(Heinz \& Sunyaev 2003). This can also explain the apparently much
weaker radio emission seen in neutron star systems. The ADAF-like
flow is inefficient in BHB, so the X-ray emission is weaker for the
same accretion flow than in neutron stars where the advected energy
is released at the surface (Kording et al.\ 2006).

However, this relation abruptly changes during transitions to the
soft state, where the radio emission is strongly suppressed
(Tananbaum et al.\ 1972; Fender et al.\ 1999b; Corbel et al.\ 2001;
Gallo et al. 2003). The very high/intermediate state is much more
complex, as it is the transition state between the low/hard to
high/soft states, i.e from a bright steady jet to strongly
suppressed radio emission (the `jet line', which seems to correspond
to the ejection of the last remaining portion of the hot inner flow:
Vadawale et al.\ 2001; 2003).  The rapid changes in the jet are not
limited to this apparent collapse.  This state is also characterized
by transient outbursts of the radio emission, some of which can be
directly resolved into bright blobs moving away from the source at
relativistic speeds (e.g. Mirabel \& Rodriguez 1994; Hjellming \&
Rupen 1995; Fender et al.\ 1999a).

All these events can be unified into the picture for the outbursts
outlined above, and given an outline theoretical basis in terms of
our current (limited) understanding of jet production. Meier (2005)
describes the two requirements for producing a jet as a strong,
ordered magnetic field, and some means of getting mass onto this
field. A large scale-height flow offers the means to do both of
these.  Thus any state in which there is a large scale-height inner
flow should produce a jet, while that from a geometrically thin disc
is expected to be much weaker, as observed.  Transient bright events
during the state transitions can be phenomenologically explained if
the jet Lorentz factor increases as the accretion disc moves inwards
(Vadawale et al.\ 2003; Fender, Belloni \& Gallo 2004; 2005). During
the rise to outburst the faster jet catches up with the slower
moving ejected material, and the resulting shocks provide local
energy dissipation and acceleration of particles. This explains why
such transient radio brightenings are {\em not} seen in intermediate
states when the source {\em declines} from high/soft to low/hard
state - the disc is moving outwards, so the jet speed decreases, so
there is no possibility for the slower jet to interact with the
faster one (Fender et al.\ 2005). Again, this can be connected to
the theoretical jet models which predict a terminal speed related to
the escape velocity of the foot-point of the magnetic field. As the
disc moves inwards, compressing the radial extent of the hot flow
and/or anchoring the large scale height magnetic field, the escape
velocity increases so the jet speed increases.

While this gives a very attractive picture, significant problems
remain.  One obvious question is what ultimately powers the jet,
whether it predominantly taps the gravitational potential or whether
it predominantly taps the spin energy, plausibly via the
Blandford--Znajek mechanism. The possibility of spin powered jets
has given rise to persistent speculation in the literature that
relativistic jets require a maximally spinning black hole. This is
open to observational constraints! Spin in BHB is measurable, though
currently controversial. If the BHB have a {\em range} of spins from
$a_*$ = 0.1--0.8 as suggested by accretion disc fitting to the disc
dominated high/soft-state spectra (Davis, Done \& Blaes 2006; Shafee
et al.\ 2006, Middleton et al.\ 2006) then this clearly rules out a
very strong dependence of jet power on spin since they are all
consistent (though with large uncertainties) with the {\em same}
$L_R\propto L_X ^{0.7}$ relation (Middleton et al.\ 2006). This
would favour a predominantly gravity powered jet.

MRI simulations of large scale-height accretion flows offer some
independent suggestions. These include the self-consistent,
magnetically generated stresses and produce jets and outflows
without additional physics. These show in general that the jet has
{\em two} components, a matter dominated, funnel wall jet and an
electromagnetic Poynting flux jet (McKinney 2005; Hawley \& Krolik
2006).  The electromagnetic jet is probably highly relativistic and
is very strongly dependent on $a_*$, indicating that this may be
partly (or perhaps even mostly) powered by the black hole spin
(McKinney \& Gammie 2004).  By contrast, the funnel wall outflow is
less relativistic and is much less dependent on black hole spin
(McKinney 2005; Hawley \& Krolik 2006). While we caution again that
the simulations do not currently include radiation, so cannot yet be
unambiguously connected to observations, it seems that the funnel
wall, matter dominated jet, powered predominantly by the gravity of
the accretion flow, matches rather well to the properties of the
jets in Galactic black hole binaries (see McKinney 2005). By
contrast, the spin powered electromagnetic jet should be strongly
beamed, making it hard to detect unless observed within
5--10$^\circ$ of their rotation axis. There are no known LMXB
systems at such low inclination angles. However, the inner disc can
be misaligned with the binary orbit, in which case the jet {\em may}
be close to the line of sight in the BHB V4641~Sgr (Maccarone 2002).

%% file: timing/timing.tex
\section{Variability Power Spectra}
\label{sec:timing}

The variability power density spectra (PDS) likewise change
dramatically as a function of $L/L_{\rm Edd}$, correlating strongly
with spectral state (see e.g. the reviews by van der Klis 2004 and
Remillard \& McClintock 2006). Here we show how the major features
of the power spectra can be explained in the truncated disc model,
though we also stress that there is much more complexity which is
not currently well understood.

\subsection{PDS from the low/hard (BHB) and island (NS) states}
\label{sec:pdsbhb}

\begin{figure}
\begin{center}
\includegraphics[width=0.9\textwidth,clip,angle=0]
{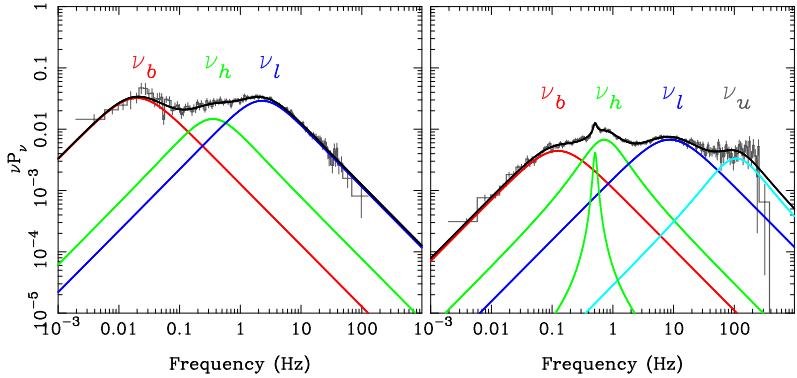}
\end{center}
\caption{The right panel shows a PDS from the hard state of GS~339--4,
together with its decomposition into 3 Lorentzian components
peaking in $\nu P(\nu )$ at $\nu_b$, $\nu_h$, and $\nu_l$,
respectively. The left panel shows a PDS from the neutron star
4U~0614+091, showing the same components (with two Lorentzians around $\nu_h$), but with a
higher frequency component, $\nu_u$, consistent with there
being additional noise power from turbulence at the surface
at the shortest timescales.}
\label{fig:pds_compare}
\end{figure}

One of the most confusing aspects of the power spectral literature
is the multiple components and variety of nomenclature used. Again,
similarly to spectral states, this reflects the evolution of the
field driven by the growing signal--to--noise and concomitant
understanding of the power spectral shapes. The left panel of
Fig.~\ref{fig:pds_compare} shows a power spectrum from GX~339--4 in
one of its hard states. With poor data this can be described as band
limited noise, with a `flat top' in $\nu P(\nu )$, so represents a
power spectrum with equal variability power per decade in frequency,
i.e. $P(\nu)\propto \nu^{-1}$. This extends between a low and high
frequency break, $\nu_{b}$, below which the PDS is $\propto \nu^0$,
and $\nu_{l}$, above which the spectrum steepens to $\propto \nu^2$.
However, with good data it is apparent that this band limited noise
is bumpy, and not well represented by a (twice broken) power law
(Belloni \& Hasinger 1990). Instead, it is much better described by
a series (generally 4--5) of peaked noise components (Lorentzians),
where the peak frequency, width and normalization are free to vary
(Psaltis, Belloni \& van der Klis 1999; Nowak 2000; Belloni, Psaltis
\& van der Klis 2002). In this description the `flat top' is made
from 2--3 Lorentzians, with the lowest and highest frequency
components typically having roughly equal power, giving a broad peak
in $\nu P(\nu)$ between $\nu_b$ (also sometimes called $\nu_{\rm
low}$ or $\nu_0$) and $\nu_l$. In between these there is another
component, peaking at $\nu_h$, which is associated with (or
sometimes replaced by) the low frequency QPO at $\nu_{\rm LF}$,
which can have very complex harmonic structure.  At the highest
frequencies there is sometimes a weak component peaking at $\nu_u$,
forming a small bump in the dimmest hard state, but this is soon
lost in the noise as the source spectrum softens.

Neutron stars in the island state have long been known to show very
similar power spectra to black holes in the hard state (e.g. Yoshida
et al 1993). In the newer language of Lorentzians as described above
then there are clear similarities to the BHB, with these systems
showing the {\em same} sorts of Lorentzian components (see the right
panel Fig.~\ref{fig:pds_compare}), which show the {\em same}
correlations between frequencies as for the black holes (Psaltis et
al.\ 1999; Wijnands \& van der Klis 1999; Belloni et al.\ 2002; van
der Klis 2004). However, there are also clear differences. Their
smaller mass gives faster timescales, but even allowing for this
there is more high frequency power in the neutron stars than in the
black holes (Sunyaev \& Revnivtsev 2000). Fig.~\ref{fig:pds_compare}
shows power spectra from a neutron star and a black hole (which are
matched in $\nu_b$, scaled by the mass difference). It is clear that
the major difference is simply in normalization of the highest
frequency component $\nu_u$. This can be explained as turbulence at
the boundary layer giving additional high frequency noise to excite
whatever resonance produces the component at $\nu_u$ (Sunyaev \&
Revnivtsev 2000).

\subsection{Evolution of the PDS during BHB transitions}

\begin{figure}
\begin{center}
\includegraphics[width=0.9\textwidth,clip,angle=0]
{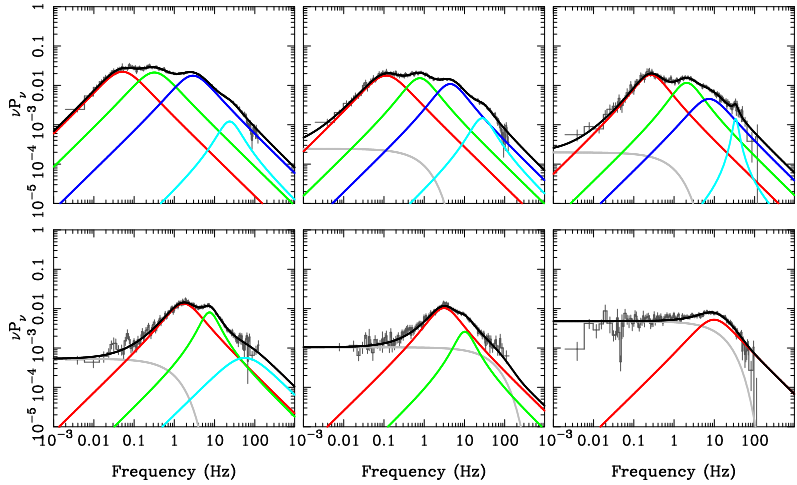}
\end{center}
\caption{The evolution of the PDS in Cyg X-1 as it makes a transition
from the hard to the soft state, together with the individual Lorentzian
components. The major feature is that all frequencies increase
together, and that each component is strongly suppressed as it
approaches $\sim 5$~Hz (after Axelsson et al.\ 2005). The additional
noise component shown as the grey line is not seen in transient BHB.}
\label{fig:pds_trans}
\end{figure}

The evolution of the power spectrum in Cyg X-1 is shown in the
panels of Fig.~\ref{fig:pds_trans} (after Axelsson, Borgonovo \&
Larsson 2005), with these individual Lorentzians superposed. It is
clear that these components are linked together, so that their
frequencies all increase as the spectrum softens from the dimmest to
the brightest hard state through to the soft state (see e.g. van der
Klis 2004). The most obvious correlations are that the LF QPO (or
$\nu_h$ if the QPO is not seen as a clear component) and low
frequency break in the continuum noise power change such that
$\nu_{LF} \sim 5\nu_{b}$ (Wijnands \& van der Klis 1999), while
$\nu_l\sim 10 \nu_{LF}$ (Psaltis et al.\ 1999; Belloni et al.\
2002). There is much less data on the weak high frequency component,
but including results from neutron star systems (see below) gives
$\nu_u\sim 30\nu_l^{1/2}$ (Belloni et al.\ 2002).  However, there
are also systematic correlations in the width and rms of these
components, such that as each component approaches $\sim 5$ Hz its
amplitude and width drops (Pottschmidt et al.\ 2003; Kalemci et al.\
2004; Belloni et al.\ 2005; Axelsson et al.\ 2005; Kalemci et al.\
2006).

Even without a clear model for the origin of the variability, these
data give strong support for the truncated disc model. The existence
of characteristic frequencies in the PDS shows that some particular
radius is picked out by the variability, and that fact these
frequencies {\em move} (by up to a factor 50; Cui et al.\ 1999)
shows that this radius must also move, as predicted by the truncated
disc/hot inner flow models. This is a significant success of these
models, since the changing geometry picture was developed to explain
the spectral transitions.

This can be made more quantitative in a variety of ways. Physically,
the truncated disc model has several characteristic frequencies. The
lowest of these would probably be the viscous timescale of the inner
edge of the thin disc. This acts as a low pass filter, suppressing any
faster fluctuations in the mass accretion rate through the disc due to
the inability of the disc to respond to these (Psaltis \& Norman 2000,
Churazov et al.\ 2001). This predicts $\nu_b\sim \nu_{visc}=(H/R)^{2}
\alpha\nu_\phi$ (see section~\ref{sec:stability}), so for $H/R\sim
\alpha\sim 0.1$ (as appropriate for the thin disc) this gives
$\nu_{LFB}\sim 0.2 (r/6)^{-3/2} (m/10)^{-1}$ Hz, where $m =
M/$M$_\odot$ is the compact object mass. This predicts that $\nu_b$
sweeps from 0.03 to 0.2~Hz as observed during transitions from the
hard to very high/soft state if $r$ decreases from 20 to 6~$r_g$.

A different way to determine the disc truncation radius is through the
low frequency QPO. There is as yet no clear consensus on the origin of
this feature, though the relativistic precession model (Stella \&
Vietri 1998; 1999; Stella, Vietri \& Morsink 1999) has many
attractions (see e.g van der Klis 2004). Nonetheless, whatever the
origin, any characteristic timescale should be longer than the
Keplerian orbital period. The QPO itself is not very
constraining due to the long timescales implied by its low frequency.
However, using $\nu_{LF}$ with the correlations described above to
predict $\nu_u$ even when it is not seen give much tighter constraints
on the disc radius. These relations again imply that the inner disc
radius moves from $\sim 20R_g$ in the faintest hard state where the LF
QPO is seen to $\sim 4-6R_g$ in the weakest very high state (almost
soft state) (Di Matteo \& Psaltis 1999).

Thus two independent ways to use the frequencies contained in the PDS
give the same radius for the inner edge of the truncated disc and
these radii are {\em quantitatively} as predicted by the changing
geometry model for {\em spectral} transitions, with the disc extending
down to the last stable orbit only in the soft state. While there is
as yet no wideranging study of these model predictions across all the
BHB, Chaty et al. (2003) note that the large radius of the truncated
disc inferred from the hard state spectrum of the dim transient XTE
J1118+480 (see Fig.~\ref{fig:ls_disc}) is qualitatively consistent
with the longer characteristic PDS timescales seen in this object.

\begin{figure}
\begin{center}
\includegraphics[width=0.9\textwidth,clip,angle=0]
{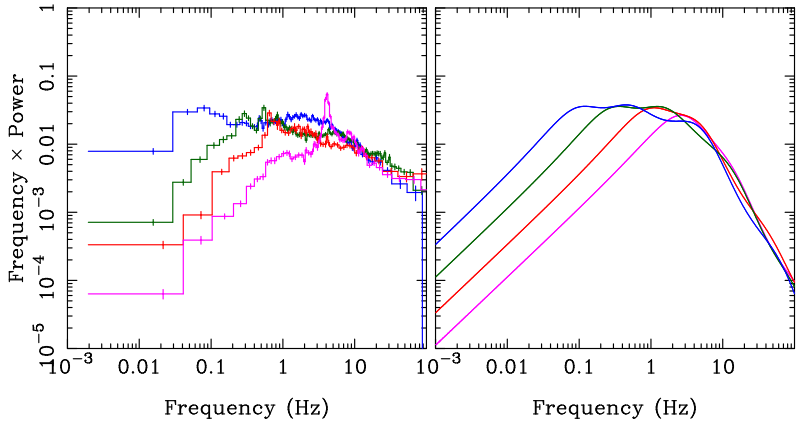}
\end{center}
\caption{Left: a range of PDS from XTE J1550--564 during a
transition. The major features are that the power spectrum narrows
as the spectrum softens, but the power above 5~Hz remains remarkably
constant. Right: a model showing how the PDS evolves if an initial
spectrum of fluctuations (consisting of 4 Lorentzians with
equal power, and correlated frequencies as described in the text) is
filtered through a hot inner flow. The viscous timescale at the last
stable orbit acts as a low pass filter at fixed frequency
$\nu_{\rm max}=5$~Hz. As the initial fluctuations move to higher
frequencies, more and more of the initial power spectrum is suppressed
by the filtering, leading to a narrowing of the observed PDS, and to a
remarkably constant spectrum above $\nu_{\rm max}$.}
\label{fig:lowpass}
\end{figure}

The truncated disc picture can even explain to some extent the
evolution of the PDS {\em shape} during the transitions, not just
the evolution of the characteristic frequencies. The left hand panel
of Fig.~\ref{fig:lowpass} shows that while the frequencies change
dramatically, the power at high frequencies remains remarkably
constant (Gierli{\'n}ski, Niko{\l}ajuk \& Czerny 2007). The
truncated disc geometry implies that whatever fluctuations are
produced at the truncation radius ($\nu_b$, $\nu_{LF}$, and possibly
$\nu_l$ and $\nu_u$) have to propagate down through the hot flow.
This is especially clear in the hard state, where the truncation
radius can be much larger than the last stable orbit, yet the
luminosity must be predominantly produced in the region of maximum
gravitational energy release, i.e.  concentrated close to the black
hole. Thus the truncation radius produces a spectrum of
fluctuations, but these are further modulated by being propagated
through the hot flow. In particular, the viscous timescale at the
inner edge of the hot flow (i.e. the last stable orbit or below)
acts as a low pass filter at frequency $\nu_{\rm max}$. The observed
power spectrum is then given by multiplying the power spectrum of
the intrinsic fluctuations with the power spectrum of the low pass
filter. This acts to form a high frequency `barrier' at $\nu_{\rm
max}$.

The right hand panel of Fig.~\ref{fig:lowpass} shows the effect of
this for $\nu_{\rm max}=5$~Hz on an intrinsic spectrum of 4 equal
rms Lorentzian components with frequencies $\nu_b$, $\nu_{h}$,
$\nu_l$ and $\nu_u$ all correlated as described above. As the disc
inner radius moves inwards, all its characteristic frequencies
increase, but fluctuations faster than the viscous timescale at the
inner radius of the hot flow are strongly damped.  Each component in
turn is affected as it approaches $\nu_{\rm max}$. First the high
frequency section of each Lorentzian is trimmed, making it
asymmetric, and then as its peak frequency moves through $\nu_{\rm
max}$, the rms of that component is strongly suppressed.

Thus filtering by a hot flow keeps the power spectrum above $\sim
$5~Hz remarkably constant, as seen in the data. Since all the other
frequencies are increasing, this leads to a narrowing of the
observed power spectrum, again as seen in the data
(Fig.~\ref{fig:lowpass}). The frequency $\nu_{\rm max}\sim 5$~Hz
also matches quantitative expectations of the viscous timescale at
the last stable orbit of a hot inner flow ($H/R\sim 0.3$). With
$\alpha\sim 0.1$ this gives a predicted dynamical timescale of $\sim
$2~ms, close to the $\sim $4~ms expected for the last stable of a 10
M$_{\odot}$ black hole (Gierli{\'n}ski et al.\ 2007; Done \&
Gierli{\'n}ski 2007).

\subsection{Propagation models}
\label{sec:prop}

The model described above, where a set of characteristic frequencies
are produced at the inner disc edge (perhaps excited by the
turbulence generated at the disc/hot flow interface) and then
propagated down across the hot flow, has yet to be modelled in
detail.  Instead, the models in the literature mostly incorporate
the propagating fluctuation model of Lyubarskii (1997) which makes
the broad range of frequencies to describe the continuum `flat top
noise' from a broad range of radii. In the original model, random
viscosity fluctuations at some radius (such as would be produced by
the time and spatially variable MRI viscosity: Balbus 2005) give
random mass accretion rate fluctuations on the viscous timescale at
that radius. These affect the emission of the ring at that time, but
this is a negligible fluctuation on the total luminosity if the
radius is large. More importantly, this fluctuation in mass
accretion rate affects the next radial zone in, modulating its
random fluctuations. This has the effect that the fluctuations at
each radius are the {\em product} of the fluctuations from all
previous radii, forming a fluctuation power spectrum $P(\nu)\propto
\nu^{-1}$ down to the inner boundary of the flow, so modulating the
region where most of the energy is released. Thus the propagation
models make the broad power spectrum from a broad range of radii,
while model described above gives a broad range of frequencies from
a single radius, and then propagates these down through the extended
hot inner flow.

Nonetheless, these pure propagation models explain another puzzling
aspect of the variability which points to a deep interconnectedness of
its properties. The size of the rms fluctuations, $\sigma$, is
linearly related to the source flux, $F$, such that that $\sigma/F$
remains constant (Uttley \& McHardy 2001). This is more or less
equivalent to saying that the fluctuations have a log-normal
distribution  (Uttley, McHardy \& Vaughan 2005).
There is no way to do this in any model of
variability which takes a {\em sum} of independent events, so all shot
noise models are ruled out (Uttley et al. 2005). This
also rules out self organized criticality models (e.g. Mineshige,
Takeuchi \& Nishimori 1994), despite them also having some radial
connections. In these models the variability is propagated inwards
only once the accretion flow at that radius crosses some critical
threshold in properties. This produces a power law rather than
log-normal distribution of fluctuations, pointing to the propagation
being truly diffusive rather than triggered abruptly as in the SOC
models.

\subsection{Propagation in a truncated disc geometry}
\label{sec:propdisc}

However, it is plain that these models do need to be coupled to the
truncated disc geometry in order to produce the noise power
(including all QPO's) in the {\em Comptonized} spectrum, not in the
disc (Rodriguez et al.\ 2004; Revnivtsev \& Gilfanov 2006;
Sobolewska \& {\.Z}ycki 2006).  Such coupling can give a qualitative
match to the changes in power spectra seen in the hard-to-soft state
transition in Cyg X-1 (Churazov, Gilfanov \& Revnivtsev 2001). It
can also explain the otherwise utterly puzzling behaviour of the
time lags between different frequency bands. Thermal Comptonization
models build up the spectrum from repeated scatterings. To produce
higher energy photons requires more scatterings, so these should lag
behind the lower energy photons by the light crossing timescale.
This is a very short timescale, of order milliseconds for the BHB,
and should be independent of variability timescale for a uniform
region. Instead, the observed lags are smaller for rapid flux
changes, and can be as long as a second (Miyamoto \& Kitamoto 1989).
A non--uniform density profile over an enormous emission region can
match these aspects of the data (Kazanas, Hua \& Titarchuk 1997).
However, as well as being physically unlikely given the small size
scale of the gravitational energy release, this can be ruled out as
it predicts longer timescales for variability at higher energies,
opposite to that observed (Maccarone, Coppi \& Poutanen 2000).
Instead the lags are much more likely to be associated with spectral
variability, where the spectrum evolves from soft to hard (Poutanen
\& Fabian 1999). This can be produced in the propagating fluctuation
model using the {\em same} geometry as sketched in
Fig.~\ref{fig:lowhard}. The fluctuations start at large radii, where
the disc and hot flow overlap. This gives a soft spectrum as there
are many seed photons from the disc (due either to intrinsic
emission or reprocessed hard X-rays) which illuminate the flow.  As
this fluctuation propagates inwards then it goes into the region
where there is no disc underneath the flow, so fewer soft photons
and hence a harder spectra with a smaller amount of reflection
(Revnivtsev, Gilfanov \& Churazov 1999). Thus the spectrum changes
from soft to hard during the rapid variability. This can match the
energy and time dependence of the time lags, as well as the broad
band power spectral shape and rms--flux relation (Kotov, Churazov \&
Gilfanov 2001; Arevalo \& Uttley 2006).

\subsection{Unsolved Problems}
\label{sec:unsolved}

This section has not touched on how the frequencies at the
truncation radius are generated, nor on the more complex behaviour
of the neutron star island--banana transition, where $\nu_l$ and
$\nu_u$ suddenly morph into (or are replaced by?) the `twin peak'
narrow kHz QPO's, nor the weak narrow high frequency (3:2 harmonic?)
QPO's occasionally seen in the BHB in the very high state (van der
Klis 2004) Nonetheless, a model where the truncated disc produces a
spectrum of characteristic frequencies which then propagate into the
hot inner flow to modulate its Comptonized emission plainly captures
the essence of the behaviour.  Further investigations, looking at
modes of this hot flow (Rezzolla et al 2003; Giannos \& Spruit 2004;
Blaes, Fragile \& Arras 2006) and how it interacts with the surface
should surely give insights into these remaining issues.

%% file: winds/winds.tex
\section{Accretion disc winds}
\label{sec:winds}

The previous sections have shown how the combination of a fairly
standard Shakura--Sunyaev disc and a hot inner accretion flow can
explain the major spectral transitions seen in BHB and the disc
accreting NS in the range $10^{-3}< L/L_{\rm Edd}<0.5$. However,
there are also indications that the disc structure {\em changes} at
high mass accretion rates, and that winds become very important.
Thus to understand systems at $L/L_{\rm Edd}>0.5$ we need more than
the simple disc and hot flow picture, we need to understand the
impact of winds and outflows on the disc structure.

In recent years, a growing number of X-ray binaries have been found
to exhibit absorption lines from highly ionized elements, most often
He and H-like Fe at 6.67 and 6.95~keV. These systems range from
microquasars such as GRO~J1655--40 (Ueda et al.\ 1998; Yamaoka et al
2001; Miller et al.\ 2006) and GRS~1915+105 (Kotani et al.\ 2000;
Lee et al.\ 2002) to atolls (see e.g. the review by Diaz Trigo et
al.\ 2006 and references therein) Where multiple absorption lines
are seen then this gives an excellent probe of the physical
conditions in the wind (e.g. Ueda et al.\ 2004; Miller et al.\
2006), and these spectra indicate the presence of significant
amounts of highly ionized material which is generally outflowing at
moderate velocities ($\sim$500~km~s$^{-1}$ in both BHB and NS).

The common features of these systems is that they are all viewed at
fairly high inclination angles: most of the atolls are also dippers
(Boirin et al.\ 2005; Diaz Trigo et al.\ 2006).  Thus the absorption
is almost certainly due to material driven from the accretion disc,
and then photoionized by the strong X-ray illumination from the
innermost regions of the accretion flow. The reprocessed emission
and scattered flux from this extended material can be seen directly
in the accretion disc-corona sources, where the intrinsic X-rays are
obscured (e.g. Kallman et al.\ 2003), but for the majority of highly
inclined sources the wind material is seen in absorption against the
much brighter intrinsic central X-ray source. This persistent
absorption, seen at all orbital phases, can be dramatically enhanced
during `dip' events, where there is more material above the disc due
to the impact between the accretion stream and disc (e.g. Boirin et
al 2005; Diaz Trigo et al.\ 2006).

Another common feature is that these objects are all in the soft or
very high states for the BHB and banana branch for the atolls, so
they all possess an inner disc. This could imply that the inner disc
is the origin of the wind material, or that radiation from the inner
disc is necessary for launching or driving the wind. A less direct
link could be through photoionization, where the soft disc spectrum
means that iron is not completely stripped, rendering the material
invisible. Alternatively, the wind and untruncated inner disc need
not have any causal connection, simply both being consequences of
high $L/L_{\rm Edd}$. Winds are certainly predicted to become
stronger at higher $L/L_{\rm Edd}$, so are most likely to be
observed from high $L/L_{\rm Edd}$ sources i.e. those which also
have an inner accretion disc.

We first review potential theoretical models for the origin of the
wind, and then use the observed properties to show that the most
likely is a thermal wind from the outer accretion disc.

\subsection{Theoretical models of winds}
\label{sec:windtheory}

Accretion discs can potentially power several different kinds of
outflow. Radiation pressure on electrons becomes dynamically
important as $L$ approaches $L_{\rm Edd}$, reducing the effective
gravity by a factor $\sim 1-L/L_{\rm Edd}$. This can be made much
more efficient if the cross-section for interaction between the
matter and radiation is enhanced by line opacity. There are multiple
line transitions in the UV region of the spectrum, so discs where
the luminosity is predominantly in the UV region of the spectrum can
drive a powerful wind at luminosities far below Eddington. Such line
driven disc winds are seen in CV's and are probably also responsible
for the broad absorption line (BAL) outflows seen in AGN (e.g.
Pereyra, Hillier \& Turnshek 2006). However, the disc temperature
for black hole binaries means their spectra peak in the soft X-ray
regime, with little luminosity in the UV region where line driving
is most important (Proga \& Kallman 2002). While these winds in
binaries are generally observed via the lines, the momentum absorbed
in these ion transitions is very small, completely insufficient to
drive the wind.

Another type of outflow from a disc is a thermally driven wind
(Begelman, McKee \& Shields 1983). Here again the central
illumination is important, but the process is less direct.  The
illumination heats the upper layers of the disc to a temperature of
order the Compton temperature, $T_{\rm IC}$. This will expand due to
the pressure gradient, and at large enough radii the thermal energy
driving the expansion is larger than the binding energy, leading to
a wind from the outer disc, while at smaller radii the material
forms an extended atmosphere (corona) above the disc.  Simple
estimates of the transition radius between the static corona and
outflowing wind give $R=3\times 10^4 \cdot (T_{\rm IC}/10^8~{\rm
K})^{-1} R_g $ (Begelman et al.\ 1983), while a more careful
analysis shows this is an overestimate, and that thermal winds are
launched at a radius a factor 5--10 smaller than this (Begelman et
al. 1983; Woods et al.\ 1996). The column density of the wind and
its outflow velocity all increase with luminosity, reaching
$\sim3\times 10^{23}$ cm$^{-2}$ and 720~km s$^{-1}$ for $L/L_{\rm
Edd}=0.3$ at an inclination of $70^\circ$ (Woods et al.\ 1996). At
even higher luminosities the combination of radiation pressure to
thermal driving results in an even more powerful wind (Proga \&
Kallman 2002).

\begin{figure}
\begin{center}
\includegraphics[clip=true,width=0.6\textwidth,angle=0]
{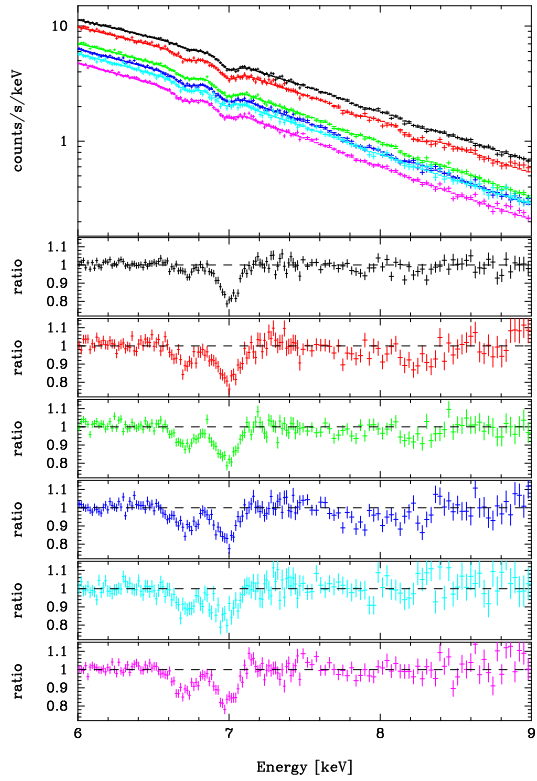}
\end{center}
\caption{Evolution of absorption lines on the outburst decay of 4U 1630--47 (Kubota et al.\ 2006) }
\label{fig:1630abs}
\end{figure}

\begin{figure}
\begin{center}
\includegraphics[clip=true,width=0.6\textwidth,angle=0]
{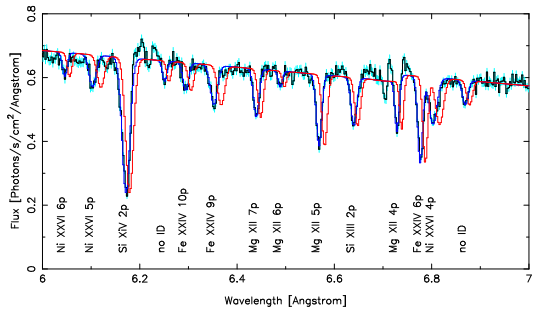}
\end{center}
\caption{Absorption lines found in GRO J1655--40 (Miller et al.~2006)}
\label{fig:1655abs}
\end{figure}

The last type of outflow is a magnetically driven wind. These are
much harder to quantitatively study as the magnetic field
configuration is not known, yet they are almost certainly present at
some level as the underlying angular momentum transport is known to
be due to magnetic fields. Winds and jets are clearly present in
magnetohydrodynamical (MHD) simulations which include these magnetic
stresses self consistently (e.g. Balbus 2005). However, as yet these
calculations generally neglect radiative cooling, so describe hot,
geometrically thick flows rather than the cool, geometrically thin
disc appropriate here. Proga (2000, 2003) make some calculations of
magnetic winds from a geometrically thin disc, but imposed an
external field geometry. The mass loss rates depend on this field
configuration, but in general the winds can be substantially
enhanced, and can be launched from any radius.  This lack of
diagnostic power means that they can be reliably identified only
when all other potential mechanisms for a wind are ruled out.

\subsection{The nature of the wind in XRB}
\label{sec:windobs}

For all sources the inferred properties of the persistent (non-dip)
ionized absorption are very close to those expected from the
detailed thermal models of Woods et al.\ (1996). These models
include calculations of the ionization state expected from the
material, and {\em predict} that iron is predominantly H and He-like
at the highest luminosity, while the absorption also includes lower
ionization species at lower $L/L_{\rm Edd}$.  Such a decrease in the
ionization state of the absorber is seen in the multiple spectra
taken during the decline from outburst of the BHB transient
4U~1630--47 (Kubota et al.\ 2007; see Fig.~\ref{fig:1630abs}). This
all makes a compelling case for this material to be predominantly
formed from irradiation driven mass loss from the disc.

However, there is one potential requirement for significant magnetic
driving from the beautiful Chandra grating data on the ionized
absorber seen in the BHB GRO J1655--40 (Miller et al.\ 2006, see
Fig.~\ref{fig:1655abs}).  Here the launching radius inferred for the
absorber is much smaller than that of the thermal wind, requiring
magnetic fields (Miller et al.\ 2006).  Given the continuity of
properties seen between the wind in GRO J1655--40 and in the other
BHB and NS systems, this would argue for magnetic fields to be
important in all these systems. However, Netzer (2006) shows that
the small launch radius is not a unique interpretation of the data,
so it seems likely that the material is predominantly thermally
driven in all systems, and that its column, outflow velocity and
ionisation increase with $L/L_{\rm Edd}$.

%% file: se/se.tex
\section{Super-Eddington accretion flows}
\label{sec:se}

Winds should become even more powerful as the luminosities approach
(and go beyond) Eddington as radiation pressure reduces the effective
gravity to $1-L/L_{\rm Edd}$. This means the outflows can be launched
from closer to the black hole and this mass loss may also be important
in terms of changing the underlying disc structure. Plainly though
this does {\em not} limit the source luminosity to $\le L_{\rm Edd}$
as both BHB and NS can show luminosities above this. The
super--Eddington BHB include V404 Cyg and V4641 Sgr near the peak of
their outbursts, together with GRS 1915+105 which probably oscillates
between 0.3 and 3~$L_{\rm Edd}$, and (probably) the HMXB SS 433 which
accretes at hyper Eddington rates ($> 1000\times$ higher than those
required to radiate at $L_{\rm Edd}$: Begelman, King \& Pringle 2006;
Poutanen et al. 2007).  For LMXB NS, all the Z source subclass
(Hasinger \& van der Klis 1989) have $L \sim L_{\rm Edd}$, except for
the peculiar Z source Cir X-1 which can reach 10~$L_{\rm Edd}$ (Done
\& Gierli{\'n}ski 2003)!

This handful of sources show some clear trends in their binary
parameters. The disc instability model predicts that the peak
luminosity is roughly proportional to the size of the disc involved in
the outburst (King \& Ritter 1998). Only systems with relatively long
orbital periods have a large enough disc to potentially reach $L_{\rm
Edd}$, as observed (Shahbaz, Charles \& King 1998), and these
generally require an evolved and/or high mass donor in order to fill
the Roche lobe at such separations (King et al.\ 1997).

\subsection{Black hole binaries}
\label{sec:sebhb}

BHB give the cleanest picture of the accretion flow itself, and these
show clearly that there is some sort of instability present above
$L/L_{\rm Edd}\sim 1$. V404 Cyg and V4641 Sgr both showed evidence for
disruption of the accretion flow near the peak of their outbursts at
$L/L_{\rm Edd}\sim 2-3$, with ejected material forming a dense outflow
which completely obscured the source (Tanaka \& Lewin 1995: {\.Z}ycki,
Done \& Smith 1999; Revnivtsev et al 2002).  This dramatic disruption
of the disc may be connected to the transient nature of these sources
rather than showing the quasi--steady nature of the accretion
flow. However the only such flow at Eddington luminosities is
GRS~1915+105 (e.g. Done, Wardzi{\'n}ski \& Gierli{\'n}ski 2004), and
this show unique variability, strongly reminiscent of limit cycle
behaviour (e.g. Belloni et al.\ 2000). If this is a true limit cycle,
then this implies the presence of another stable branch at $L> L_{\rm
Edd}$. One obvious candidate for this is the optically thick advective
branch (see Section~\ref{sec:radpress}). However, the clear signatures
of mass loss seen from this system (Kotani et al.\ 2000; Lee et al.\
2002) also point to additional cooling from winds being important, as
expected for $L \ge L_{\rm Edd}$ (e.g. Shakura \& Sunyaev 1973;
Begelman et al 2006; Ohsuga 2006; 2007; Poutanen et al. 2007). There
are also powerful jet ejections linked to the limit cycle variability,
which can act as another cooling channel (Janiuk, Czerny \&
Siemiginowska 2002). At even higher quasi steady mass accretion rates
(SS~433), the outflow forms an optically thick shroud around the
source, which may control the angle of the precessing jet (Begelman et al.\
2006). We caution that the accretion structure at such high $L/L_{\rm
Edd}$ is almost certainly impacted by the effect of jets and outflows
rather than just being described by standard (with or without
advection) disc equations.

\subsection{Neutron stars}
\label{sec:sens}

The super Eddington NS systems (Z sources) look subtly different from
the atoll systems. Firstly their evolution on a colour--colour diagram
is different. The atolls show a transition to the hard (island) state
at low luminosities. This is not present in the Z sources, as their
high luminosity means they are always in the soft state.  Thus they
can be generally be fit by two components, a disc at low energies and
(optically thick) Comptonization with seed photons from the (unseen)
NS star surface at higher energies. Due to this spectral similarity, Z
sources occupy similar area in the colour-colour diagram to the upper
banana branch of atolls (Done \& Gierli{\'n}ski 2003). However, the
way the two categories of sources move in the diagram is markedly
different. Z sources move faster and make a small (mostly) Z-shaped
track while the atolls move more slowly along the upper banana branch
(e.g.  Hasinger \& van der Klis 1989).

The topological similarity of the Z source track to the full atoll
track makes some sort of truncated disc model very attractive.
However, the accretion flow cannot be optically thin at such high
luminosities, so the disc cannot be truncated by the usual
`evaporation to a hot inner flow' mechanism. Instead the disc could be
truncated by a residual magnetic field which is finally overcome by
the increasing ram pressure of the flow at around Eddington
(Gierli{\'n}ski \& Done 2002a). The idea that Z sources differ from
atolls in more than just accretion rate was suggested at the onset by
Hasinger \& van der Klis (1989).  However, closer examination of the Z
track shows that it can sometimes have a further branches at high
luminosity which look like another Z off the flaring branch
(e.g. Gilfanov et al.\ 2003). If the horizontal to flaring branch
transition is driven by the disc reaching the NS surface, then what
causes the additional branches?  A better model for these very high
mass accretion rate flows might be where the inner disc is truncated
by mass loss processes in a wind and/or jet (Takahashi \& Makishima
2006). Some evidence for this model comes for the fact that spectral
decomposition shows that while the accretion disc luminosity increases
monotonically along the Z track, the boundary layer contribution
decreases along the normal branch, from roughly equal to that of the
disc at the top (close to the horizontal branch) to around zero on the
flaring branch (Done, {\.Z}ycki \& Smith 2002; Revnivtsev \&
Gilfanov 2006; Takahashi \& Makishima 2006). Alternatively, this could
instead indicate that the boundary layer is progressively obscured by
the accretion disc thickening.

The link between Z sources and atolls will soon become much clearer,
with the recent discovery of the first {\em transient} Z source, XTE
J1701--462 (Homan et al.\ 2007). As this source declines from its peak
(it was still bright at the time of writing this review) then it
should transform into a bright atoll (banana branch) and then into a
hard (island) state if mass accretion rate is the only distinction
between Z and atoll sources. The peculiar Z source Cir X-1 is
similarly fading (Saz Parkinson et al. 2003), so again should show
whether the only real difference between atolls and Z sources is their
mean accretion rate, rather than additional complexity such as a
surface magnetic field. This source reached an unprecedented
$\sim$10~$L_{\rm Edd}$ (Done \& Gierli{\'n}ski 2003), where its
spectrum was extremely soft, much softer than expected from the NS
models. This is plausibly due to the emission being thermalized in a
wind from the disc which is so strong that it becomes optically
thick. The observed $\sim 1$~keV temperature from Cir X-1 at
10~$L_{\rm Edd}$ requires reprocessing from a large region, with
radius around 100~$R_g$.  Direct evidence for a wind in this system is
seen through the detection of X-ray P Cygni profiles (Brandt \& Schulz
2000). 

Irrespective of what the Z source spectra imply about the nature of
the accretion flow, it is obvious that these sources {\em do not} show
the same limit cycle instability as GRS~1915+105. However, their
variability on the flaring branch is very rapid and complex, hinting
at some sort of instability. Our understanding of such high mass
accretion rate flows is extremely rudimentary, yet it is such
super-Eddington flows which most probably power the ULX, and certainly
power the growth of supermassive black holes in the early Universe.

%% file: conclusions/conclusions.tex
\section{Conclusions}
\label{sec:conc}

This review illustrates how the multitude of observations of
accreting black holes and neutron stars can fit into a model of a
changing accretion flow structure as a function of mass accretion
rate. The long term light curves of these binary systems give clear
evidence for an outer disc with properties like those predicted by
the time dependent version of the Shakura--Sunyaev disc structure
equations, but the spectra only show evidence for such a disc in the
inner regions at (generally) high luminosity $L/L_{\rm Edd}\ge 0.1$.
At lower luminosities, there is extensive evidence that the inner
disc is simply not present, with the outer disc making a transition
into a hot inner accretion flow. Decreasing this disc truncation
radius with increasing mass accretion rate until the disc extends
down to the last stable orbit gives the major hard--to--soft
spectral transitions seen in these systems. This geometry can also
explain the correlated variability power spectral evolution, with
the decreasing disc radius giving  increasing characteristic
frequencies which are then propagated down through the hot flow, and
hence filtered on the viscous timescale of the hot flow at its
(fixed) inner radius  at (or below) the last stable orbit. Even the
jet behaviour can be (qualitatively) understood in this picture, as
a large scale height flow is probably required for jet formation, so
the collapse of the inner flow as the disc reaches its minimum
radius at the last stable orbit triggers a similar collapse of the
radio emission.

While much can be explained in such models, there is also much left
to be understood. A proper model for the variability, including QPO
formation, is perhaps soon within reach, while detailed predictions
of the jet properties and their impact on the accretion flow may
take somewhat longer. Nonetheless, the full numerical simulations of
the magnetic stresses which form the physical basis for viscosity
are making tremendous progress, and starting to incorporate
radiative processes which mean they can be applied to observations.
Such simulations (the full MRI stresses, plus full radiation
processes) may also point to the nature of the accretion flow at
high luminosities, $\ge 0.5L/L_{\rm Edd}$, where the disc is no
longer a simple Shakura-Sunyaev structure but powers substantial
winds. Such flows are important not only to explain the observations
of high mass accretion rate flows in the local universe (GRS
1915+105, ULX's, narrow line Seyfert 1's) but also have cosmological
significance. These are the flows required in the early Universe to
quickly build up mass into the first quasars, and these winds can
play an important role in quasar feedback on galaxy formation.

While such an optimistic assessment of the potential progress may
seem unrealistic, the tremendous breakthroughs outlined in this
review seemed equally unlikely 10 years ago. Advances in
instrumentation coupled to new theoretical models and improved
numerical simulations hold out a tantalizing glimpse of
understanding accretion flows in strong gravity.

%% file: thanks/thanks.tex
\begin{acknowledgements}
We would like to thank everyone who gave comments on the first draft,
especially Thierry Courvoisier, Tom Maccarone and Andrzej Zdziarski. 
CD and MG acknowledge support from a Particle Physics and Astronomy
Research Council Senior and StandardFellowship, respectively. 
\end{acknowledgements}